\documentclass[aps,prb,floatfix,superscriptaddress,showpacs,amsmath,twocolumn,amssymb,shortbibliography]{revtex4-1}

\usepackage{bm,enumerate,amssymb,amsmath,bbold}
\usepackage{graphicx}
\usepackage{ulem,siunitx}
\usepackage{color}
\usepackage{url,comment,csquotes}
\usepackage[english]{babel}

\newcommand\EatDot[1]{}

\def\normOrd#1{\mathop{:}\nolimits\!#1\!\mathop{:}\nolimits}
\def\tr{\Theta}
\def\opx{{\hat x}}
\def\opy{{\hat y}}
\def\txt12{1/2}

\definecolor{darkgreen}{rgb}{0,0.6,0}
\definecolor{orange2}{rgb}{0.8,0.8,0.2}
\definecolor{grey}{rgb}{0.6,0.6,0.6}
\definecolor{darkblue}{rgb}{0,0,0.7}
\usepackage{subfigure}

\usepackage{hyperref}
\usepackage[all]{hypcap}

\renewcommand{\emph}{\textit}
\date{December 1, 2015}

\begin{document}

\title{Projective symmetry group classification of chiral spin liquids}

\author{Samuel Bieri}

\email{samuel.bieri@alumni.epfl.ch}

\affiliation{Laboratoire de Physique Th\'eorique de la Mati\`ere Condens\'ee, CNRS UMR 7600, Universit\'e Pierre et Marie Curie, Sorbonne Universit\'es, 75252 Paris, France}

\affiliation{Institute for Theoretical Physics, ETH Z\"urich, 8099 Z\"urich, Switzerland}

\author{Claire Lhuillier}

\affiliation{Laboratoire de Physique Th\'eorique de la Mati\`ere Condens\'ee, CNRS UMR 7600, Universit\'e Pierre et Marie Curie, Sorbonne Universit\'es, 75252 Paris, France}

\author{Laura Messio}

\affiliation{Laboratoire de Physique Th\'eorique de la Mati\`ere Condens\'ee, CNRS UMR 7600, Universit\'e Pierre et Marie Curie, Sorbonne Universit\'es, 75252 Paris, France}

\begin{abstract}
We present a general review of the projective symmetry group classification of fermionic quantum spin liquids for lattice models of spin $S=1/2$. We then introduce a systematic generalization of the approach for {\it symmetric} $\mathbb{Z}_2$ quantum spin liquids to the one of {\it chiral} phases (i.e., singlet states that break time reversal and lattice reflection, but conserve their product). We apply this framework to classify and discuss possible chiral spin liquids on triangular and kagome lattices. We give a detailed prescription on how to construct quadratic spinon Hamiltonians and microscopic wave functions for each representation class on these lattices. Among the chiral $\mathbb{Z}_2$ states, we study the subset of U(1) phases variationally in the antiferromagnetic $J_1$-$J_2$-$J_d$ Heisenberg model on the kagome lattice. We discuss spin structure factors and symmetry constraints on the bulk spectra of these phases.
\end{abstract}


\maketitle

\tableofcontents

\section{Introduction}

Ideas for chiral quantum liquids in two-dimensional spin $S=\txt12$ Heisenberg models go back to the early days of frustrated magnetism,\cite{KalmeyerLaughlin87_PRL_59_2095, *KalmeyerLaughlin89_PRB.39.11879, WWZ89_PRB_39_11413, Yang93_PRL.70.2641} and they were motivated to a large extent by the physics of the quantum Hall effect.\cite{PrangeGirvin, Kitaev2003, *Kitaev2006, Schroeter07_PRL.99.097202, Bieri11_CRP_12_332, *Bieri12_ANP_327_959} These exotic spin states respecting spin rotation and lattice translation, but breaking time-reversal and mirror symmetries, are expected to show very unusual physical properties such as chiral edge modes, quantized thermal or spin Hall effects,\cite{Lee10_PRL.104.066403} and bulk excitations with anyon statistics. While historically, a ``macroscopic'' time-reversal breaking was believed to be necessary for exotic phases to emerge (e.g., a uniform magnetic field as in the quantum Hall effect), it was soon realized by Haldane\cite{Haldane88_PRL.61.2015} that net magnetic flux is not a mandatory ingredient. This line of thinking cumulated in the now very active research on topological quantum phases.\cite{HasanKane11_RMP.82.3045, GrafPorta13, ElliottFranz15_RMP.87.137} Recently, chiral spin liquids have regained a lot of attention,\cite{Messio12_PRL.108.207204, *Messio13_PRB_87_125127, Gong2014, HuBeccaSheng_PRB.91.041124, Gong15_PRB.91.075112, SterLauchli15_PRB.92.125122, Bieri15_PRB.92.060407, HeShengChen14_PRL.112.137202, Iqbal15_PRB.92.220404, Nielsen2013, GreiterSchroeterThomale14_PRB.89.165125, Bauer2014, Glazman15_PRL.114.037203, GorohovskyPereiraSela15_PRB.91.245139, MengThomale15_PRB.91.241106, KumarFradkin15_PRB.92.094433, Hickey15_PRB.91.134414, Hickey15_arXiv150908461} partly due to potential realization of such exotic phases in kapellasite and related materials.\cite{Janson08_PRL_101_106403, *Janson09_confProc, Fak12_PRL_109_037208, Bernu13_PRB_87_155107, Kermarrec14_PRB.90.205103, Boldrin_PRB.91.220408}

At the heart of the spin liquid construction is fractionalization of spin in terms of {\it spinons}, i.e., effective low-energy quasiparticles carrying a fractional spin quantum number.\cite{AffleckZouAnderson_PRB.38.745, AffleckMarson_PRB.37.3774, *MarstonAffleck89_PRB.39.11538, Lee_RMP_78_2006, Balents10_Nature_464_199} This is in contrast to {\it magnon} excitations in spin wave theory\cite{MikeCherny13_RMP.85.219} of more conventional long-range ordered phases, which carry integer spin. At a formal level, spin can be written in terms of spinon operators,\cite{partons} as we will discuss below. A physically interesting and highly nontrivial question is whether -- in a concrete spin model -- fractionalized spinons can emerge as quasi-free (i.e., deconfined) excitations at low energy. In the case of confinement (i.e., local binding of spinon pairs), the bound state is nothing but a magnon excitation, and a conventional phase is realized. Deconfined spinons are known to emerge in one-dimensional spin chains,~\cite{ColdeaTennantTsvelik02_PRL.86.1335, Kohno2007, Mourigal_NatPhys2013, GiamarchiBook} but the central question remains whether this effect can carry over to higher dimensions. Two-dimensional quantum spin models\cite{Ran07_PRL.98.117205, *Hermele08_PRB.77.224413, YanHuseWhite_03062011, Depenbrock12_PRL.109.067201, JiangWengSheng08_PRL.101.117203, Jiang12_PRB.86.024424, Iqbal11_PRB.83.100404, Iqbal11_PRB.84.020407, Iqbal13_PRB.87.060405, *Iqbal14_PRB.89.020407, Iqbal15_PRB.91.020402, Motrunich05_PRB.72.045105, *Motrunich06_PRB.73.155115, Lee05_PRL_95_036403, Grover10_PRB_81_245121, Mishmash13_PRL.111.157203, Thomale14_PRB.89.020408, Starykh15_RMP75} and materials\cite{Shimuzi03_PRL_91_107001, Mendels07_PRL.98.077204, deVries08_PRL.100.157205, Han2012, Lee08_Science_321_5894, MendelsBert10, *MendelsBert11} with geometric frustration are strong contenders for such exotic phases. Recently, an interesting study found evidence for spin fractionalization in an organic square-lattice compound at high energy.\cite{DallaPiazza2015} Quantum spin liquids have also been proposed in three-dimensional hyperkagome systems.\cite{Yoshihiko07_PRL.99.137207, Zhou08_PRL.101.197201, Lawler08_PRL.101.197202}

The projective symmetry group~(PSG) classification was introduced by Wen\cite{Wen02_PRB.65.165113, *wenOrig} on the square lattice for so-called {\it symmetric} liquids, i.e., spin phases that do not break any lattice symmetry, spin rotation, nor time reversal. In essence, the PSG classification seeks to list all possible classes of lattice symmetry representations in the enlarged Hilbert space of fractionalized spinons.\cite{mps_spt} The discrete (and finite) number of these symmetry representations is then an enumeration and characterization of possible QSL phases. Wen's original work used a fractionalization in terms of {\it fermionic} spinons (also known as ``Abrikosov fermions'').\cite{Abrikosov67} An extension to the anisotropic triangular lattice was later addressed,\cite{ZhouWen02} but only recently PSG classifications of symmetric liquids were published for honeycomb\cite{LuRan_PRB.84.024420, YouKimchiVishw12_PRB.86.085145} and kagome lattices.\cite{LuRanLee11_PRB.83.224413}

Other extensions have appeared in the literature.\cite{srbroken} Wang {\it et al.}\cite{WangVish06_PRB.74.174423, Wang10_PRB.82.024419} performed a classification of symmetric spin phases in the case of {\it bosonic} fractionalization\cite{Sachdev92_PRB.45.12377, Tay11_PRB.84.020404, LiBeccaHuSorella12_PRB.86.075111} (so-called ``Schwinger bosons") on triangular, kagome, and honeycomb lattices. To some extent, this problem is simpler than the fermionic one, because the emergent gauge symmetry is U(1) instead of SU(2). When bosonic spinons condense, they give rise to conventional N\'eel phases. Otherwise, bosonic liquids always exhibit a spin gap and $\mathbb{Z}_2$ gauge fluctuations. A PSG classification of chiral spin liquids within Schwinger boson theory has been published by two of us and Misguich.\cite{Messio13_PRB_87_125127, bfrelation}

The principal goal of our paper is to present the general theory of the projective symmetry group (PSG) classification using fractionalization with fermionic spinons, in the case when lattice symmetries and spin rotation are preserved, except possibly some point group symmetries and time reversal. Here, we generalize in a systematic way the notion of {\it symmetric} spin liquids to the one of {\it chiral} spin liquids~(CSLs) within the parton construction.\cite{partons} In the latter case, we distinguish between {\it Kalmeyer-Laughlin} CSLs that break all reflection symmetries of the lattice, and {\it staggered flux} CSLs that break lattice rotation, up to time reversal. As an application of the general formalism, we list all quantum spin liquids for the triangular and kagome lattices. In these examples, we consider phases potentially realized in Heisenberg models with exchange interactions up to {\it third} lattice neighbors. Results from various approaches have recently suggested that novel chiral spin states can be expected in the presence of such long range interactions on those lattices.\cite{Domenge05_PRB.72.024433, Janson08_PRL_101_106403, Messio11_PRB_83_184401, Messio12_PRL.108.207204, Mishmash13_PRL.111.157203, Imada14_JPSJ.83.093707, Gong2014, SterLauchli15_PRB.92.125122, Bieri15_PRB.92.060407, Gong15_PRB.91.075112, HuBeccaSheng_PRB.91.041124, LiBishopCampbell15_PRB.91.014426, Saadatmand15_PRB.91.245119, ShuWhite15_PRB.92.041105, HuSheng15_PRB.92.140403} We hope that our exhaustive listing may trigger further investigations of microscopic spin models, potentially identifying some of the classified states as viable ground state candidates.

This paper is meant to be largely self-contained in its core results. Some of the presented material (especially in Sec.~\ref{sec:frac}) may therefore be known to specialists, and is sometimes tacitly assumed in publications. To our knowledge, however, the explicit and general presentation of this paper is new, and can therefore be useful to a wider audience. We also provide a list of references for further reading, and we comment on recent developments in the field. For example, {\it symmetric} quantum spin liquids are merely special cases in our general framework, and their classification is included here. We take this opportunity to correct incomplete PSG classifications of symmetric quantum spin liquids on the triangular lattice that have recently appeared in the literature.\cite{Lu15, Mei15}

We also discuss some general properties of the SU(2) gauge fluxes characterizing a spin-rotation symmetric QSL that seem to be new. In the fully gauge invariant formalism, these fluxes have a far richer structure than the familiar U(1) gauge fluxes (e.g., of electromagnetism). In Sec.~\ref{sec:Pop}, we derive the spin order parameter corresponding to the SU(2) gauge flux. For three-site loops, the same result is obtained in the U(1) formalism,\cite{WWZ89_PRB_39_11413} but it differs for higher-order loops. In Sec.~\ref{sec:theoryPSG}, we further discuss symmetry constraints on the gauge flux, which turn out to depend on the projective representation class. In contrast to the simpler U(1) case, SU(2) gauge fluxes are not trivially additive. That is, the total flux angle through a large lattice loop is, in general, not the sum of fluxes through elementary plaquettes. This fact is responsible for the absence of a ``CPT Theorem'' in $\mathbb{Z}_2$ quantum spin liquids, i.e., reflection symmetry combined with time reversal may be broken (even if spin rotation is respected). We also comment on the Chern number that can be nontrivial only in the case of Kalmeyer-Laughlin, but must vanish in staggered flux CSL states.

This paper is organized as follows. In the next section, we review some notation and results on the fermionic fractionalization of spin $S=\txt12$. We introduce quadratic spinon Hamiltonians, and their characterization by SU(2) gauge fluxes and the invariant gauge group~(IGG). In Sec.~\ref{sec:theoryPSG}, we present the general theory of projective symmetry representations, and the constraints they impose on quadratic spinon Hamiltonians and on fluxes. In Secs.~\ref{sec:trgPSG} and \ref{sec:kagPSG}, we exemplify these theoretical results in the case of triangular and kagome lattices, respectively. We list all possible symmetric and chiral spin liquids, and we give concrete recipes on how to construct corresponding quadratic Hamiltonians. We discuss some special cases that are known in the literature. The reader primarily interested in the list of chiral spin liquids on these lattices may directly go to Secs.~\ref{sec:invPSGtrg} or \ref{sec:invPSGkag}, respectively. Finally, in the remainder of Sec.~\ref{sec:kagPSG}, we present a microscopic quantum phase diagram for the antiferromagnetic $J_1$-$J_2$-$J_d$ Heisenberg spin model on the kagome lattice, and we relate to known results. We discuss equal-time spin structure factors and symmetry constraints on the spinon spectra for some of the found QSL phases.

\section{Spin fractionalization}\label{sec:frac}

Let us introduce the fermionic fractionalization of spin $S=\txt12$ operators, and discuss the resulting emergent SU(2) gauge structure. Related fractionalization schemes have been discussed for higher values of spin,\cite{Lui10_PRB.82.144422, *Liu10_PRB.81.224417, ZhangWang06, Bieri12_PRB_86_224409, Cenke12_PRL_108_087204, Wang15_PRB.91.195131} but systematic classification are open problems in these cases.

The spin one-half operator $S^a$ ($a=1,2,3$; or $a = x,y,z$) can be written in terms of {\it two} flavors of complex fermions, ${\bm f} = (f_\alpha) = (f_\uparrow, f_\downarrow)^T$,
as
\begin{equation}\label{eq:spinrep}
  2 S^a = {\bm f}^\dagger \sigma_a {\bm f}\,,
\end{equation}
where $\sigma_a$ are Pauli matrices.\cite{AffleckZouAnderson_PRB.38.745, AffleckMarson_PRB.37.3774, *MarstonAffleck89_PRB.39.11538} The fermions $f_\alpha$ are called {\it spinon operators}. Note that Eq.~\eqref{eq:spinrep} is only formal, meaning that the operators on both sides follow the same SU(2) commutation relations.
However, they act in different Hilbert spaces: spin space is $\mathbb{C}^2$, while the fermionic Fock space is four dimensional. We call
\begin{equation}
  {\bm f} = (f_\uparrow, f_\downarrow)^T
\end{equation}
a {\it spin doublet} because of its transformation under SU(2) spin rotation as ${\bm f} \mapsto U {\bm f}$.

It follows from Eq.~\eqref{eq:spinrep} that ${\bm S}^2 = \frac{3}{4}\, n (2-n)$,
where $n = {\bm f}^\dagger {\bm f}$ is the spinon occupation number. For spin one-half, we therefore see that the filling must be $n = 1$. States with other fillings, $n=0$ or $n=2$, lead to ${\bm S}^2 = 0$. Henceforth, we will call these fermionic states {\it unphysical}, because they do not correspond to spin states.

The requirement $n=1$ is only one of three equivalent ways to specify the physical spin space. They are
\begin{subequations}\label{eq:constr}
\begin{align}
  {\bm f}^\dagger {\bm f} - 1 = 0\,,\\
  {\bm f}^\dagger \varepsilon {\bm f}^* - {\bm f}^T\varepsilon{\bm f} = 0\,,\\
  i ({\bm f}^\dagger \varepsilon {\bm f}^* + {\bm f}^T\varepsilon{\bm f}) = 0\,,
\end{align}
\end{subequations}
with $\varepsilon = i\sigma_2$ the antisymmetric tensor. In the following it will be convenient to introduce a {\it gauge doublet}
\begin{equation}
  {\bm \psi} = (f_\uparrow, f_\downarrow^\dagger)^T\,.
\end{equation}
In analogy with Eq.~\eqref{eq:spinrep}, we define
\begin{equation}\label{eq:constrpsi}
  2 G^a = {\bm\psi}^\dagger\sigma_a{\bm\psi}\,,
\end{equation}
and the constraints \eqref{eq:constr} are elegantly written as $G^a = 0$.

\subsection{Emergent SU(2)$_\text{g}$ symmetry}

The enlarged fermionic Hilbert space leads to additional internal symmetries that are not present in spin space.\cite{AffleckZouAnderson_PRB.38.745} For example, the U(1) phase of the spinon is clearly arbitrary, and $f_\alpha\mapsto e^{i\theta}f_\alpha$ does not affect the spin operator in \eqref{eq:spinrep}. However, in the fermionic representation of spin $S=\txt12$, there is a further particle-hole redundancy. Due to anticommutation, it is easy to see that a transformation $f_\sigma \mapsto f_\sigma\cos\varphi + \sigma f_{\bar\sigma}^\dagger \sin\varphi$ does not affect the form \eqref{eq:spinrep} of the spin operator. Note that this symmetry is absent in bosonic fractionalization schemes.

Since these transformations do not commute, a particle-hole transformation can be preceded and followed by a phase change. This is compactly written in terms of the gauge doublet ${\bm \psi}$ as
\begin{equation}\label{eq:gaugetr}
    {\bm \psi} \mapsto e^{i\theta\sigma_3} e^{i\varphi\sigma_2} e^{i\psi\sigma_3} {\bm \psi} = g {\bm \psi}\,,
\end{equation}
and $g$ is an SU(2) matrix. We call this a ``gauge transformation'' or SU(2)$_\text{g}$, since it is {\it local}, i.e., it can be performed independently on each site of a lattice. Note that the constraint $(G^a)$ in Eq.~\eqref{eq:constrpsi} transforms as a real vector under SU(2)$_\text{g}$, while the spin $(S^a)$ is gauge invariant. Conversely, it is easy to see that $(G^a)$ is spin-rotation invariant, while $(S^a)$ transforms as a vector.

The additional gauge redundancy in spinon space means that there is some freedom in how physical (spin) symmetries act in the spinon Hilbert space. A symmetry transformation -- say $x$ -- may be accompanied by an SU(2) gauge transformation $g_x$. However, this choice is not arbitrary, since the gauge transformations must respect the algebraic relations among symmetry transformations. In mathematical terms, we say that the symmetry group is represented (projectively) in the spinon Hilbert space. This is at the core of the PSG classification and we will discuss it in more details later. In the following, we introduce the representations of symmetries that act on a single site.

\subsection{Time reversal and spin rotation}

The antiunitary time reversal transformation $\tr$ inverts the spin direction, ${\bm S}\mapsto -{\bm S}$. For spin-$\frac{1}{2}$ operators, time-reversal is implemented as $\tr\!: {\bm f}\mapsto \varepsilon {\bm f}$ or ${\bm \psi}\mapsto \varepsilon {\bm \psi}^*$ in terms of the gauge doublet ($\varepsilon = i\sigma_2$). However, in the present context it is convenient\cite{Wen02_PRB.65.165113} to supplement time reversal by a (particle-hole) gauge transformation $g = \varepsilon^T$, such that
\begin{equation}\label{eq:TRbare}
  \tr\!: {\bm \psi}\mapsto {\bm \psi}^*
\end{equation}
or ${\bm f}\mapsto {\bm f}^*$. An advantage of this choice is that time reversal and gauge transformations \eqref{eq:gaugetr} manifestly commute: $\tr\circ g = (\varepsilon^T g^* \varepsilon) \circ \tr = g\circ \tr$; (acting to the right).\citep{gaugeTransfBoson} Note that the choice \eqref{eq:TRbare} is only a convenient starting definition, and additional gauge transformations (denoted by $g_\tr$) may be associated with time reversal. However, for chiral spin liquids, we find that there is generally no relevant freedom in the representation of time reversal. This point is discussed in more detail below.

As discussed before, spin rotation is implemented in spinon space as ${\bm f}\mapsto U{\bm f}$, where $U$ is the SU(2) rotation matrix. Recently, it was realized that spin rotation may also be implemented {\it projectively}, with an associated gauge transformation $g = U$ that is ``locked in'' with the SU(2) spin rotation.\cite{Chen12_PRB.85.094418} This interesting possibility leads to a new class of {\it Majorana} spin liquids and may shed light on alternative fractionalization schemes.\cite{Biswas11_PRB.83.245131, HerfurthKopietz13_PRB.88.174404} In the present paper, we restrict ourselves to the case when spin rotation is represented linearly in spinon Hilbert space (i.e., with trivial gauge transformation).

\subsection{Quadratic spinon Hamiltonians}\label{sec:h0}

A main goal of the PSG construction is to investigate quantum Heisenberg models $H = \sum_{i,j} J_{ij} {\bm S}_i\cdot{\bm S}_j$ on frustrated lattices, where fractionalized quantum phases are expected to arise. However, replacing the spin representation Eq.~\eqref{eq:spinrep} in the Heisenberg model results in {\it quartic} spinon interaction terms, and not much is gained. Progress can be made by mean-field decoupling the spinons through a Hubbard-Stratonovich transformation and using a path-integral approach.\cite{AffleckMarson_PRB.37.3774, *MarstonAffleck89_PRB.39.11538, AuerbachLarson91_PRB.43.7800} To lowest order (i.e., at a saddle point), these approximations produce quadratic spinon theories that are then solvable. In this paper, we do not want to put emphasis on this approach. Instead, we directly go to the quadratic spinon theory. {\it A posteriori}, such a theory may be justified to describe the low-temperature phase of a particular microscopic spin model, e.g., by using variational wave functions, as we will describe below. Alternatively, the procedure can be viewed as a classification of possible symmetric saddle points for Heisenberg models.

The PSG allows to classify and construct quadratic spinon Hamiltonians $H_0$ that respect all or some symmetries of a given spin lattice model. Such a spinon Hamiltonian is conveniently written in terms of the gauge doublet ${\bm \psi}_j$ as\cite{sumConvention}
\begin{equation}\label{eq:h0}
  H_0 = \sum_{i,j} {\bm \psi}_i^\dagger u_{ij} {\bm \psi}_j + \text{H.c.} + \sum_j \lambda^a_j {\bm \psi}_j^\dagger \sigma_a {\bm \psi}_j\,.
\end{equation}
In the path-integral approach, the three real parameters $\lambda^a_j$ are Lagrange multipliers, enforcing the constraints \eqref{eq:constr}. In the present context, they correspond to on-site spinon chemical potentials ($\lambda^z$) and complex $s$-wave pairing terms ($\lambda^x + i \lambda^y$).

In general, the link matrices can be written as $u_{ij} = u^\mu_{ij} \tau_\mu$, with $(\tau_\mu) = (i \mathbb{1}_2, \sigma_a)$ and $u^\mu_{ij}$ are four complex parameters on each link. Without loss of generality, we choose $[u_{ij}]^\dagger = u_{ji}$. Equation~\eqref{eq:h0} is the most general quadratic Hamiltonian invariant under global spin rotations around $S^z$. Such a rotation acts as $\psi_j\mapsto e^{i\alpha}\psi_j$, and this is obviously a symmetry of \eqref{eq:h0}. In fact, {\it real} parameters $u^\mu_{ij}$ correspond to singlet, while {\it imaginary} $u^\mu_{ij}$ correspond to triplet hopping and pairing terms.\cite{Momoi09_PRB.80.064410, *Momoi11_PRB.84.134414, DoddsBhattaYBK13_PRB.88.224413, Reuther14_PRB.90.174417}

In this paper, we focus on the case when the full SU(2) spin rotation symmetry is unbroken. To see that real $u^\mu_{ij}$ conserve spin rotation, we may consider the generator around $S^y$, ${\bm f}_j\mapsto\varepsilon{\bm f}_j$. Under this transformation, the gauge doublet goes ${\bm\psi}_j\mapsto\varepsilon{\bm\psi}^*_j$, hence ${\bm\psi}_i^\dagger u_{ij}{\bm\psi}_j + \text{H.c.} = u^\mu_{ij}({\bm\psi}_i^\dagger\tau_\mu {\bm\psi}_j + {\bm\psi}_j^\dagger\tau_\mu^\dagger{\bm\psi}_i) \mapsto {\bm\psi}_i^\dagger u_{ij} {\bm\psi}_j + \text{H.c.}$ is invariant. Here, we have used the fermionic anticommutation and $\tau_\mu^* = \varepsilon \tau_\mu\varepsilon$. Similarly, it is clear that imaginary $u^\mu_{ij}$ change sign under this spin rotation, so they correspond to triplet terms.

Particular sets of link and on-site parameters $u = [u_{ij}, \lambda_j] = [u^\mu_{ij}\tau_\mu, \lambda^a_j\sigma_a]$ are called a mean-field {\it ansatz} (or {\it ans\"atze} for plural). From now on, we restrict ourselves to real $u^\mu_{ij}$ with full SU(2) spin rotation symmetry.\cite{tripletU} In the widely used notation, the real parameters $u^\mu_{ij}$ are written as $(u^\mu_{ij}) = (\xi_{ij}^2, \Delta^1_{ij}, \Delta^2_{ij}, \xi^1_{ij})$, where $\xi_{ij} = \xi^1_{ij} + i \xi^2_{ij}$ are complex hopping, and $\Delta_{ij} = \Delta^1_{ij} + i \Delta^2_{ij}$ singlet pairing amplitudes on the link $(i,j)$. In this language, the ansatz reads
\begin{equation}\label{eq:umf}
  u_{ij} = \left(\begin{array}{cc}
    \xi_{ij} & \Delta_{ij}\\
    \Delta_{ij}^* & -\xi_{ij}^*
  \end{array}\right)\,,
\end{equation}
and we have $\text{det}[u_{ij}] = - |\xi_{ij}|^2 - |\Delta_{ij}|^2$.

There are two important aspects of the spinon Hamiltonian $H_0$ in Eq.~\eqref{eq:h0}. It is either viewed as a low-energy effective theory for quantum spin phases, or it may serve as a tool for constructing microscopic wave functions for rigorous variational energy calculations in spin models. Either way, {\it physical} properties of a spin phase specified by $H_0$ are always independent of the chosen gauge. We denote the set of lattice gauge transformations by ${\mathcal G} = \{ g \}$ with $g = \otimes g_j$. The SU(2)$_\text{g}$ transformations $g_j$ act independently on each site as ${\bm\psi}_j \mapsto g_j{\bm\psi}_j$. In terms of the ansatz $u$, the elements of ${\mathcal G}$ act as
\begin{equation}\label{eq:gaugetru}
  g\!: u = [u_{ij}; \lambda_j] \mapsto g(u) = [g_i^\dagger u_{ij} g_j; g_j^\dagger \lambda_j g_j]\, .
\end{equation}
Different ans\"atze are therefore unitary equivalent under gauge transformations. For example, the spectrum of $H_0$ is gauge invariant and therefore a characteristic of an equivalence class. 

To construct microscopic spin wave functions from $H_0$, we proceed by taking the ground state $|\psi_0(u)\rangle$ of $H_0$ (or excited states) and remove unphysical components by applying the {\it Gutzwiller projector} $P_G = \prod_j n_j[2 - n_j]$,
\begin{equation}\label{eq:psi}
  |\psi(u)\rangle = P_G|\psi_0(u)\rangle\, .
\end{equation}
In the case of fermionic spinons, expectation values in such wave functions can efficiently be computed numerically using variational Monte Carlo~(VMC) techniques.\cite{Gros88_PRB_38_931, *Gros89_AnnPhys_189_53, EdeggerMuthuGros07} This works best for singlet wave functions, because only Slater determinants need to be evaluated. In the case of triplet pairing terms, more resourceful calculation of Pfaffians is generally required.\cite{Bouchaud88_jpa-00210729, *Bouchaud87_0295-5075-3-12-005, Bajdich06_PRL.96.130201, *Bajdich08_PRB.77.115112, WeberGiamarchi06_PRB.73.014519, Bieri12_PRB_86_224409} In analogy with Laughlin states for the quantum Hall effect, these wave functions can be used in variational investigations of actual lattice spin models. Apart from energetics, various other physically interesting properties may be calculated from the projected wave function, such as static or dynamic spin structure factors, excitation gaps, modular matrices, etc.\cite{GiamarchiLhuillier93_PRB.47.2775, Yunoki05_PRB.72.092505, Bieri07_PRB_75_035104, TanWang08_PRL.100.117004, LiYang10_PRB.81.214509, TaoLi13, Chou20131589, DallaPiazza2015, Zhang12_PRB.85.235151, HuBeccaSheng_PRB.91.041124}

The {\it invariant gauge group}~(IGG) is an important concept in the phenomenology of quantum spin liquid phases when we view $H_0$ as a low-energy effective theory. It is defined as the subgroup of gauge transformations $\mathcal{G}$ that leave the spinon Hamiltonian $H_0$ invariant, i.e., $g(u) = u$ for all $g\in$~IGG$_u$. IGG$_u$ always contains $\mathbb{Z}_2$ as a subgroup since global transformations $g_j = \pm\mathbb{1}_2$ leave any ansatz invariant. However, IGG$_u$ may be bigger and contain global U(1) or even SU(2) transformations. The IGG$_u$ characterizes the emergent low-energy gauge fluctuations in the effective theory. For example, if $\mathcal{G}$ (sometimes called ``high-energy'' gauge group) is completely broken to $\mathbb{Z}_2$ in the mean-field state, the emergent gauge bosons are gapped and expected to be irrelevant at low energy. However, in liquids with IGG$_u =$~U(1) or SU(2), gapless gauge bosons (``photons'' or ``gluons'') are present and may strongly affect the low-energy physics. Depending on the IGG$_u$ of its ansatz, a spin liquid is said to have a $\mathbb{Z}_2$, U(1), or SU(2) gauge structure.\cite{Wen02_PRB.65.165113, Lee_RMP_78_2006} In the first case, it is called a ``$\mathbb{Z}_2$ QSL state'' or simply ``$\mathbb{Z}_2$ liquid'', etc.

As mentioned previously, a spinon Hamiltonian may respect space group symmetries or time reversal when those transformations are accompanied by appropriate SU(2) lattice gauge transformations in $\mathcal{G}$. The symmetries of an ansatz $u$ along with gauge transformations, SG$\ltimes\mathcal{G}$, is called the {\it invariant} projective symmetry group, and denoted by PSG$_u$. The PSG$_u$ is a way to distinguish between phases ($H_0$) that have the same physical symmetries. In Sec.~\ref{sec:theoryPSG}, we will explain the classification of those symmetry representations, without making reference to any ansatz $u$ (this is the {\it algebraic} PSG). The corresponding ans\"atze are subsequently constructed.

\subsection{SU(2) gauge flux}\label{sec:Pmf}

A useful way to characterize quadratic spinon Hamiltonians $H_0$ is by their SU(2) {\it gauge fluxes}.\cite{AffleckZouAnderson_PRB.38.745, Wen02_PRB.65.165113, Lee_RMP_78_2006, Lawler08_PRL.101.197202, Iqbal11_PRB.83.100404, Clark13_PRL.111.187205, LiYuLi15_NJP} Given an ansatz $u$, we associate the SU(2) flux with oriented lattice (Wilson) loops $\mathcal{C}$ starting from a base site $j$. The SU(2) flux is defined as the matrix product of $u_{ij}$ over the sites of the loop,
\begin{equation}\label{eq:Pmf}
  P_j = \prod_{\mathcal{C}} u_{kl} = u_{j j_2} u_{j_2 j_3}\ldots u_{j_{q} j}\,.
\end{equation}
Lattice gauge transformations \eqref{eq:gaugetru} cancel out on the intermediate sites, but the SU(2) flux depends on the gauge of the base site as $P_j \mapsto g_j P_j g_j^\dagger$. However -- as we will discuss in more details later -- trace and determinant of $P_j$ are gauge invariant and independent of the base site.

An interesting use of the SU(2) flux is the determination of the invariant gauge group of an ansatz. As discussed previously, IGG$_u$ contains important information about the low-energy degrees of freedom and gauge structure of the theory. To determine the IGG$_u$, we may proceed in the following way. Let us pick a field $u_{12}$ on the link $(1,2)$. For gauge transformations in IGG$_u$, we have $g_1 u_{12} g_2^\dagger = u_{12}$ by definition. This equation always has the solution $g_2 = u_{21} g_1 u_{12}$, where we assume $u^\dagger u = \mathbb{1}_2$ for simplicity. The same argument on a third site gives $g_3 = u_{32} g_2 u_{23} = u_{32} u_{21} g_1 u_{12} u_{23}$, etc. We can propagate the gauge transformation in IGG$_u$ to any site, $g_q = u_{q q-1}\ldots u_{32} u_{21} g_1 u_{12} u_{23}\ldots u_{q-1 q}$. When the path is closed to a loop, the SU(2) flux matrix appears. For consistency reason, the gauge transformation on the first site must again be the same, and we have the constraint $g_1 = P_1^\dagger g_1 P_1$, or
\begin{equation}\label{eq:Pcommut}
  [g_1, P_1] = 0\,,
\end{equation}
for all $g_1\in\text{IGG}_u$. In principle, all flux matrices $P_1$ can be calculated for a given ansatz. The IGG$_u$ on this site is then the subgroup of SU(2) that commutes with the flux matrices for all paths starting from that site.\cite{fluxLambda}

A sufficient condition for all flux matrices to commute with a given gauge transformation is of course that all matrices $u_{ij}$ commute with this gauge transformation. It is easy to see that there is always such a gauge. This condition is simpler to check than Eq.~\eqref{eq:Pcommut}, and it is what we do in practice.

Next, we discuss some properties of the ansatz matrix $u_{ij} = u^\mu_{ij} \tau_\mu$. When $u^\mu_{ij}$ are real for an SU(2) spin rotation invariant ansatz, it can be written as
\begin{equation}\label{eq:umfphase}
  u_{ij} = \rho_{ij}\, \sigma_3 \exp\{i\varphi\, {\bm n}\cdot{\bm \sigma}\}\,,
\end{equation}
where $\rho_{ij}$ and $\varphi$ are real numbers, and ${\bm n}$ is a unit vector. Hence, $\text{det} [u_{ij}] = -\rho_{ij}^2$ and $u_{ij}^\dagger u_{ij} = \rho_{ij}^2 \mathbb{1}_2$. Using these properties of $u_{ij}$, we see that $P_j^\dagger P_j =  \mathbb{1}_2\,|\text{det}[P_j]|$, and $P_j$ is (proportional to) a unitary matrix. It can therefore be written as
\begin{equation}\label{eq:Pfluxmf}
  P_j = \rho\, g_j [\cos\theta + i\sigma_3\sin\theta] (\sigma_3)^q g_j^\dagger\,,
\end{equation}
with $g_j$ some gauge transformation. Here, $\rho$ is real, and $q$ is the number (or parity) of sites in the loop $\mathcal{C}$.
It follows that $\text{det}[P_j] = (-)^q\rho^2$, and the trace of the SU(2) flux is given by
\begin{equation}\label{eq:tracetheta}
  \text{Tr} P_j = \left\{ \begin{array}{ll}
    2\rho \cos\theta, & q \text{ even,}\\
    2\rho\, i \sin\theta, & q \text{ odd.}
  \end{array}
  \right.
\end{equation}
Hence, the trace is real for even-site loops, while it is imaginary for odd loops. For a given loop, the parameter $\rho$ (as long as $\rho\neq 0$) can be changed by an irrelevant scaling $u\mapsto \alpha u$, so it has no intrinsic physical meaning. However, the angle $\theta$ is an important gauge-invariant characteristics of the SU(2) flux.

Let us contrast some properties of the SU(2) flux with the more familiar case of a U(1) gauge flux. When the spinon Hamiltonian $H_0$ only contains hopping terms and no pairing, we obviously have IGG$_u=$~U(1). In fact, whenever IGG$_u$ is U(1), there is always a gauge in which the ansatz is pure hopping. In this case, we write $\prod_\mathcal{C} \xi_{kl}\propto \exp(i\phi)$, where $\xi_{ij}$ are hopping amplitudes. The flux $\phi$ is invariant under local U(1) transformations, and it may be identified with $\theta$ in Eq.~\eqref{eq:tracetheta}. However, we emphasize that only $\cos\phi$, resp.\ $\sin\phi$ are invariant under the full SU(2)$_{\text{g}}$ lattice gauge group, and not the U(1) flux $\phi$ itself. For example, a global particle-hole transformation changes the sign of $\phi$ for even-site loops, and $\phi \mapsto \pi - \phi$ for odd loops.

Another interesting property of the SU(2) flux angle $\theta$ in Eq.~\eqref{eq:tracetheta} is that it is {\it nonadditive} in general. Consider two fluxes $P^\mathcal{A}_1$ and $P^\mathcal{B}_1$ on the loops $\mathcal{A} \neq \mathcal{B}$ starting from the same base site. Then, the flux on the combined loop $\mathcal{C}$ is the matrix product $P^\mathcal{C}_1 = P^\mathcal{B}_1 P^\mathcal{A}_1$. In general, the flux angle $\theta$ for the combined loop is not the sum of the ones on $\mathcal{A}$ and $\mathcal{B}$, $\theta_\mathcal{C} \neq \theta_\mathcal{A} + \theta_\mathcal{B}$. The angles are additive only in the special case when the flux matrices commute, $P^\mathcal{A}_1 P^\mathcal{B}_1 = P^\mathcal{B}_1 P^\mathcal{A}_1$. As expected, the flux angles are additive in U(1) liquids, i.e., when IGG$_u=$~U(1) and all $P_j$ obviously commute. In the absence of this additivity (i.e., in $\mathbb{Z}_2$ liquids), it is certainly {\it not} sufficient to specify flux angle patterns on elementary plaquettes of the lattice to characterize quadratic spinon Hamiltonians.

Since we are primarily interested in {\it chiral} spin liquids, i.e., states that break time-reversal symmetry, we need to understand how time reversal affects the SU(2) gauge fluxes. Our familiarity with the electromagnetic U(1) flux may mislead us to think that a nontrivial flux angle $\theta$ implies breaking of time reversal. However, this is only correct for odd-site loops. For the gauge choice \eqref{eq:TRbare}, it is straightforward to show that the ansatz changes sign under time reversal,
\begin{equation}\label{eq:TRu}
  \tr\!: u_{ij}^\mu \mapsto -u_{ij}^\mu, \lambda^a_j \mapsto -\lambda^a_j\, .
\end{equation}
From the definition \eqref{eq:Pmf}, it then follows that the SU(2) flux on {\it even-site} loops is time-reversal invariant, while it changes sign for {\it odd-site} loops.\cite{trRep} As a result, an ansatz breaks time reversal if the flux angle $\theta$ is nontrivial ($\neq 0, \pi$) on odd-site loops. A nontrivial SU(2) flux can be threaded through even-site loops without necessarily breaking time-reversal symmetry. A well-known example of this rather surprising fact is the ``staggered flux'' state on the square lattice\cite{AffleckMarson_PRB.37.3774, *MarstonAffleck89_PRB.39.11538, EdeggerMuthuGros07} which is believed to be relevant to the pseudogap phase of cuprate high-temperature superconductivity. In this case, a U(1) flux $\pm\phi$ is threaded through the elementary plaquettes. An SU(2) gauge rotation brings the staggered-flux ansatz to a pairing state with $d$-wave symmetry, which is manifestly time-reversal invariant.\cite{LeeWen01_PRL.63.224517, IvanovLee04, *Ivanov04, Bieri09_PRB_79_174518, u1liqsf} A more detailed discussion of the relation between time-reversal symmetry and physical observables will be given in Sec.~\ref{sec:invPSGgen}.

\subsection{Flux operators}\label{sec:Pop}

In this section, we want to explore which physical {\it spin operator} -- or order parameter -- the SU(2) gauge flux corresponds to. Equation~\eqref{eq:Pmf} is the property of an ansatz $u$, and it is not clear a priori what spin expectation value it stands for, if any. A closely related question has been addressed in the classic paper by Wen, Wilczek, and Zee,\cite{WWZ89_PRB_39_11413} where the authors considered the U(1) flux in the pure hopping formalism (see Appendix~\ref{app:U1flux}). However, to our knowledge, the question has not been addressed in the general SU(2) invariant framework. In the following, we present a formalism that includes both hopping and pairing terms on the same footing.

The spinon hopping and pairing operator corresponding to the ansatz matrix \eqref{eq:umf} is
\begin{equation}\label{eq:uhat}
  \hat u_{ij} = \frac{1}{2}\left(\begin{array}{cc}
    {\bm f}_i^\dagger{\bm f}_j & {\bm f}_i^T\varepsilon{\bm f}_j\\
    {\bm f}_j^\dagger\varepsilon{\bm f}_i^* & {\bm f}_j^\dagger{\bm f}_i
  \end{array}\right)\,.
\end{equation}
In a mean-field decoupling of the Heisenberg term in the spinon singlet channel, one would write ${\bm S}_i\cdot{\bm S}_j \sim \text{Tr}[\hat u_{ij}^\dagger u_{ij}] + \text{H.c.}$

As discussed before, a gauge doublet is written as ${\bm\psi} = (f_\uparrow, f_\downarrow^\dagger)^T$. In fact, a second gauge doublet is given by $\tilde{\bm\psi} = \varepsilon{\bm\psi}^*$. Since $\varepsilon g^* \varepsilon = g$ for any SU(2) matrix $g$, $\tilde{\bm\psi}$ transforms in the same way as ${\bm \psi}$ under gauge transformations, $\tilde{\bm\psi} \mapsto g \tilde{\bm\psi}$. Similarly, $\tilde{\bm f} = \varepsilon{\bm f}^*$ is a second spin doublet. It is now convenient to introduce the spinon matrix\cite{AffleckZouAnderson_PRB.38.745, WenZee90, Chen12_PRB.85.094418}
\begin{equation}\label{eq:mspinor}
  \Psi = ({\bm \psi}, \tilde{\bm \psi}) = ({\bm f}, \tilde {\bm f})^T =
  \left(\begin{array}{cc}
    f_\uparrow & f_\downarrow\\
    f_\downarrow^\dagger & -f_\uparrow^\dagger\\
  \end{array}\right)\,.
\end{equation}
By construction, $\Psi$ transforms by left multiplication under gauge transformations, $\Psi\mapsto g\Psi$. Since the spin doublets are rows of $\Psi$, it transforms by right multiplication under spin rotation, $\Psi\mapsto \Psi U^T$. Furthermore, one can easily show that \eqref{eq:uhat} and \eqref{eq:mspinor} are related by
\begin{equation}\label{eq:upsi}
  2 \hat u_{ij} = \Psi_i \Psi_j^\dagger\, .
\end{equation}
In this form, the operator $\hat u_{ij}$ is manifestly spin-rotation invariant, and it transforms in the same way as the field $u_{ij}$ under gauge transformations, namely, Eq.~\eqref{eq:gaugetru}.

We are now equipped to evaluate the SU(2) flux operator $\hat P_j$. To obtain $\hat P_j$, we replace the field $u_{ij}$ in \eqref{eq:Pmf} by the operator $\hat u_{ij}$, Eqs.~\eqref{eq:uhat} or \eqref{eq:upsi}. We have
\begin{equation}\label{eq:Pop}
  \hat P_1 = \prod_\mathcal{C}\hat u_{kl} = \frac{1}{2^q} \Psi_1\Psi_2^\dagger\Psi_2\Psi_3^\dagger\ldots \Psi_{q-1}^\dagger\Psi_q \Psi_q^\dagger \Psi_1\, .
\end{equation}
It is useful to note that $\Psi^\dagger\Psi = \mathbb{1}_2 + 2 S^a \sigma_a^*$. Furthermore, one can show that $\Psi\Psi^\dagger = \mathbb{1}_2 - 2 G^a \sigma_a$ and $\Psi\sigma_a^*\Psi^\dagger = - 2 S^a\mathbb{1}_2$, in the notation introduced at the beginning of Sec.~\ref{sec:frac}. Next, we normal-order the flux operator with respect to the spinon vacuum as $\normOrd{\hat P} = \hat P - \langle 0|\hat P |0\rangle$. Using these facts, the result is
\begin{equation}\label{eq:Phat}
\begin{split}
  2 \normOrd{\hat P_1} = -\text{Tr}[&\underline{S}_1\underline{S}_2\ldots\underline{S}_q]\, \mathbb{1}_2\\ &- \text{Tr}[\underline{S}_2\underline{S}_3\ldots\underline{S}_q]\, G_1^a\sigma_a\,,
\end{split}\end{equation}
where $\underline{S} = S^a\sigma_a^*$ and the traces are over $2\times 2$ matrices. Note that the first term on the right-hand side of Eq.~\eqref{eq:Phat} is gauge invariant, while the last term depends on the gauge at the base site due to $G_1^a$ which transforms as a vector. Finally, taking the trace over spin indices yields
\begin{equation}\label{eq:trPhat}
  \text{Tr}[\normOrd{\hat P_1}] = -\text{Tr}[\underline{S}_1\underline{S}_2\ldots\underline{S}_q]\,.
\end{equation}
This expression is both gauge invariant {\it and} independent of base site, consistent with Eqs.~\eqref{eq:Pfluxmf} and \eqref{eq:tracetheta}. Furthermore, the trace is purely real or imaginary on even- or odd-site loops, respectively, analogous to \eqref{eq:tracetheta}. Note that we have not used the constraint $G^a_j=0$ anywhere in this calculation.

For $q=2$, we have $\text{Tr}[\normOrd{\hat P}] = -2{\bm S_1}\cdot{\bm S}_2$; for $q=3$, $\text{Tr}[\normOrd{\hat P}] = 2 i {\bm S_1}\cdot ({\bm S}_2\wedge{\bm S}_3)$, for $q=4$, $\text{Tr}[\normOrd{\hat P}] = 2[({\bm S_1}\cdot{\bm S}_3)({\bm S_2}\cdot{\bm S}_4) - ({\bm S_1}\cdot{\bm S}_2)({\bm S_3}\cdot{\bm S}_4) - ({\bm S_1}\cdot{\bm S}_4)({\bm S_2}\cdot{\bm S}_3)]$, etc. The case $q=3$ is the {\it scalar chirality}. It corresponds to the imaginary part of this flux in the U(1) formalism.\cite{WWZ89_PRB_39_11413} In general, however, the SU(2) flux is different from the U(1) result. Note that in the U(1) formalism, the constraint has to be imposed on every site. This is not necessary in the general SU(2)-invariant context presented here. See Appendix~\ref{app:U1flux} for more details on the U(1) formalism.

\subsection{Topological degeneracy}

Similar to quantum Hall states,\cite{FrohlichZee91} (gapped, Kalmeyer-Laughlin) chiral spin liquids are expected to be described by effective Chern-Simons theories in the low-energy, long-wavelength limit.\cite{Wen90_IJMP} Such topological field theories (and their generalizations) imply a ground-state degeneracy that depends on the genus of compactified space. Inspired by this fact, ``topological order'' was postulated to characterize exotic spin phases, and strongly correlated states in general, beyond the paradigm of the conventional Landau theory of symmetry breaking.

Within the parton construction of quantum spin liquids discussed in this paper, one can construct locally indistinguishable degenerate states by threading additional gauge flux through the holes of the lattice torus. In continuous gauge theories, such a flux threading changing the vacuum sector is done by performing ``singular'' or ``large'' gauge transformations.\cite{rajaraman82}

In the present case of a lattice gauge field, the flux insertion procedure is slightly different. To do so, we introduce a ``cut'' that winds around the lattice torus and that avoids all vertices. For a short-range ansatz $u_{ij}$, it is possible to consider the links $(ij)$ that cross this cut, and to modify the ansatz on those links as $u_{ij}\mapsto g u_{ij}$ or $u_{ji}\mapsto u_{ji}g^\dagger$, depending on the link direction, where $g$ is an SU(2) matrix. It is easy to see that the SU(2) gauge flux for lattice loops winding around the torus is changed as $P_i\mapsto g P_i$. However, the modification of $u$ must only affect observables $P_i$ on Wilson loops $\mathcal{C}$ that wind around the torus, while fluxes through local loops are unchanged. Our flux insertion procedure, however, may affect local loops, and this can even break translation symmetry: a local loop crossing the cut twice changes as $P_i = u_{ij} P_{jk} u_{kl} P_{li} \mapsto g u_{ij} P_{jk} u_{kl} g^\dagger P_{li}$. Hence, local observables remain unchanged only if the ``large'' gauge transformation $g$ can be pulled through any local Wilson matrix $P_{ik}$ that starts and ends at the cut. Clearly, this is only the case for $g\in\text{IGG}_u$.

We therefore see that a $\mathbb{Z}_2$ QSL state only allows topological flux insertion with $g=-1$, or $\theta_\mathcal{C}\mapsto\theta_\mathcal{C}+\pi$. Such a ``$\pi$-flux'' insertion can be done through any hole of the compactified lattice torus, and it corresponds to a change in spinon boundary condition from periodic to antiperiodic. This construction inserting sign flips on a cut is similar to other instances of toy models for topological order, such as quantum dimer models\cite{Kivelson87_PRB.35.8865, *Rokhsar88_PRL.61.2376, Moessner01_PRL.86.1881, *Moessner01_PRB.63.224401, MisguichSerbanPasquier02_PRL.89.137202, *MisguichSerbanPasquier03_PRB.67.214413, Ralko05_PRB.71.224109, *Ralko06_PRB.74.134301, Ribeiro07_PRB.76.172301} or the toric code.\cite{Kitaev2003, *Kitaev2006}

On a torus, the number of topologically degenerate parton wave functions for $\mathbb{Z}_2$ quantum spin liquids is therefore four, $\{|\phi_1, \phi_2\rangle\}$, with $\phi_n=0,\pi$. On a general compact space of genus $g>0$, this number is $2^{g+1}$. However, the simplest abelian Chern-Simons theory potentially describing a CSL has a topological degeneracy of $2^g$ on a genus $g$ surface.\cite{Wen90_IJMP} Therefore, it can occur that the four degenerate parton states of a $\mathbb{Z}_2$ CSL on the 2-torus are not linearly independent, and they span only a two-dimensional space.\cite{Zhang12_PRB.85.235151, HuBeccaSheng_PRB.91.041124}

\section{Theory of PSG classification}\label{sec:theoryPSG}

A primary goal of the PSG approach is the construction and classification of {\it quadratic} spinon Hamiltonians $H_0$ that respect all or some symmetries of a given lattice spin model. However, even beyond quadratic spinon Hamiltonians, the PSG allows to distinguish phases that have the same symmetries. It therefore provides a classification scheme that goes beyond the conventional Landau theory of symmetry breaking. 

In this section, we perform a rather formal and general discussion of the PSG construction. This may help to elucidate some core concepts of the approach. In subsequent sections, we perform this program in the concrete examples of triangular and kagome lattices.

\subsection{Algebraic PSG}\label{sec:algPSG}

Before imposing symmetry constraints (such as translation, etc) on the spinon Hamiltonian $H_0$ in Eq.~\eqref{eq:h0}, the group of symmetry transformations SG must be represented in the spinon Hilbert space via a gauge transformation $g\in \mathcal{G}$. Let us introduce the group $\mathcal{G}\rtimes$SG. We define the action of an element $Q_x = (g, x)\in\mathcal{G}\rtimes$SG on an ansatz $u = [u_{ij}, \lambda_j]$ as
\begin{equation}\label{eq:gt} 
  Q_x(u) = [g_i u_{x^{-1}(i,j)} g_j^\dagger; g_j \lambda_{x^{-1}(j)} g_j^\dagger]\,.
\end{equation}
The multiplication law in this group is therefore
\begin{equation}\label{eq:Qmult}
  Q_x Q_y = (g_x, x)(g_y, y) = (g_x x g_y x^{-1}, x y)\,,
\end{equation}
where we use the notation $x g x^{-1} = x (\otimes g_j) x^{-1} = \otimes g_{x^{-1}(j)}\in \mathcal{G}$. The inverse of an element is given by
\begin{equation}
Q_x^{-1} = (g, x)^{-1} = (x^{-1} g^{-1} x, x^{-1})\,.
\end{equation}

$Q\!: $~SG~$\rightarrow\mathcal{G}\rtimes$SG is required to be a {\it representation} of the symmetry group in the gauge group $\mathcal{G}$. Let $e\in$~SG be the identity element. Two representations $Q$ and $\tilde Q$ are {\it equivalent} if and only if there exists a gauge transformation $G = (g, e)$ such that $\tilde Q = G^{-1} Q G$; (i.e., the same gauge transformation is applied to {\it all} elements of $Q$). 
This equivalence relation is natural in view of our discussion of quadratic spinon Hamiltonians in Sec.~\ref{sec:h0}. In Sec.~\ref{sec:invPSGgen}, we will introduce the systematic construction of such Hamiltonians, and it will be clear that this equivalence between representations translates into gauge equivalence of ans\"atze. With this in mind, we may call a particular representative of a class of representations a ``gauge choice''.

Furthermore, $Q$ is a {\it projective} representation of the symmetry group, i.e., the algebraic relations in SG must be respected up to certain gauge transformations. We have
\begin{equation}\label{eq:cocycle}
   Q_x Q_y = \omega(x, y) Q_{x y}\,,
\end{equation}
where $\omega = (\omega, e)$ are elements of some subgroup of $\mathcal{G}$, called the invariant gauge group (IGG) or {\it factor set}.\cite{EssinHermele13_PRB.87.104406} In this paper, we restrict our discussion to global $\mathbb{Z}_2$ transformations, i.e., $\omega\in\text{IGG}=\{\pm 1\}$. In this case, we call the representations $\mathbb{Z}_2$ PSG classes.
We will see that the IGG introduced here is related to the invariant gauge group IGG$_u$ of the ansatz we are going to construct. However, the factor set IGG is generally a subgroup of IGG$_u$ of the resulting ansatz $u$, and IGG$_u$ can be larger.

The elements $\omega$ of the factor set in \eqref{eq:cocycle} transform to $g^\dagger\omega g$ under gauge transformations. Since we focus on $\omega\in$~IGG~$=\mathbb{Z}_2$, the signs $\omega$ are gauge invariant and, therefore, provide a characterization of PSG classes. However, as we will see, these signs are not always sufficient to distinguish $\mathbb{Z}_2$ PSG classes. The set of equivalence classes of projective representations of SG in $\mathcal{G}$ is called {\it algebraic~PSG}.

\subsection{Symmetry group}

In principle, the algebraic PSG can be worked out for the full symmetry group of a system. However, it is not necessary to solve this problem in generality. In fact, we only need representations of the subgroup of symmetries that we want to be respected in the phase. For {\it symmetric} quantum spin liquids,\cite{Wen02_PRB.65.165113} the full space group as well as time reversal and spin rotation is required to be respected. To generalize this notion, we define the {\it chiral spin liquid} as a state that respects spin rotation, but the lattice space group is respected only {\it up to time reversal} $\tr$.  For two-dimensional Bravais lattices, the space group generators are translations $T_\opx$ and $T_\opy$, a reflection symmetry $\sigma$, and the lattice rotation $R$ (e.g., $\pi/3$ rotation for the triangular lattice, etc). We are therefore interested in the group generated by
\begin{equation}\label{eq:sg}
  \text{SG} = \{T_\opx\, \tr^{\tau_{t}}, T_\opy\, \tr^{\tau_{t}}, \sigma\, \tr^{\tau_\sigma}, R\, \tr^{\tau_R}\}\,.
\end{equation}
The signatures $\tau_{t}, \tau_\sigma, \tau_R \in\{0, 1\}$ specify different ways in which time reversal can be broken in the chiral spin liquid. For triangular-based lattices at the focus of this paper, only $\tau_{t}=0$ is possible.\cite{Messio13_PRB_87_125127} Henceforth, we will set $\tau_{t} = 0$. 
Liquids that break all reflection symmetries of the lattice are labeled by $\tau = (\tau_\sigma, \tau_R) = (1, 0)$; such liquids may be called of {\it Kalmeyer-Laughlin} type.\cite{KalmeyerLaughlin87_PRL_59_2095, *KalmeyerLaughlin89_PRB.39.11879} Chiral liquids can also break lattice rotation $R$, in which case $\tau_R = 1$. As we will see, this implies that the SU(2) flux changes sign under rotation. We call them {\it staggered flux} states.\cite{staggeredFlux}

In the case all $\tau_x=0$, the full lattice space group is respected. To have a fully symmetric liquid, however, time reversal $\tr$ has to be added to SG. As we will discuss later, its representation can sometimes be relevant in the construction of symmetric $\mathbb{Z}_2$ spin liquids.

In analogy with quantum liquids that respect the space group up to time reversal as in Eq.~\eqref{eq:sg}, N\'eel states of classical spins can have similar symmetry properties (supplemented by rotations of spin). For time reversal to be broken, the arrangement of classical spins ${\bm S}_j$ must be {\it nonplanar} such that ${\bm S}_1\cdot({\bm S}_2\wedge{\bm S}_3)\neq 0$ on some triangles. Classical spin states that respect the space group up to time reversal and spin rotation are called ``regular magnetic orders''. They have been classified and discussed in Ref.~[\onlinecite{Messio11_PRB_83_184401}] for several two-dimensional lattices. Examples are {cuboc-1}, {cuboc-2}, and octahedral states on kagome, or tetrahedral states on triangular or honeycomb lattices.\cite{Jolicoeur90_PRB.42.4800, Chubukov92_PRB.46.11137, Korshunov93_PRB_47_6165}

For the PSG construction of chiral spin liquids, a simplification in Eq.~\eqref{eq:sg} stems from the fact that we do not need to know how time reversal $\tr$ is represented in spinon space. For example, if $\tau_\sigma=1$, only the representation of $\sigma\tr$ will be relevant. Since $\tr$ commutes with all space group symmetries, its presence does not affect the representation classes. So the algebraic PSG of lattice symmetries are the same for chiral and for time-reversal conserving (symmetric) spin liquids. To alleviate our notations, we will often write $\sigma$ instead of $\sigma\tr^{\tau_\sigma}$, and $R$ instead of $R\tr^{\tau_R}$ in the following.

In fact, the algebraic PSG can also be used to construct spin liquids with broken spin rotation (``triplet'' or ``nematic'' QSLs),\cite{Momoi09_PRB.80.064410, *Momoi11_PRB.84.134414, DoddsBhattaYBK13_PRB.88.224413, Reuther14_PRB.90.174417}. However, in this paper we focus on the spin-rotation invariant case.

\subsection{Invariant ansatz}\label{sec:invPSGgen}

Once all projective representations $Q$ of SG (or equivalence classes thereof) are listed, it remains to find ans\"atze $u$ that respect those symmetries for each PSG representation. For a space group symmetry $x$ to be respected (up to time reversal), the ansatz $u$ must satisfy
\begin{equation}\label{eq:lg}
  Q_x( u ) = (-)^{\tau_x} u\,,
\end{equation}
where $\tau_x=1$ if $x$ includes time reversal, and $\tau_x=0$ otherwise [see Eq.~\eqref{eq:TRu}]. The action of $Q_x$ on the ansatz was defined in \eqref{eq:gt}. On the one hand, for elements of the point group, Eq.~\eqref{eq:lg} imposes {\it constraints} on sites and links that are left invariant by the action of $x$. On the other hand, this equation can be used to {\it propagate} fields on a given site or link to another location on the lattice.

For example, the on-site field $\lambda$ satisfies
\begin{equation}\label{eq:clambda}
  \lambda_j = (-)^{\tau_x} g_{x j} \lambda_{x^{-1}(j)} [g_{x j}]^\dagger\,.
\end{equation}
Here, the action of the space group element $x$ goes along with rotations of the vector $(\lambda^a)$ by the representation $g_x$. If the site is left invariant and $x(j) = j$, then this is a constraint on $\lambda_j$. Otherwise, the equation can be used to propagate $\lambda$ from one site to another. In general, on a two-dimensional Bravais lattice, there are at most two independent elements of the point group that leave a site invariant. Therefore, there can be no more than two constraint equations.

Pairs of sites (links) may be left invariant, or they may be exchanged by a nontrivial element of the point group. In the first case, the constraint on that link is
\begin{equation}\label{eq:cu}
  u_{ij} = (-)^{\tau_x} g_{x i} u_{ij} [g_{x j}]^\dagger\, .
\end{equation}
In the latter case, the link direction is inverted and the ansatz must satisfy
\begin{equation}\label{eq:cinvert}
  [u_{ij}]^\dagger = (-)^{\tau_x} g_{x i} u_{ij} [g_{x j}]^\dagger\, .
\end{equation}
The constraints and their number (0, 1, or 2) must be determined on a case-by-case basis for each type of link (first-, second-neighbor, etc). 

Note that the constraints \eqref{eq:clambda} and \eqref{eq:cu} must also be imposed for the trivial transformation $x = e$, and for all elements $g_e\in$~IGG. This ensures that IGG$_u$ of the constructed ansatz contains IGG as a subgroup. However, in our case of IGG~$=\mathbb{Z}_2$, this does not restrict the ansatz in any way.

The sites and links of a given type (e.g., first-neighbor links, etc) in a unit cell are usually mapped to one another by elements of the point group. In this case, it is sufficient to pick $\lambda_j$ on one site, and the fields $u_{ij}$ on one link of each type. All fields in the unit cell are then obtained by propagation using the point group representation. The ansatz on sites and links of one unit cell can subsequently be propagated to the entire lattice by translation.

Finally, it is clear that an ansatz $u$ constructed for a PSG representation $Q$ using Eqs.~\eqref{eq:lg} is gauge equivalent to an ansatz $\tilde u$ constructed from another representative $\tilde Q$ in the same PSG class. When the representative of the PSG class is changed from $Q$ to $(g, e)\cdot Q \cdot(g^\dagger, e)$, then the constructed ansatz is $\tilde u = g(u)$. Note that the symmetry constraints usually fix some phases in $u_{ij}$ [direction of ${\bm n}$ in Eq.~\eqref{eq:umfphase}], but there remain free parameters in the ansatz.

\subsubsection{Properties of SU(2) gauge flux}\label{sec:propFlux}

So far, we have discussed how symmetry affects the ansatz $u$. This point of view is very useful for constructing concrete quadratic spinon theories on the basis of a given algebraic PSG representation. However, the ansatz is gauge dependent, and it is not a physical observable. A gauge invariant characterization of the theory is provided by the SU(2) flux introduced in Sec.~\ref{sec:Pmf}. Next, we discuss properties and restrictions imposed by symmetries and their PSG representation on the SU(2) gauge flux.

In Sec.~\ref{sec:trgPSG}, we will see that $\mathbb{Z}_2$ PSG representations of the translation generators for a Bravais lattices can always be chosen as $g_x = \pm\mathbb{1}_2$. This gauge is very convenient and interesting, as it implies that the SU(2) fluxes $P_j$ are {\it uniform} on the lattice: by virtue of Eqs.~\eqref{eq:gt} and \eqref{eq:lg}, we have $Q_x(P_j) = P_{x^{-1}(j)} = P_j$. 

As discussed in Sec.~\ref{sec:Pmf}, the gauge flux through even- and odd-site loops can be written as
\begin{subequations}\label{eq:evenoddP}
\begin{align}
  P_\text{even}&= e^{i\theta ({\bm n}\cdot {\bm\sigma}) } = \cos\theta +i ({\bm n}\cdot {\bm\sigma}) \sin\theta\,,\\
  P_\text{odd} &= -i\partial_\theta e^{i\theta ({\bm n}\cdot {\bm\sigma}) } = ({\bm n}\cdot {\bm\sigma}) \cos\theta +i \sin\theta\,.
\end{align}
\end{subequations}
Here, we neglect an unimportant scale $\rho\neq0$. ${\bm\sigma} = (\sigma_a)$ are Pauli matrices, and ${\bm n}$ is a real unit vector (flux director). Hence, the SU(2) flux $P(\theta, {\bm n})$ is parameterized by an angle $\theta$ and a director ${\bm n}$, and, as discussed, $\text{Tr}[P_\text{even}] = 2\cos\theta$, resp.\ $\text{Tr}[P_\text{odd}] = 2i\sin\theta$ are gauge invariant. ${\bm n}$ rotates like a vector under gauge transformations on the base site of the loop. However, {\it relative} orientations of directors for different loops starting from the same site provide gauge invariant information, e.g., on the invariant gauge group (IGG$_u$).

A constraint on the SU(2) flux comes from reflection symmetries $\sigma$ that map the loop $\mathcal{C}$ to itself, leaving at least one site invariant. Next, we discuss the restriction on the flux resulting from such symmetries. Again, we anticipate that the property $\sigma^2=1$ of any reflection symmetry implies that its representation $g_\sigma = \mathbb{1}_2$ or $g_\sigma = i\sigma_a$ (up to a gauge and unimportant signs). 
In addition, reflection inverts the loop direction, and we have $\sigma\!: P(\theta)\mapsto g_\sigma [P(\theta)]^\dagger  [g_\sigma]^\dagger = g_\sigma P(-\theta) [g_\sigma]^\dagger$.

Let us first discuss the case of {\it even-site} loops. Since even-site loops are insensitive to time-reversal signatures [see discussion around Eq.~\eqref{eq:TRu}], $\tau_\sigma$ does not enter the constraint in this case. Reflection symmetry therefore imposes
\begin{equation}
  g_\sigma  [P_\text{even}]^\dagger [g_\sigma]^\dagger = P_\text{even}\, .
\end{equation}
We see that for a trivial (linear) representation $g_\sigma = \mathbb{1}_2$, the flux through even-site loops is trivial, and $\theta=0$ or $\pi$. Only for nontrivial representations $g_\sigma = i\sigma_a$ can the angle $\theta$ be arbitrary. In this case, the flux director ${\bm n}$ must be in the plane {\it perpendicular} to $g_\sigma$, i.e., $n_a=0$.

For {\it odd-site} loops, the time-reversal signature does enter the constraint, and we have
\begin{equation}
  g_\sigma  [P_\text{odd}]^\dagger [g_\sigma]^\dagger = (-)^{\tau_\sigma} P_\text{odd}\, .
\end{equation}
In fact, regardless of the representation $g_\sigma$, one can immediately conclude that the flux angle $\theta = 0$ on odd-site loops when $\tau_\sigma = 0$, since $\text{Tr} P_\text{odd}(-\theta) = -\text{Tr} P_\text{odd}(\theta)$. Hence, $P_\text{odd}={\bm n}\cdot{\bm\sigma}$, with ${\bm n}$ arbitrary if $g_\sigma = \mathbb{1}_2$, otherwise ${\bm n}\parallel g_\sigma$. For $\tau_\sigma = 1$, a trivial reflection representation $g_{\sigma} = \mathbb{1}_2$ fixes the SU(2) flux angle to $\theta = \pm\pi/2$. Only $g_{\sigma} = i\sigma_a$ allows an arbitrary $\theta$, and the flux director ${\bm n}$ must be perpendicular to $g_{\sigma}$ in this case.

The meaning of the restrictions on the flux angle for odd-site loops (e.g., $\theta=0$ if $\tau_\sigma=0$, or $\theta=\pm\pi/2$ for $\tau_\sigma=1$ and $g_\sigma=\mathbb{1}_2$) are clear. However, restrictions on the directors ${\bm n}$ are more subtle, since ${\bm n}$ is not observable on a single loop. These restrictions only manifest as {\it addition rules} for flux angles $\theta$ on different loops. Let us discuss the case $\tau_\sigma=0$. Here, the restriction ${\bm n}\parallel g_\sigma$ on odd loops, and ${\bm n}\perp g_\sigma$ on even-site loops means that we can join any even number of odd-site loops to an even-site loop, without changing the total flux angle $\theta$. In particular, joining two reflection-symmetric odd-site loops must result in an even-site loop with $\theta=0$.

Let us summarize this section. A trivial (linear) PSG representation of a reflection symmetry strongly restricts the possible SU(2) fluxes through lattice loops that have this symmetry. For even-site loops, the flux angle $\theta$ is fixed to $0$ or $\pi$; for odd-site loops, it is fixed to $\pm\pi/2$. Only a nontrivial representation $g_{\sigma}=i\sigma_a$ allows general SU(2) flux angles, while the flux director $\bm n$ is restricted (see Table~\ref{tab:sym}).

\subsubsection{Time-reversal constraint}\label{sec:TRconstraint}

In this paper, we are primarily interested in chiral spin liquids with broken time reversal. Nevertheless, let us briefly discuss the additional constraints that are imposed on time-reversal {\it symmetric} liquids.\cite{Wen02_PRB.65.165113, wenOrig} For the construction of symmetric liquids, we add time reversal $\tr$ to the symmetry group SG in \eqref{eq:sg}. Its representation must satisfy $(g_\tr)^2 = -\mathbb{1}_2$,\cite{gTchoice} which implies that $g_\tr = i\sigma_a$ (up to a gauge). Here we always assume a uniform gauge where $g_\tr$ is independent of lattice site. Next, one imposes time reversal on the ansatz via Eq.~\eqref{eq:lg},
\begin{equation}\label{eq:trconstr}
  Q_\tr(u) = - u\,,
\end{equation}
where $Q_\tr(u) = g_\tr u [g_\tr]^\dagger$. In the usual expansion $u = u^\mu \tau_\mu$ with $(\tau_\mu) = (i\mathbb{1}_2, \sigma_a)$, it is clear that time reversal forces the {\it temporal} component $u^0$ to vanish, independent of its (uniform) representation $g_\tr$ (note that $u^0$ corresponds to imaginary singlet hopping). In addition to that, Eq.~\eqref{eq:trconstr} forces one spatial component $u^a=0$, corresponding to the choice of $g_\tr$.

Conversely, a nonzero temporal component $u^0$ breaks time reversal, independent of its (uniform) representation.\cite{imagHoppingT} However, $u^0=0$ is only a necessary, but not sufficient condition for time-reversal symmetry. The spatial components $(u^a)$ of the ansatz are conveniently understood as real vectors on the links of the lattice. An ansatz respects time reversal if, in addition to $u^0=0$, the components $(u^a)$ on all links are coplanar. If they are nonplanar, time reversal is generally broken.

Similar to the space group constraints discussed previously, the restriction on the ansatz $u$ due to time reversal is a convenient practical device, but it does not provide any physical or gauge-invariant insight. To this end, it is more useful to consider the time-reversal constraints on the SU(2) gauge flux. Again, we need to separately consider even- and odd-site loops. In a time-reversal symmetric QSL, the gauge fluxes satisfy
\begin{subequations}\label{eq:evenoddPtr}
\begin{align}
  g_\tr  P_\text{even} [g_\tr]^\dagger &= P_\text{even}\,,\\
  g_\tr  P_\text{odd} [g_\tr]^\dagger  &= -P_\text{odd}\,.
\end{align}
\end{subequations}
As discussed, the representation $g_\tr$ must be nontrivial, $g_\tr\neq\mathbb{1}_2$, in a parton theory. Again, we use expressions \eqref{eq:evenoddP} to solve these constraints. For {\it even-site} loops, the flux angles $\theta$ are unconstrained, provided that all flux directors ${\bm n}$ are {\it collinear} and parallel to $g_\tr = i\sigma_a$. For {\it odd-site} loops, on the other hand, time reversal imposes $\theta=0$, i.e., $P_\text{odd} = {\bm n}\cdot {\bm \sigma}$, with directors ${\bm n}$ {\it perpendicular} to $g_\tr$, i.e., $n_a=0$.

Thus we see that the gauge-invariant content of time-reversal symmetry is a flux $\theta=0$ and ${\bm n}\perp g_\tr$ on odd-site loops. On even-site loops, the flux angle $\theta$ is unrestricted by time reversal, but all directors ${\bm n}$ must be parallel. Physically, this means that the SU(2) flux \mbox{angles} are {\it additive} on even-site loops in time-reversal symmetric liquids. As a corollary, we can conclude that a time-reversal invariant QSL with only even-site loops is always a U(1) state [i.e., IGG$_u$=U(1)]: as discussed in Sec.~\ref{sec:Pmf}, collinearity of SU(2) flux directors is a sufficient condition for a U(1) state. However, if there are also odd-site loops, the symmetric state generally has a $\mathbb{Z}_2$ gauge structure. An example for the latter scenario is the ``sublattice pairing state'' (SPS) on the honeycomb lattice.\cite{LuRan_PRB.84.024420, FlintLee13_PRL.111.217201}

\subsubsection{PT theorem}\label{sec:PTtheorem}

Having discussed time-reversal and reflection symmetries in (singlet) quantum spin liquids, a natural question arises about the status of their mutual relationships. Do these symmetries imply each other, i.e., is there a ``(C)PT theorem''? For example, the full lattice space group is respected when all $\tau_x = 0$ in SG, Eq.~\eqref{eq:sg}. Does this imply time-reversal invariance and that we automatically have a symmetric spin liquid? The combination CPT is necessarily conserved in relativistic quantum field theories,\cite{wightman89} but this need not be the case in our nonrelativistic framework.

\begin{table}
\begin{tabular}{c|c|c|c}
\hline\hline
loop & $\tr$ & $\sigma$ & $\sigma\tr$ \\
\hline
even & ${\bm n}\parallel g_\tr$ & ${\bm n}\perp g_\sigma$ & ${\bm n} \perp g_{\sigma\tr}$\\
odd & $\theta=0, {\bm n}\perp g_\tr$ & $\theta=0, {\bm n}\parallel g_\sigma$ & ${\bm n}\perp g_{\sigma\tr}$\\
\hline\hline
\end{tabular}
\caption{SU(2) gauge flux $P$ through even- and odd-site loops as given in Eqs.~\eqref{eq:evenoddP}, restricted by time reversal $\tr$, reflection $\sigma$, and their combination $\sigma\tr$. 
\label{tab:sym}}
\end{table}

In Table~\ref{tab:sym}, we summarize the restrictions imposed by time reversal $\tr$, reflection symmetry $\sigma$, and their combination $\sigma\tr$ on the SU(2) gauge flux. Here, we only consider the nontrivial cases when $g_\sigma \neq \mathbb{1}_2$; (otherwise the flux angle is completely fixed, see previous sections). The constraint from $\sigma\tr$ is listed for completeness. As discussed, this symmetry does not restrict the flux angle~$\theta$ for any loop parity. However, the directors ${\bm n}$ are forced to be coplanar for all reflection-symmetric loops if $g_{\sigma\tr}$ is nontrivial.

Next, we focus on columns $\tr$ and $\sigma$ in Table~\ref{tab:sym}. A PT theorem would require the equivalence of symmetries $\tr \Leftrightarrow \sigma$ for the SU(2) flux on all loops of the lattice. We can always choose $g_\tr$ and $g_\sigma$ to be orthogonal, i.e., $\text{Tr}[g_\tr g_\sigma] = 0$. However, even with this gauge, we see that a PT theorem does not hold in general. For even-site loops, time-reversal does imply reflection symmetry, but not so for odd-site loops: time-reversal symmetric fluxes generally violate the additivity property required by reflection symmetry. The situation is exactly reversed for the inverse relation. Reflection on odd-site loops implies time-reversal invariance for these loops, but not so on even-site loops: reflection-symmetric fluxes on even-site loops do not generally commute, and flux angles are not additive, as required by time reversal.

For U(1) liquids, the situation is much simpler than in the general case of $\mathbb{Z}_2$ gauge structure discussed above. In the U(1) case, all flux directors ${\bm n}$ are collinear, and the angles $\theta$ are therefore additive. As a result, additivity properties that may lead to violation of the PT theorem in $\mathbb{Z}_2$ states are automatically avoided. Therefore, the PT theorem generally holds in U(1) quantum spin liquids.

\subsubsection{Chern number}

Finally, we discuss some simple general properties of the Chern number of chiral spin liquid states constructed in this paper. We consider the (first) Chern number of occupied spinon bands at half filling for an infinite lattice, assuming that there is a gap in the spinon spectrum. A nontrivial Chern number for an ansatz $u$ implies chiral edge modes in the quadratic theory. These topological properties are likely to be robust with respect to interactions and Gutzwiller projection.

It is well known that the Chern number $C$ changes sign under time reversal, $\tr( C ) = - C$. Furthermore, one can easily check that it also changes sign under lattice reflections, $\sigma(C) = - C$. Therefore, an ansatz with nontrivial Chern number has to break both time reversal and all lattice reflection symmetries. However, the combination $\tr \sigma$ must be respected. Furthermore, one can check that the Chern number is invariant under lattice rotation, $R(C) = C$.

In terms of the time-reversal signatures $\tau$ that we introduced in the symmetry group \eqref{eq:sg}, these considerations mean that the Chern number can be nontrivial only when $\tau_\sigma=1$ and $\tau_R=0$. In other words, the Chern number is zero in the symmetric spin liquids and in the staggered-flux CSL states. It can only be nonzero in the case of Kalmeyer-Laughlin chiral spin liquids, $(\tau_\sigma,\tau_R) = (1,0)$.

\section{Triangular PSG}\label{sec:trgPSG}

The discussions in the previous sections were quite general. As a concrete example, we now explicitly do the PSG classification for the triangular lattice. In the next section, we will consider the kagome lattice, which is similar to the triangular case in many respects.

The first step consists in finding the gauge representations of the lattice symmetry group (algebraic PSG). Following that, the ansatz compatible with those symmetry representations will be constructed (invariant PSG). More technical details on these calculations can be found in Appendix~\ref{app:trgPSG}.

\begin{figure}
\includegraphics[width=.4\textwidth]{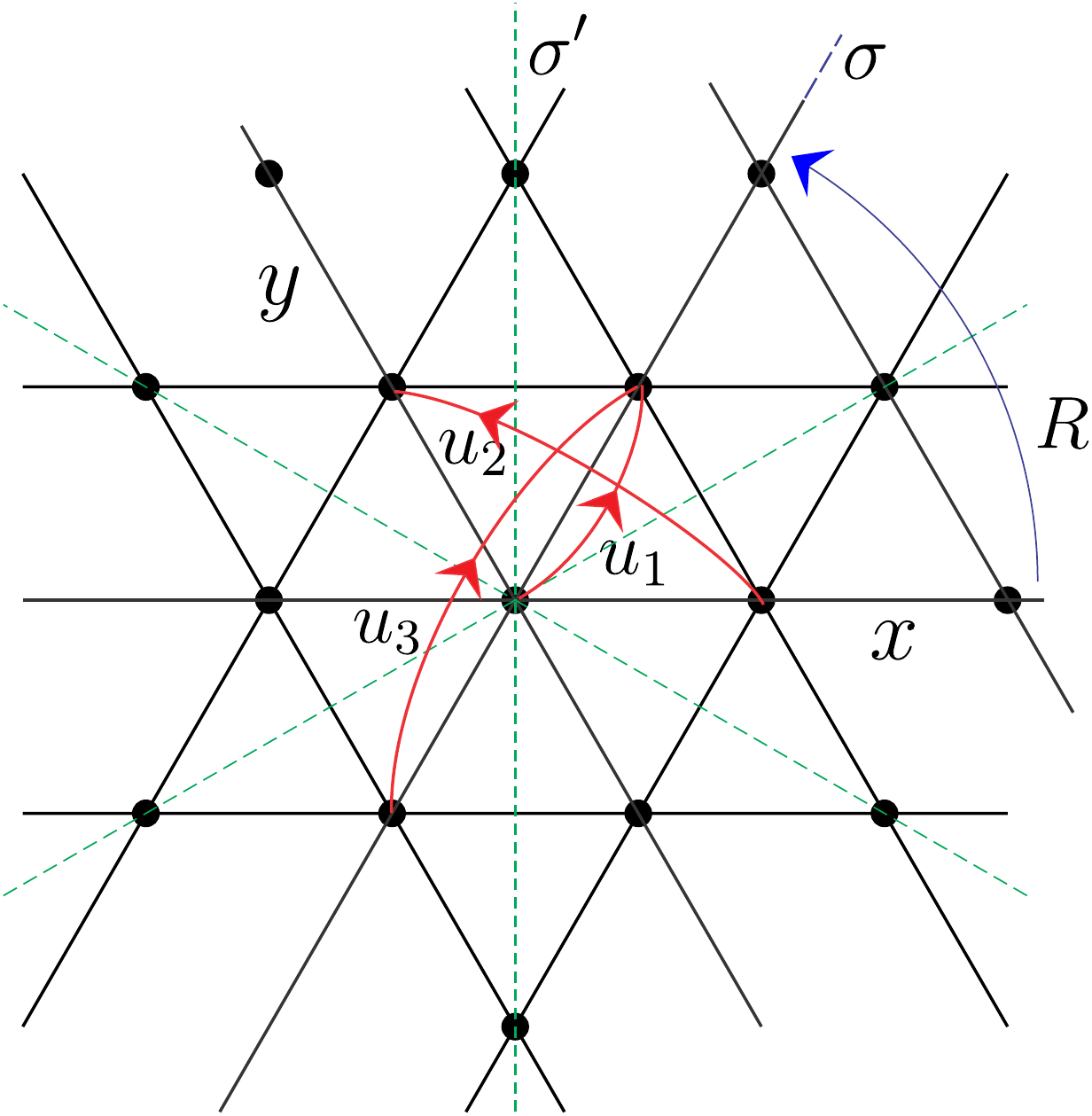}\caption{Symmetry generators and ansatz parameters $u$ for the triangular lattice.\label{fig:trgsym}}
\end{figure}

\subsection{Translation group}

We are interested in Bravais lattices where the translation group is generated by $T_\opx$ and $T_\opy$. These generators commute,
\begin{equation}\label{eq:Txy}
  T_\opx T_\opy = T_\opy T_\opx\,,
\end{equation}
and their order is infinite on the infinite lattices under consideration [i.e., $(T_a)^n\neq e\,, \forall n\,; a=\opx,\opy$]. Equation~\eqref{eq:Txy} constrains the possible representations $Q_a$. Using Eq.~\eqref{eq:cocycle}, we have
\begin{equation}\label{eq:Cxy}
  Q_\opx Q_\opy = (\epsilon_2) Q_\opy Q_\opx
\end{equation}
with $\epsilon_2 = \omega(\opx,\opy) \omega(\opy,\opx)^{-1} \in$~IGG.

First, we can choose a gauge where $g_\opx$ and $g_\opy$ are particularly simple. Upon changing the gauge, we have
\begin{equation}
  Q_a^{(g)} = (g^\dagger g_a T_a g T_a^{-1},T_a)\,.
\end{equation}
Hence, $g_\opx^{(g)} = g^\dagger g_\opx T_\opx g T_\opx^{-1}$, or $g_\opx^{(g)}(x,y) = g^\dagger(x,y) g_\opx(x,y) g(x-1,y)$, and we can always use $g(x-1, y) = [g_\opx(x,y)]^\dagger g(x,y)$ to set $g_\opx^{(g)}(x-1,y) = \mathbb{1}_2$. Starting this from $x\rightarrow +\infty$, we find $g_\opx(x,y) = \mathbb{1}_2$
without loss of generality. Note that it is not possible to also diagonalize $g_\opy$ in this way, because it would affect $g_\opx$.\cite{landauGauge}

For $g_\opx = \mathbb{1}_2$, Eq.~\eqref{eq:Cxy} becomes $T_\opx g_\opy T_\opx^{-1} = (\epsilon_2) g_\opy$, or $g_\opy(x-1,y) = (\epsilon_2) g_\opy(x,y)$. Therefore, $g_\opy(x,y) = (\epsilon_2)^x g_{0\opy}(y)$. A gauge transformation $g(x,y) = g_{0\opy}^{\dagger}(y+1) g_{0\opy}^{\dagger}(y+2)\ldots$ makes $g_\opy(x,y) = (\epsilon_2)^x$ and leaves $g_\opx$ invariant. Hence,
\begin{subequations}\label{eq:g12}
\begin{align}
  g_\opx(x,y) &= \mathbb{1}_2\,,\\
  g_\opy(x,y) &= (\epsilon_2)^x\, \mathbb{1}_2\,,\label{eq:g12b}
\end{align}
\end{subequations}
with $\epsilon_2\in$~IGG, are the most general projective representations of the translation group of a Bravais lattice.

Note that the choices leading to Eq.~\eqref{eq:g12} do not completely fix the gauge. We are still free to do global gauge transformations (i.e., constant or sublattice-dependent transformations). However, depending on $\epsilon_2$, there may be even more unfixed space-dependent gauge transformations. A transformation $g$ that leaves $g_\opx$ invariant satisfies $g(x,y) = I g(x-1,y)$, or one that conserves $g_\opy$ is $g(x,y) = I g(x,y-1)$, with $I\in$~IGG. Since IGG always contains $-1$, staggered transformations $g(x,y) = (-)^x \mathbb{1}_2$ and $g(x,y) = (-)^y \mathbb{1}_2$ are still possible.\cite{WangVish06_PRB.74.174423}

\begin{table}
\begin{tabular}{c|cc|ccc|c}
\hline\hline
No. & $g_{\sigma}$ & $g_{R}$ & $\epsilon_{\sigma}$ & $\epsilon_{R\sigma}$ & $\epsilon_{R}$ & sym\\
\hline
1 & $\mathbb{1}_2$ & $\mathbb{1}_2$ & $+$ & $+$ & $+$ & SU(2)\\
2 & $i\sigma_3$ & $\mathbb{1}_2$ & $-$ & $-$ & $+$ & U(1)\\
3 & $\mathbb{1}_2$ & $i\sigma_3$ & $+$ & $-$ & $-$ & U(1)\\
4 & $i\sigma_3$ & $i\sigma_3$ & $-$ & $+$ & $-$ & U(1)\\
5 & $i\sigma_2$ & $i\sigma_3$ & $-$ & $-$ & $-$ & $\mathbb{Z}_2$\\
6 & $i\sigma_2$ & $a$ & $-$ & $-$ & $+$ & $\mathbb{Z}_2$\\
7 & $i\sigma_2$ & $b$& $-$ & $-$ & $-$ & $\mathbb{Z}_2$\\
\hline\hline
\end{tabular}
\caption{PSG representations of the point group for the triangular lattice. $a=\exp(i\sigma_3 \pi/3 )$ and $b=\exp(i\sigma_3 \pi/6 )$. The signs $\epsilon_{(\cdot)}$ specify the $\mathbb{Z}_2$ representation of Eq.~\eqref{eq:pg2}. $g_\sigma(x,y)$ and $g_R(x,y)$ further depend on the sign $\epsilon_2$ as given in Eq.~\eqref{eq:pRep}. The column ``sym'' indicates the global gauge freedom that remains unfixed in the algebraic PSG.
\label{tab:PSGtrg}}
\end{table}

\subsection{Point group}\label{sec:trgPTgroup}

Next, we discuss the PSG representations of the triangular-lattice point group. Our definition of the generators $\sigma$ (reflection) and $R$ (lattice rotation), as well as translation $T_\opx$ and $T_\opy$ are given in Fig.~\ref{fig:trgsym}.

In addition to Eq.~\eqref{eq:Txy}, the following algebraic relations among the generators define the space group of a triangular Bravais lattice:\cite{Messio11_PRB_83_184401, Messio13_PRB_87_125127}
\begin{subequations}\label{eq:pg1}
\begin{align}
  \sigma T_\opx &= T_\opy \sigma\,,\label{eq:pg1a}\\
  T_\opx R T_\opy &= R\,,\label{eq:pg1b}\\
  T_\opy R &= R T_\opy T_\opx\,,\label{eq:pg1c}
\end{align}
\end{subequations}
and
\begin{subequations}\label{eq:pg2}
\begin{align}
  \sigma^2 &= e\,,\label{eq:pg2a}\\
  (R\sigma)^2 &= e\,,\label{eq:pg2b}\\
  R^6 &= e\,,\label{eq:pg2c}
\end{align}
\end{subequations}
where $e$ is the identity transformation. Similar to the discussion and calculations in the last section, we need to find the $\mathbb{Z}_2$ gauge representation of the point group generators that are consistent with Eqs.~\eqref{eq:pg1} and \eqref{eq:pg2}. In Appendix~\ref{app:trgPSGalg}, we show that they can be brought to the form
\begin{subequations}\label{eq:pRep}
\begin{align}
  g_\sigma(x,y) &= (\epsilon_2)^{x y}\, g_{\sigma}\,,\label{eq:pReps}\\
  g_R(x,y) &= (\epsilon_2)^{x y + y(y+1)/2} g_{R}\,,\label{eq:pRepR}
\end{align}
\end{subequations}
where $g_\sigma$ and $g_R$ are translation-invariant SU(2) matrices.

As we discuss in more detail in Appendix, the point group relations \eqref{eq:pg2} translate into the following constraints on the constant matrices in Eq.~\eqref{eq:pRep}: $[Q_\sigma]^2 = ((\epsilon_{\sigma}) \mathbb{1}_2, e)$, $[Q_{R \sigma}]^2 = (\epsilon_{R\sigma})$, and $[Q_R]^6 = (\epsilon_{R})$. The solutions to these equations are given in Table~\ref{tab:PSGtrg}. The signs $\epsilon_\sigma, \epsilon_{R\sigma}, \epsilon_R$ are gauge invariant, so they obviously distinguish equivalence classes of projective representations of the point group. Note that they are identical for PSG Nos.~2 and 6, and for PSG Nos.~5 and 7, respectively. One can check that these PSGs are not gauge equivalent on the triangular lattice.\cite{tgular_ineq} The signs $\epsilon_{(\cdot)}$ are therefore not sufficient to distinguish PSG classes, and, in turn, $\mathbb{Z}_2$ quantum spin liquid phases.

In Table~\ref{tab:PSGtrg}, we choose particular gauges (or class representatives) for the point group representations. The column ``sym'' displays the remaining global gauge freedom after this gauge fixing. This freedom will be useful, as it can help to simplify the corresponding ansatz (invariant PSG) that will be constructed in the next section.

In the chosen gauge, the representation $g_R$ in Table~\ref{tab:PSGtrg} determines how the complex phase of spinon pairing changes under a $\pi/3$ lattice rotation [up to signs due to $\epsilon_2$ and $\tau_R$; see Eqs.~\eqref{eq:pRepR} and \eqref{eq:lg}]. Therefore, we anticipate that PSG Nos.~1 and 2 give rise to $s$-wave pairing. Similarly, Nos.~3 -- 5 potentially lead to $f$-wave, No.~6 to $d+id$-wave, and No.~7 to $p+ip$-wave parings. However, the paring amplitudes may vanish by symmetry in the ansatz (see next section). It is interesting to note that the highest possible angular momentum of spinon pairing for a QSL state on the triangular lattice is $f$-wave.

PSG representations on the triangular lattice were also discussed in two recent preprints.\cite{Mei15, Lu15} However, the classification in these papers is incomplete, as they missed the triangular PSG classes 6 and 7 in Table~\ref{tab:PSGtrg}. This leads to the missing of higher angular momentum spinon pairing, e.g., of type $d+i d$, as found in the time-reversal symmetric quadratic-band-touching~(QBT) state discussed in Ref.~[\onlinecite{Mishmash13_PRL.111.157203}].

Taking into account the sign $\epsilon_2$ and Table~\ref{tab:PSGtrg}, there are thus $2\times 7 = 14$ PSG representation classes for the space group of the triangular lattice. Note that the choice of time-reversal representation $g_\tr$ formally doubles the number of PSG classes to 28.\cite{trgSymPSG} As discussed previously, the gauge representation of time reversal does not play a role in the construction of chiral spin liquids. If, however, we want to impose time reversal on an ansatz (not in combination with a point-group symmetry), the choice of $g_\tr$ is sometimes relevant.

\subsection{Invariant ans{\"a}tze}\label{sec:invPSGtrg}

\begin{table}
\begin{tabular}{c|cc|cccc|cccc}
\hline\hline
No. & $\tau_{\sigma}$ & $\tau_{R}$ & $\epsilon_2$ & PSG & $g_\sigma$ & $g_R$ & $\lambda [\sigma_a]$ & $u_1 [\tau_\mu]$ & $u_2 [\tau_\mu]$ & $u_3 [\tau_\mu]$\\

\hline
1 & 0 & 0 & $+$ & 1 & $\mathbb{1}_2$ & $\mathbb{1}_2$ & $1,2,3$ & $3$ & $1,3$ & $1,2,3$\\
1a & 0 & 0 & $+$ & 2 & $i\sigma_3$ & $\mathbb{1}_2$ & 3 & 3 & 3 & 3\\
2 & 0 & 0 & $+$ & 6 & $i\sigma_1$ & $a$ & x & $1$ & $1$ & $1$\\
3 & 0 & 0 & $-$ & 4 & $i\sigma_3$ & $i\sigma_3$ & $3$ & $1$ & x & $3$\\
4 & 0 & 0 & $-$ & 3 & $\mathbb{1}_2$ & $i\sigma_3$ & $3$ & x & $1$ & $3$\\
5 & 0 & 0 & $-$ & 5 & $i\sigma_2$ & $i\sigma_3$ & x & $1$ & $2$ & x\\
6 & 0 & 0 & $-$ & 7 & $i\sigma_2$ & $b$ & x & 1 & 2 & x\\
\hline
7 & 1 & 0 & $+$ & 6 & $i\sigma_2$ & $a$ & $3$ & $1,3$ & $1,3$ & $1,3$\\
8 & 1 & 0 & $-$ & 7 & $i\sigma_1$ & $b$ & $3$ & $0,1$ & $0,2$ & $3$\\
9 & 1 & 0 & $-$ & 6 & $i\sigma_2$ & $a$ & $3$ & $0$ & $0$ & $1,3$\\
10 & 1 & 0 & $-$ & 5 & $i\sigma_1$ & $i\sigma_3$ & $3$ & $0,1$ & $0,2$ & $3$\\
10a & 1 & 0 & $-$ & 2 & $i\sigma_2$ & $\mathbb{1}_2$ & $3$ & $0$ & $0$ & $3$\\
10b & 1 & 0 & $-$ & 3 & $\mathbb{1}_2$ & $i\sigma_2$ & x & $0,3$ & $0$ & x\\
10c & 1 & 0 & $-$ & 4 & $i\sigma_2$ & $i\sigma_2$ & x & $0$ & $0,3$ & x\\
10d & 1 & 0 & $-$ & 1 & $\mathbb{1}_2$ & $\mathbb{1}_2$ & x & $0$ & $0$ & x\\

\hline\hline
\end{tabular}
\caption{Quantum spin liquids on the triangular lattice respecting rotation symmetry ($\tau_R=0$). All lattice symmetries are respected for $\tau_\sigma=0$; states No.~1a to 6 also respect time reversal, so they are {\it symmetric} QSLs.\cite{Wen02_PRB.65.165113} The reflection symmetries are broken in the Kalmeyer-Laughlin CSLs No.~7 to 10d ($\tau_\sigma=1$). Column ``PSG'' refers to the point group representations in Table~\ref{tab:PSGtrg}. $\lambda$ is the on-site field, and $u_a$ is the ansatz on links shown in Fig.~\ref{fig:trgsym}, in the notation of allowed real components $(\tau_\mu) = (i\mathbb{1}_2, \sigma_a)$; ``x'' means that the field must vanish by symmetry.  $a=\exp(i\pi\sigma_3/3)$ and $b = \exp(i\pi\sigma_3/6)$.
\label{tab:invPSGtrgZ2}}
\end{table}

In the last sections, we presented the algebraic PSG classes for the triangular lattice. We now introduce the corresponding ans\"atze $u$. As one can see from Eq.~\eqref{eq:sg}, the time-reversal signatures $\tau_\sigma$ and $\tau_R$ enter at this stage.

We restrict our discussion to first-, second-, and third-neighbor links of the ansatz. For our choice of symmetry generators $\sigma$ and $R$, it is convenient to impose the constraints on the links $u_1$, $u_2$, and $u_3$ shown in Fig.~\ref{fig:trgsym}. For each of these links, there are two constraint equations, and they are discussed in Appendix~\ref{app:trgPSGinv}. The solutions to the symmetry constraints are given in Table~\ref{tab:invPSGtrgZ2} for quantum spin liquids with unbroken rotation ($\tau_R=0$), and in Table~\ref{tab:invPSGtrgZ2sf} for staggered flux states ($\tau_R=1$). {\it A priori}, the number of ans\"atze for each of the four time-reversal signatures ($\tau_\sigma$, $\tau_R$) is the same as the number of algebraic PSG classes (i.e., seven). However, sometimes the resulting ans\"atze are redundant or trivial. In Tables~\ref{tab:invPSGtrgZ2} and \ref{tab:invPSGtrgZ2sf}, we only list ans\"atze that allow nonzero parameters on at least first- or second-neighbor links. Among the chiral states (i.e., at least one of $\tau_\sigma, \tau_R \neq 0$), we further omit the ones that cannot generally break time reversal. However, we include states that are special cases of others, and we denote them by a, b, etc., in column ``No.''.

\begin{table}
\begin{tabular}{c|cc|cccc|cccc}
\hline\hline
No. & $\tau_{\sigma}$ & $\tau_{R}$ & $\epsilon_2$ & PSG & $g_\sigma$ & $g_R$ & $\lambda [\sigma_a]$ & $u_1 [\tau_\mu]$ & $u_2 [\tau_\mu]$ & $u_3 [\tau_\mu]$\\

\hline
11 & 0 & 1 & $+$ & 3 & $\mathbb{1}_2$ & $i\sigma_2$ & $1,3$ & $0,3$ & $1,3$ & $0,1,3$\\
11a & 0 & 1 & $+$ & 5 & $i\sigma_3$ & $i\sigma_1$ & $3$ & $0,3$ & $3$ & $0,3$\\
12 & 0 & 1 & $+$ & 7 & $i\sigma_1$ & $b$ & x & $0,1$ & $1$ & $0,1$\\
12a & 0 & 1 & $+$ & 1 & $\mathbb{1}_2$ & $\mathbb{1}_2$ & x & $0$ & x & $0$\\
13 & 0 & 1 & $-$ & 5 & $i\sigma_3$ & $i\sigma_1$ & $3$ & $1$ & x & $0,3$\\
14 & 0 & 1 & $-$ & 3 & $\mathbb{1}_2$ & $i\sigma_2$ & $1,3$ & x & $2$ & $0,3$\\
15 & 0 & 1 & $-$ & 2 & $i\sigma_1$ & $\mathbb{1}_2$ & x & $3$ & $1$ & $0$\\
16 & 0 & 1 & $-$ & 7 & $i\sigma_2$ & $b$ & x & 3 & x & $0,1$\\
17 & 0 & 1 & $-$ & 6 & $i\sigma_2$ & $a$ & x & $1,3$ & $2$ & $0$\\
17a & 0 & 1 & $-$ & 1 & $\mathbb{1}_2$ & $\mathbb{1}_2$ & x & x & 3 & 0\\
\hline

18 & 1 & 1 & $+$ & 4 & $i\sigma_2$ & $i\sigma_2$ & $1,3$ & $3$ & $0,1,3$ & $1,3$\\
18a & 1 & 1 & $+$ & 5 & $i\sigma_1$ & $i\sigma_2$ & $3$ & $3$ & $0,3$ & $3$\\
19 & 1 & 1 & $+$ & 7 & $i\sigma_2$ & $b$ & x & $1$ & $0,1$ & $1$\\
19a & 1 & 1 & $+$ & 1 & $\mathbb{1}_2$ & $\mathbb{1}_2$ & x & x & $0$ & x\\
20 & 1 & 1 & $-$ & 6 & $i\sigma_1$ & $a$ & x & $1$ & $2,3$ & x\\

\hline\hline
\end{tabular}
\caption{Chiral spin liquids on the triangular lattice with broken lattice rotation ($\tau_R=1$), i.e., staggered flux phases. The notation is the same as in Table~\ref{tab:invPSGtrgZ2}.
\label{tab:invPSGtrgZ2sf}}
\end{table}

\begin{figure*}
\includegraphics[width=.8\textwidth]{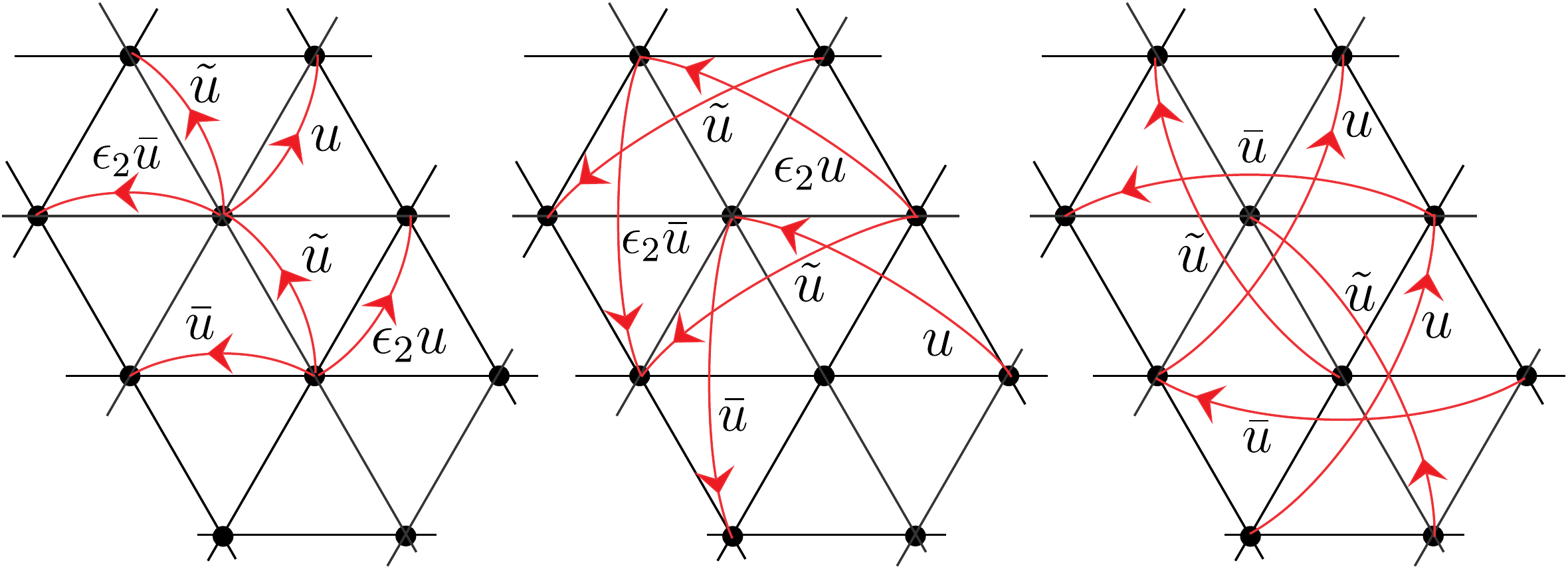}\caption{First-, second-, and third-neighbor mean-fields propagated to the doubled unit cell of the triangular lattice. The sign $\epsilon_2 = \pm 1$ labels the translation representation; $\tilde u = (-)^{\tau_R} g_R u [g_R]^\dagger$ and $\bar u = (g_R)^2 u [(g_R)^2]^\dagger$ are rotated mean fields, depending on $\tau_R\in\{0,1\}$ and on the gauge representation $g_R$. The allowed components of $u$ are specified in Tables~\ref{tab:invPSGtrgZ2} and \ref{tab:invPSGtrgZ2sf} for each PSG.\label{fig:dcelltrg}}
\end{figure*}

As described in Sec.~\ref{sec:invPSGgen}, the ansatz on a link given in Tables~\ref{tab:invPSGtrgZ2} and \ref{tab:invPSGtrgZ2sf} is propagated to the entire lattice by rotation and translation. In our (Landau) gauge \eqref{eq:g12}, the translation representation in $x$-direction is uniform, while translation in $y$-direction may lead to additional signs if $\epsilon_2=-1$. It is therefore sufficient to double the unit cell of the lattice in $y$-direction. The rotation and translation of the mean field to the doubled unit cell of the triangular lattice is explicitly shown in Fig.~\ref{fig:dcelltrg}, in terms of the allowed $u$ given in the tables. Here, the corresponding representations of rotation and translation must be used for the propagation.

Note that the spinon unit cell doubling for $\epsilon_2=-1$ can lead to global $\pi$-fluxes through holes of the lattice torus when the linear system size is an odd multiple of two. In this case, suitable antiperiodic spinon boundary conditions have to be chosen in order to restore lattice rotation symmetry. These subtleties do not arise when the linear system size is a multiple of four.\cite{mambriniNote}

In our gauge, the on-site fields $\lambda_a$ are uniform, i.e., independent of lattice site. As discussed previously, they correspond to chemical potential and complex on-site pairing terms for the spinon. In Tables~\ref{tab:invPSGtrgZ2} and \ref{tab:invPSGtrgZ2sf}, we give the on-site fields allowed by symmetry. In actual calculations (mean-field or projection), they must be adjusted such that the three constraints $\langle G_a\rangle = 0$ are satisfied (on average or at every site, respectively). If possible, we simplify the allowed fields $u_a$ using the remaining global gauge symmetry given in Table~\ref{tab:PSGtrg} in the column ``sym''.

The ans\"atze for quantum spin liquids on the triangular lattice up to third neighbors (Tables~\ref{tab:invPSGtrgZ2} and \ref{tab:invPSGtrgZ2sf}) will be analyzed in more detail elsewhere. In the remainder of this section, we discuss some general properties and relations with known phases.

Since $\tau_\sigma = \tau_R = 0$, states 1 through 6 in Table~\ref{tab:invPSGtrgZ2} conserve all lattice symmetries. Among those states, only No.~1 can break time reversal. All others have coplanar spatial ansatz components $u$, so they automatically respect also time reversal, in accordance with a ``PT theorem'' (see Secs.~\ref{sec:TRconstraint} and \ref{sec:PTtheorem}).

State No.~1 in Table~\ref{tab:invPSGtrgZ2} is the conventional, linear representation of the space group, allowing uniform real hopping and $s$-wave pairing amplitudes. Fixing the global SU(2) gauge symmetry, we can simplify the first neighbor to pure hopping, and the second neighbor to hopping and real pairing. The third-neighbor mean field (or the on-site field) can then have both hopping and complex pairing, thus breaking time reversal. Imposing time reversal in the symmetric liquid would limit the third neighbor to real hopping and pairing.

Restricting No.~1 (or 1a) to first-neighbor hopping results in a U(1) state with a large circular spinon Fermi surface. This state is known to yield low variational energies, and a good description of the ground state of the Heisenberg model with quite large ring-exchange term.\cite{Motrunich05_PRB.72.045105, *Motrunich06_PRB.73.155115, Lee05_PRL_95_036403, Grover10_PRB_81_245121}

The symmetric phase No.~2 in Table~\ref{tab:invPSGtrgZ2} is the so-called $d + i d$ ``quadratic-band-touching''~(QBT) state which has recently been found to yield competitive variational energy in the first-neighbor Heisenberg model with positive, but not too strong ring-exchange term.\cite{Mishmash13_PRL.111.157203} It is gapless with quadratic spinon bands touching at momentum ${\bm k}=0$ in the Brillouin zone. Note that an additional real hopping leads to the more familiar, fully gapped chiral topological $d + i d$ state, No.~7 in this table.\cite{Mishmash13_PRL.111.157203, Grover10_PRB_81_245121} Both the U(1) state with a large spinon Fermi surface discussed above, as well as the QBT state are strong contenders for the physics realized in organic spin liquid candidate materials.\cite{Shimuzi03_PRL_91_107001}

The QBT state No.~2, as well as No.~6 in Table~\ref{tab:invPSGtrgZ2} are symmetric QSL ans\"atze on the triangular lattice that were missed in two recent preprints.\cite{Lu15, Mei15} 
As discussed in Sec.~\ref{sec:trgPTgroup}, the triangular-lattice PSG classification in these preprints is incomplete, as they did not find point group representations 6 and 7 in Table~\ref{tab:PSGtrg}.

The symmetric state with a real first-neighbor pairing and doubling of the spinon unit cell (No.~3 in Table~\ref{tab:invPSGtrgZ2}) was recently discussed.\cite{Lu15, Mei15} This state (dubbed ``$\pi$ flux'') has a Dirac spectrum,\cite{Lu15} and it was found to yield low variational energy in the triangular $J_1$-$J_2$ Heisenberg antiferromagnet.\cite{Mei15} 

We see that some particular states among the ones obtained through our exhaustive classification have previously been discussed in the literature. However, a systematic mean-field or variational investigation of all chiral states on the triangular lattice is an open problem.

\section{Kagome PSG}\label{sec:kagPSG}

\begin{figure}
\includegraphics[width=.4\textwidth]{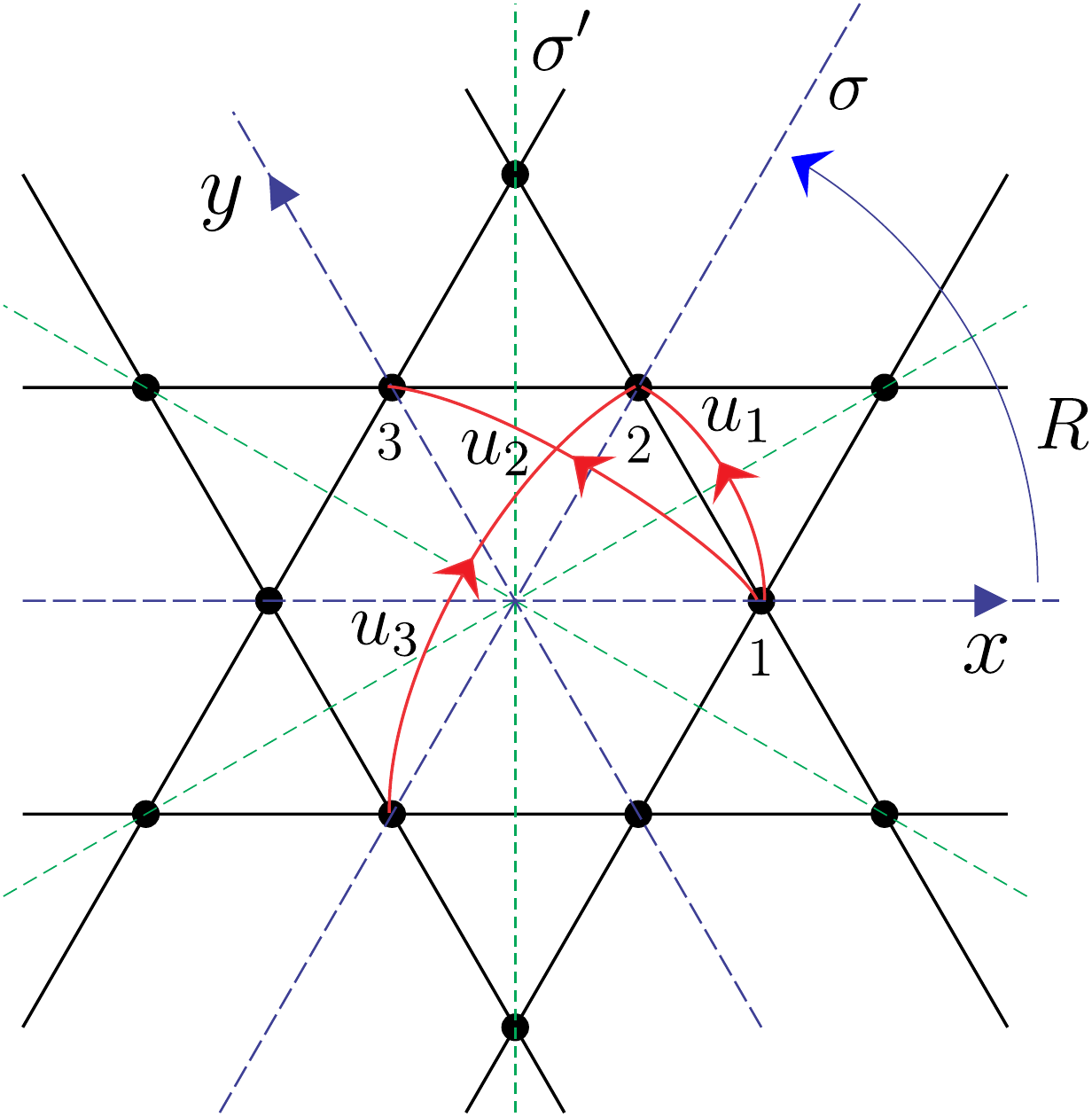}\caption{Symmetry generators and ansatz parameters $u$ for the kagome lattice.\label{fig:kagsym}}
\end{figure}

\begin{table}
\begin{tabular}{c|cc|ccc|c}
\hline\hline
No. & $g_{\sigma}$ & $g_{R}$ & $\epsilon_{\sigma}$ & $\epsilon_{R\sigma}$ & $\epsilon_{R}$ & sym\\
\hline
1 & $\mathbb{1}_2$ & $\mathbb{1}_2$ & $+$ & $+$ & $+$ & SU(2)\\
2 & $i\sigma_3$ & $\mathbb{1}_2$ & $-$ & $-$ & $+$ & U(1)\\
3 & $\mathbb{1}_2$ & $i\sigma_3$ & $+$ & $-$ & $-$ & U(1)\\
4 & $i\sigma_3$ & $i\sigma_3$ & $-$ & $+$ & $-$ & U(1)\\
5 & $i\sigma_2$ & $i\sigma_3$ & $-$ & $-$ & $-$ & $\mathbb{Z}_2$\\
\hline\hline
\end{tabular}
\caption{PSG representations of the point group for the kagome lattice. The notation is the same as in Table~\ref{tab:PSGtrg}.
\label{tab:PSGkag}}
\end{table}

Next, we discuss the PSG construction for the kagome lattice. On the kagome lattice, the same triangular Eqs.~\eqref{eq:Txy}, \eqref{eq:pg1}, and \eqref{eq:pg2} define the space group. However, the analysis is slightly more complicated because the unit cell contains {\it three} sites instead of just one. We choose the sublattice indices $\{1,2,3\}$ shown in Fig.~\ref{fig:kagsym}. Equations~\eqref{eq:Txy} and \eqref{eq:pg1} do not act on the sublattice index, but Eqs.~\eqref{eq:pg2} do. Fortunately, one can show (see Appendix~\ref{app:kagPSG}) that there is always a canonical gauge where the representations $g_R$ and $g_\sigma$ are independent of sublattice site. Therefore, the functional form of the space group representations given in Eqs.~\eqref{eq:g12} and \eqref{eq:pRep} also apply for the kagome lattice.

Since the algebraic relations among the space group generators are identical, the kagome lattice naively has the same PSG classes as the triangular lattice, listed in Table~\ref{tab:PSGtrg}. However, a translation-invariant gauge transformation exists on the kagome lattice that identifies the triangular PSG No.~6 with No.~2, and No.~7 with No.~5. When $g_R = \exp(i\beta\sigma_3)$, the sublattice gauge transformation
\begin{equation}\label{eq:sublGT}
  [g_s] = [e^{-i\pi\sigma_3/3}, \mathbb{1}_2, e^{i\pi\sigma_3/3}]
\end{equation}
changes $\beta\mapsto\beta+\pi/3$, but leaves $g_\sigma$ invariant; ($s$ is the sublattice index in Fig.~\ref{fig:kagsym}). Therefore, there are only {\it five} point group representations for the kagome lattice, listed in Table~\ref{tab:PSGkag}. Finally, including unit cell doubling $\epsilon_2$, there are in total $2\times 5 = 10$ $\mathbb{Z}_2$ PSG classes for the space group of the kagome lattice.\cite{kagSymPSG}

As we discussed in the last section, the complex roots of unity for the triangular-lattice rotation representation $g_R$ in PSGs No.~6 and 7 in Table~\ref{tab:PSGtrg} leads to ans\"atze with $d+id$ and $p+ip$ pairing symmetry, respectively. The existence of the sublattice transformation \eqref{eq:sublGT} therefore has an interesting interpretation: we can say that $d+id$-wave pairing is gauge equivalent to $s$-wave spinon pairing, while $p+ip$-wave is gauge equivalent to $f$-wave spinon pairing on the kagome lattice.

\subsection{Invariant ans{\"a}tze}\label{sec:invPSGkag}

\begin{figure*}
\includegraphics[width=.8\textwidth]{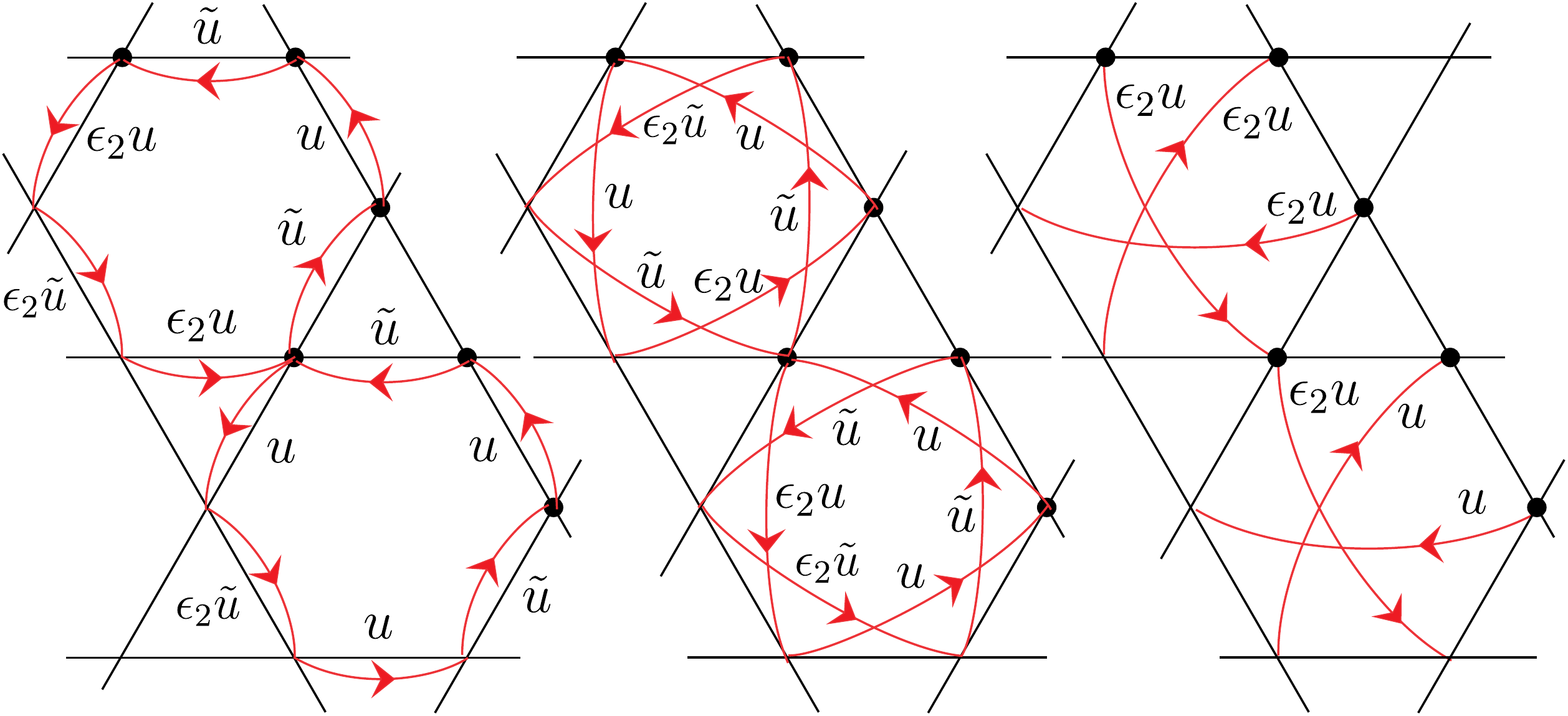}\caption{First-, second-, and diagonal mean fields propagated to the doubled unit cell of the kagome lattice. The sign $\epsilon_2 = \pm 1$ labels the translation representation, and $\tilde u = (-)^{\tau_R} g_R u [g_R]^\dagger$ is the rotated mean field, depending on rotation breaking $\tau_R\in\{0,1\}$ and on the representation $g_R$. The allowed components of $u$ are given in Tables~\ref{tab:invPSGkagZ2} and \ref{tab:invPSGkagZ2sf}.
\label{fig:dcellkag}}
\end{figure*}

\begin{table}
\begin{tabular}{c|cc|cccc|cccc}
\hline\hline
No. & $\tau_{\sigma}$ & $\tau_{R}$ & $\epsilon_2$ & PSG & $g_\sigma$ & $g_R$ & $\lambda[\sigma_a]$ & $u_1[\tau_\mu]$ & $u_2[\tau_\mu]$ & $u_3[\tau_\mu]$\\

\hline
1 & 0 & 0 & $+$ & 1 & $\mathbb{1}_2$ & $\mathbb{1}_2$ & $1,2,3$ & 3 & $1,3$ & $1,2,3$\\
1a & 0 & 0 & $+$ & 2 & $i\sigma_3$ & $\mathbb{1}_2$ & 3 & 3 & $3$ & 3\\
2 & 0 & 0 & $+$ & 4 & $i\sigma_3$ & $i\sigma_3$ & 3 & $1,3$ & 3 & 3\\
3 & 0 & 0 & $+$ & 3 & $\mathbb{1}_2$ & $i\sigma_3$ & 3 & 3 & $1,3$ & 3\\
4 & 0 & 0 & $+$ & 5 & $i\sigma_1$ & $i\sigma_2$ & x & $3$ & $1$ & x\\
5 & 0 & 0 & $-$ & 1 & $\mathbb{1}_2$ & $\mathbb{1}_2$ & $1,3$ & 3 & $1,3$ & x\\
6 & 0 & 0 & $-$ & 4 & $i\sigma_3$ & $i\sigma_3$ & 3 & $1,3$ & 3 & $1,2$\\
7 & 0 & 0 & $-$ & 3 & $\mathbb{1}_2$ & $i\sigma_3$ & 3 & 3 & $1,3$ & x\\
7a & 0 & 0 & $-$ & 2 & $i\sigma_3$ & $\mathbb{1}_2$ & 3 & 3 & 3 & x\\
8 & 0 & 0 & $-$ & 5 & $i\sigma_1$ & $i\sigma_2$ & x & 3 & 1 & 3\\
\hline

9 & 1 & 0 & $+$ & 2 & $i\sigma_2$ & $\mathbb{1}_2$ & $1,3$ & $0,3$ & $0,1,3$ & $1,3$\\
10 & 1 & 0 & $+$ & 5 & $i\sigma_1$ & $i\sigma_3$ & 3 & $0,1,3$ & $0,2,3$ & 3\\
10a & 1 & 0 & $+$ & 3 & $\mathbb{1}_2$ & $i\sigma_2$ & x & $0,3$ & $0$ & x\\
10b & 1 & 0 & $+$ & 4 & $i\sigma_2$ & $i\sigma_2$ & x & $0$ & $0,3$ & x\\
10c & 1 & 0 & $+$ & 1 & $\mathbb{1}_2$ & $\mathbb{1}_2$ & x & $0$ & $0$ & x\\
11 & 1 & 0 & $-$ & 2 & $i\sigma_2$ & $\mathbb{1}_2$ & $1,3$ & $0,3$ & $0,1,3$ & 0\\
12 & 1 & 0 & $-$ & 5 & $i\sigma_1$ & $i\sigma_3$ & 3 & $0,1,3$ & $0,2,3$ & $0,1$\\
13 & 1 & 0 & $-$ & 3 & $\mathbb{1}_2$ & $i\sigma_2$ & x & $0,3$ & 0 & $0,1,3$\\
12a & 1 & 0 & $-$ & 4 & $i\sigma_2$ & $i\sigma_2$ & x & $0$ & $0,3$ & $0$\\
12b & 1 & 0 & $-$ & 1 & $\mathbb{1}_2$ & $\mathbb{1}_2$ & x & 0 & 0 & 0\\

\hline\hline
\end{tabular}
\caption{Quantum spin liquids on the kagome lattice respecting rotation symmetry ($\tau_R=0$). Nos.~1 to 8 are liquids that do not break any lattice symmetry ($\tau_\sigma=0$). Nos.~9 to 13 are Kalmeyer-Laughlin CSL states that break all reflections ($\tau_\sigma=1$). Column ``PSG'' refers to the point group representations in Table~\ref{tab:PSGkag}. $\lambda$ is the on-site field, and the last three columns specify the ansatz $u$ on links shown in Fig.~\ref{fig:kagsym}, in the notation of allowed real components $(\tau_\mu) = (i\mathbb{1}_2, \sigma_a)$; ``x'' means that the field must vanish by symmetry.
\label{tab:invPSGkagZ2}}
\end{table}

\begin{table}
\begin{tabular}{c|cc|cccc|cccc}
\hline\hline
No. & $\tau_{\sigma}$ & $\tau_{R}$ & $\epsilon_2$ & PSG & $g_\sigma$ & $g_R$ & $\lambda[\sigma_a]$ & $u_1[\tau_\mu]$ & $u_2[\tau_\mu]$ & $u_3[\tau_\mu]$\\
\hline

15 & 0 & 1 & $+$ & 3 & $\mathbb{1}_2$ & $i\sigma_2$ & $1,3$ & $0,3$ & $1,2,3$ & $0,1,2$\\
16 & 0 & 1 & $+$ & 5 & $i\sigma_3$ & $i\sigma_1$ & 3 & $0,1,3$ & 3 & $0,3$\\
14 & 0 & 1 & $+$ & 2 & $i\sigma_3$ & $\mathbb{1}_2$ & x & $0,1$ & 3 & 0\\
14a & 0 & 1 & $+$ & 1 & $\mathbb{1}_2$ & $\mathbb{1}_2$ & x & $0$ & 3 & 0\\
14b & 0 & 1 & $+$ & 4 & $i\sigma_3$ & $i\sigma_3$ & x & $0$ & 3 & 0\\
18 & 0 & 1 & $-$ & 3 & $\mathbb{1}_2$ & $i\sigma_2$ & $1,3$ & $0,3$ & $1,2,3$ & x\\
19 & 0 & 1 & $-$ & 5 & $i\sigma_3$ & $i\sigma_1$ & 3 & $0,1,3$ & 3 & 1\\
17 & 0 & 1 & $-$ & 2 & $i\sigma_3$ & $\mathbb{1}_2$ & x & $0,1$ & 3 & $1,2$\\
17a & 0 & 1 & $-$ & 1 & $\mathbb{1}_2$ & $\mathbb{1}_2$ & x & $0$ & 3 & x\\
17b & 0 & 1 & $-$ & 4 & $i\sigma_3$ & $i\sigma_3$ & x & $0$ & 3 & x\\
\hline

21 & 1 & 1 & $+$ & 4 & $i\sigma_2$ & $i\sigma_2$ & $1,3$ & $2,3$ & $0,1,3$ & $1,3$\\
22 & 1 & 1 & $+$ & 5 & $i\sigma_1$ & $i\sigma_2$ & 3 & 3 & $0,2,3$ & 3\\
20 & 1 & 1 & $+$ & 2 & $i\sigma_3$ & $\mathbb{1}_2$ & x & 3 & $0,1$ & x\\
20a & 1 & 1 & $+$ & 1 & $\mathbb{1}_2$ & $\mathbb{1}_2$ & x & 3 & 0 & x\\
20b & 1 & 1 & $+$ & 3 & $\mathbb{1}_2$ & $i\sigma_3$ & x & 3 & 0 & x\\
25 & 1 & 1 & $-$ & 4 & $i\sigma_2$ & $i\sigma_2$ & $1,3$ & $2,3$ & $0,1,3$ & 2\\
26 & 1 & 1 & $-$ & 5 & $i\sigma_2$ & $i\sigma_1$ & 3 & 3 & $0,1,3$ & x\\
24 & 1 & 1 & $-$ & 2 & $i\sigma_3$ & $\mathbb{1}_2$ & x & 3 & $0,1$ & 3\\
23 & 1 & 1 & $-$ & 1 & $\mathbb{1}_2$ & $\mathbb{1}_2$ & x & 3 & 0 & $1,3$\\
23a & 1 & 1 & $-$ & 3 & $\mathbb{1}_2$ & $i\sigma_3$ & x & 3 & 0 & $3$\\

\hline\hline
\end{tabular}
\caption{Chiral spin liquids on the kagome lattice breaking rotation symmetry ($\tau_R=1$), i.e., staggered flux states. 
The notation is the same as in Table~\ref{tab:invPSGkagZ2}.
\label{tab:invPSGkagZ2sf}}
\end{table}

In this section, we present the invariant PSG (ans\"atze) for the kagome lattice. The constraint equations are solved for the links shown in Fig.~\ref{fig:kagsym}. They are discussed in Appendix~\ref{app:kagPSGinv}, and the solutions are given in Tables~\ref{tab:invPSGkagZ2} and \ref{tab:invPSGkagZ2sf}. The fields $u_1$, $u_2$, and $u_3$ are explicitly propagated to the doubled unit cell in Fig.~\ref{fig:dcellkag}. In this figure, we denote $\tilde u_a = (-)^{\tau_R} g_R u_a [g_R]^\dagger$, where $g_R$ depends on the particular PSG representation. In contrast to the triangular lattice, there is only {\it one} symmetry constraint on first- and second-neighbor links of the kagome lattice. For this reason, there are no trivial or redundant ans\"atze to omit in this case. When a state is a special case of another, we indicate this by a, b, etc., in the column ``No.'', as was also done for the triangular case in Sec.~\ref{sec:trgPSG}.

In Table~\ref{tab:invPSGkagZ2}, we list ans\"atze with unbroken rotation symmetry ($\tau_R = 0)$. No.~1 through 8 (with $\tau_\sigma=0$) respect the full lattice space group. In most cases, time reversal is automatically respected, so they are {\it symmetric} liquids.\cite{LuRanLee11_PRB.83.224413} However, similar to the triangular lattice, there are some states that generally break time reversal, thus violating a ``PT theorem'' (see Sec.~\ref{sec:PTtheorem}). They are Nos.~1 and 6 in this table. For $\tau_\sigma=1$, all reflection symmetries of the lattice are broken, and both small (first neighbor) and large (second neighbor) triangles can have nontrivial fluxes.

The state resulting from trivial (linear) representation of the space group, No.~1 in Table~\ref{tab:invPSGkagZ2}, has simple uniform hopping and pairing terms. The first-neighbor U(1) state has a large Fermi surface. No spin model has been found so far where this state yields low variational energy.\cite{HsuSchofield91, Iqbal11_PRB.84.020407} The first-neighbor state No.~7 with a doubled spinon unit cell ($\epsilon_2=-1$) is the ``Dirac spin liquid'' that has been found to yield excellent ground state energy for the nearest-neighbor Heisenberg model.\cite{Ran07_PRL.98.117205, *Hermele08_PRB.77.224413, Iqbal13_PRB.87.060405, Iqbal14_PRB.89.020407, Iqbal15_PRB.91.020402} This gapless U(1) state with Dirac spectrum is a strong candidate for explaining the physics of the herbertsmithite QSL material.\cite{Han2012}

The chiral state No.~13 in Table \ref{tab:invPSGkagZ2} (first neighbor) has uniform U(1) fluxes $\theta$ through elementary triangles, and $\pi-2\theta$ through hexagons, respectively. This chiral spin liquid was discussed by Marston and Zheng,\cite{MarstonZheng91} and by Hastings.\cite{Hastings00_PRB.63.014413} It has recently been found provide a good description of the ground state of the antiferromagnetic $J_1$-$J_2$-$J_d$ Heisenberg in some parameter range.\cite{HuBeccaSheng_PRB.91.041124, SterLauchli15_PRB.92.125122} 

\begin{figure}
  \includegraphics[width=0.35\textwidth]{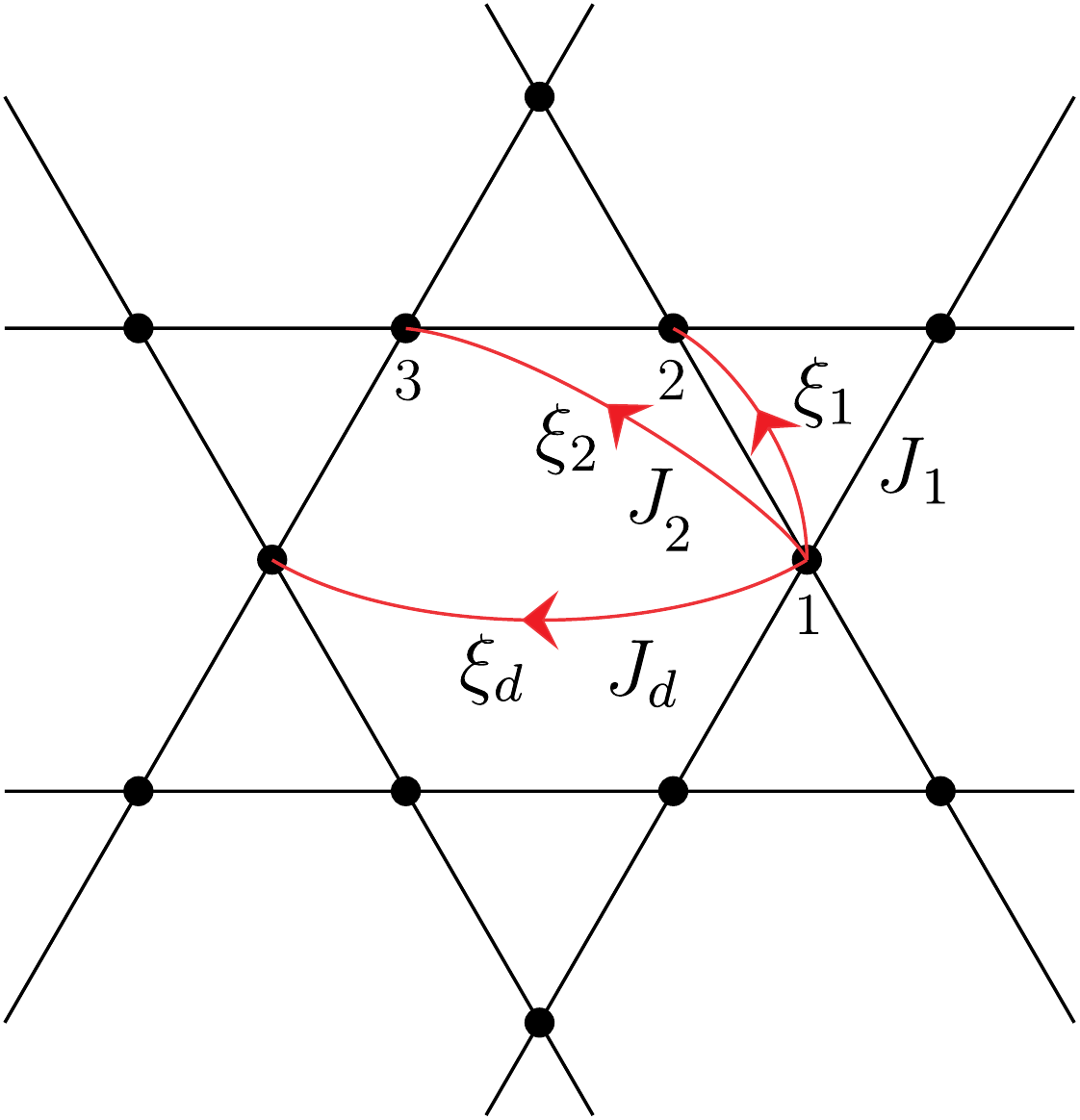}\caption{
Hopping parameters $\xi_a$ on first, second, and diagonal lattice links as used in the main text. The propagation of these parameters to the entire lattice is then done using the algebraic PSG. The exchange interactions $J_1$, $J_2$, and $J_d$ as used in Eq.~\eqref{eq:model} are also given.
\label{fig:hoppings}}
\end{figure}

In Ref.~[\onlinecite{Bieri15_PRB.92.060407}], we found the U(1) states dubbed CSL~A and CSL~B to yield low variational energies in the $J_1$-$J_2$-$J_d$ Heisenberg model with a dominant antiferromagnetic $J_d$ interaction across the diagonals of the hexagon. CSL~A and B are Nos.~12a and 13, respectively, in Table~\ref{tab:invPSGkagZ2}. These chiral states with spinon Fermi surfaces have no U(1) flux through the hexagons of the lattice, but variable fluxes through the elementary triangles. CSL~A can explain several of the intriguing physical properties observed in the kapellasite QSL candidate material.\cite{Fak12_PRL_109_037208}

In Table~\ref{tab:invPSGkagZ2sf}, we list staggered flux states, i.e., those with broken rotation symmetry ($\tau_R=1$). The kagome lattice has six reflection axes: $\sigma$ and $\sigma' = R\sigma$ shown in Fig.~\ref{fig:kagsym}, and the ones rotated by $R$ and $R^2$. In the staggered flux states, three out of these six symmetry axes are broken. For $\tau_\sigma=0$, nontrivial gauge flux is allowed on small (first neighbor) triangles of the lattice, while for $\tau_\sigma=1$, nontrivial flux is allowed on large (second neighbor) triangles.

\subsection{Phase diagram}

\begin{figure}
\includegraphics[width=.96\columnwidth]{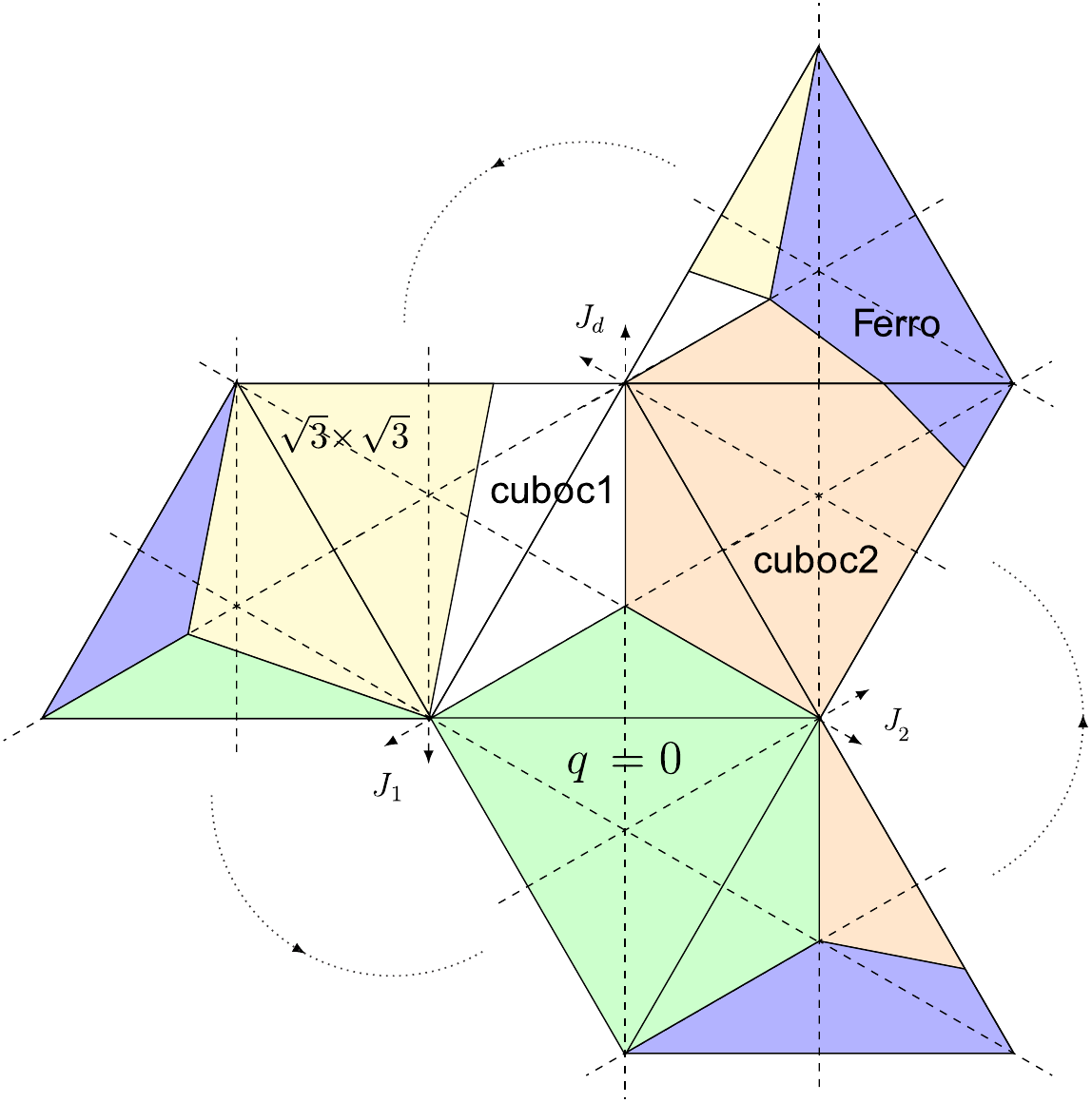}\caption{Ternary phase diagrams of the classical $J_1$-$J_2$-$J_d$ Heisenberg model on the kagome lattice for all signs of exchange interactions $J_n$ (except the fully ferromagnetic case). The parameters are normalized to $|J_1|+|J_2|+|J_d| = 1$. The triangle in the center shows the fully antiferromagnetic case ($J_n\geq0$).
\label{fig:classicalPD}}
\end{figure}

Recently, the kagome-lattice Heisenberg model with exchange interactions on first, second, and diagonal neighbors across the hexagons has gained attention because of potential realization of chiral spin liquid ground states.\cite{Messio12_PRL.108.207204, Gong2014, Gong15_PRB.91.075112, HuBeccaSheng_PRB.91.041124, Bieri15_PRB.92.060407, Gong15_PRB.91.075112, SterLauchli15_PRB.92.125122, HeShengChen14_PRL.112.137202, Iqbal15_PRB.92.220404} Here, we present a quantum phase diagram for the fully antiferromagnetic case, using Gutzwiller projected wave functions for the subset of classified U(1) CSL states. The phase diagram for {\it ferromagnetic} first- and second-neighbor interactions, relevant for the kapellasite material, was presented in Ref.~[\onlinecite{Bieri15_PRB.92.060407}].

The Heisenberg model we want to study is
\begin{equation}\label{eq:model}
H = J_1\sum_{\langle i,j\rangle} \mathbf{S}_i \cdot \mathbf{S}_j +
J_2 \sum_{\langle\langle i,j\rangle\rangle} {\bf S}_i \cdot {\bf S}_j
+ J_d \sum_{\langle i,j\rangle_d} {\bf S}_i \cdot {\bf S}_j
\end{equation}
where the exchange interactions $J_n$ on first, second, and diagonal links are defined in Fig.~\ref{fig:hoppings}.

In Fig.~\ref{fig:classicalPD}, we show the phase diagram for {\it classical} spins on this lattice, using states with regular magnetic orders.\cite{Messio11_PRB_83_184401} The seven ternary phase diagrams represent all combinations of signs for the three exchange couplings (except the fully ferromagnetic case). The interactions are normalized to $|J_1|+|J_2|+|J_d| = 1$. The central triangle is the fully antiferromagnetic case with $J_n\geq 0$. The top right triangle has ferromagnetic first- and second-neighbor interactions, $J_1, J_2\leq0$, and a frustrating diagonal interaction $J_d \geq 0$.\cite{Bieri15_PRB.92.060407}

\begin{table}
\begin{tabular}{c|cc|ccc|cccc|c}
\hline
\hline
No.& $\tau_{\sigma}$ & $\tau_{R}$ & $\epsilon_2$ & $g_{\sigma}$& $g_R$ & $\mu$ & $\beta_1$ & $\beta_2$ & $\beta_d$ & Description\\
\hline
1  & 0 & 0 & $+$ & $\mathbb{1}_2$ & $\mathbb{1}_2$ & $\mu$ & $0$ & $0$ & $0$ & large FS\\
2  & 0 & 0 & $+$ & $i\sigma_2$ & $i\sigma_2$ & x & $0$ & x & x & flat band\\
3  & 0 & 0 & $+$ & $\mathbb{1}_2$ & $i\sigma_2$ & x & x & $0$ & x & flat band\\
4  & 0 & 0 & $-$ & $\mathbb{1}_2$ & $\mathbb{1}_2$ & $\mu$ & $0$ & $0$ & x & Dirac\cite{Ran07_PRL.98.117205,*Hermele08_PRB.77.224413}\\
5  & 0 & 0 & $-$ & $i\sigma_2$ & $i\sigma_2$ & x & $0$ & x & $0$ & line FS\\
6  & 0 & 0 & $-$ & $\mathbb{1}_2$ & $i\sigma_2$ & x & x & $0$ & x & flat band\\
\hline
7 & 1 & 0 & $+$ & $i\sigma_2$ & $\mathbb{1}_2$ & $\mu$ & $\beta_1$ & $\beta_2$ & $0$ & FS\\
8 & 1 & 0 & $+$ & $\mathbb{1}_2$ & $i\sigma_2$ & x & $\beta_1$ & $\pi/2$ & x & flat band\\
9 & 1 & 0 & $+$ & $i\sigma_2$ & $i\sigma_2$ & x & $\pi/2$ & $\beta_2$ & x & flat band\\
10 & 1 & 0 & $-$ & $i\sigma_2$ & $\mathbb{1}_2$ & $\mu$ & $\beta_1$ & $\beta_2$ & $\pi/2$ & CSL~C\cite{MarstonZheng91,Hastings00_PRB.63.014413}\\
11 & 1 & 0 & $-$ & $\mathbb{1}_2$ & $i\sigma_2$ & x & $\beta_1$ & $\pi/2$ & $\beta_d$ & CSL~B\\
12 & 1 & 0 & $-$ & $i\sigma_2$ & $i\sigma_2$ & x & $\pi/2$ & $\beta_2$ & $\pi/2$ & CSL~A\\
\hline\hline
\end{tabular}
\caption{
QSL phases on the kagome lattice with U(1) gauge structure and unbroken rotation symmetry ($\tau_R=0$). The states with $\tau_\sigma=0$ respect all symmetries, including time reversal, while $\tau_\sigma=1$ are Kalmeyer-Laughlin chiral spin liquids. $\epsilon_2=-1$ indicates doubling of the spinon unit cell. $\mu$ is the chemical potential, and $\beta_a = \text{arg}(\xi_a)$ are the allowed hopping phases on the links shown in Fig.~\ref{fig:dcellkag}; ``x'' means $\xi_a=0$.
\label{tab:U1}}
\end{table}

\begin{table}
\begin{tabular}{c|cc|ccc|cccc|c}
\hline
\hline
No.& $\tau_{\sigma}$ & $\tau_{R}$ & $\epsilon_2$ & $g_{\sigma}$& $g_R$ & $\mu$ & $\beta_1$ & $\beta_2$ & $\beta_d$ & Description\\
\hline
9  & 0 & 1 & $+$ & $\mathbb{1}_2$ & $i\sigma_2$ & $\mu$ & $\beta_1$ & $0$ & $\beta_d$ & FS\\
10  & 0 & 1 & $+$ & $i\sigma_2$ & $\mathbb{1}_2$ & x & $\beta_1$ & x & $\pi/2$ & line FS\\
11  & 0 & 1 & $+$ & $\mathbb{1}_2$ & $\mathbb{1}_2$ & x & $\pi/2$ & $0$ & $\pi/2$ & line FS\\
12 & 0 & 1 & $-$ & $\mathbb{1}_2$ & $i\sigma_2$ & $\mu$ & $\beta_1$ & $0$ & x & FS/Dirac\\
13  & 0 & 1 & $-$ & $i\sigma_2$ & $\mathbb{1}_2$ & x & $\beta_1$ & x & $0$ & FS/Dirac\\
14  & 0 & 1 & $-$ & $\mathbb{1}_2$ & $\mathbb{1}_2$ & x & $\pi/2$ & $0$ & x & line FS\\
\hline
15 & 1 & 1 & $+$ & $i\sigma_2$ & $i\sigma_2$ & $\mu$ & $0$ & $\beta_2$ & $0$ & FS\\
16  & 1 & 1 & $+$ & $i\sigma_2$ & $\mathbb{1}_2$ & x & x & $\beta_2$ & x &line FS\\
17  & 1 & 1 & $+$ & $\mathbb{1}_2$ & $\mathbb{1}_2$ & x & $0$ & $\pi/2$ & x & line FS\\
18  & 1 & 1 & $-$ & $i\sigma_2$ & $i\sigma_2$ & $\mu$ & $0$ & $\beta_2$ & x & FS/Dirac\\
19  & 1 & 1 & $-$ & $i\sigma_2$ & $\mathbb{1}_2$ & x & x & $\beta_2$ & x & line FS\\
20  & 1 & 1 & $-$ & $\mathbb{1}_2$ & $\mathbb{1}_2$ & x & $0$ & $\pi/2$ & $0$ & FS/Dirac\\
\hline\hline
\end{tabular}
\caption{
Staggered flux U(1) CSL phases ($\tau_R=1$) on the kagome lattice. Notations are the same as in Table~\ref{tab:U1}.
\label{tab:U1sf}}
\end{table}

The classical phase diagram in Fig.~\ref{fig:classicalPD} hosts phases with coplanar spins, dubbed ``$q=0$'', ``$\sqrt{3}\times\sqrt{3}$'', and the simple ferromagnet. Furthermore, it exhibits the nonplanar states ``{cuboc-1}'' and ``{cuboc-2}''.\cite{Messio11_PRB_83_184401, Messio12_PRL.108.207204, Domenge05_PRB.72.024433, Janson08_PRL_101_106403} These N\'eel phases spontaneously break time reversal, and they have chiral orders ${\bm S}_1\cdot ({\bm S}_2\wedge{\bm S}_3) = \pm \frac{1}{3\sqrt{2}}$ on small (first-neighbor) kagome triangles for {cuboc-2}, and on large (second-neighbor) triangles for {cuboc-1}. These chiral orders break three out of the six lattice reflection symmetries, as well as the $\pi/3$ lattice rotation $R$, up to time reversal. With respect to our classification scheme discussed in Sec.~\ref{sec:theoryPSG}, the cuboc states therefore have similar symmetry properties as staggered-flux CSL states (though, in contrast to CLSs, the cuboc states also break continuous spin rotation symmetry). At the phase boundaries of Fig.~\ref{fig:classicalPD}, extensive classical degeneracies generally arise.

In order to simplify the problem, and to restrict the number of parameters, we consider only the subset of U(1) QSL states from the full list of $\mathbb{Z}_2$ states given in Tables~\ref{tab:invPSGkagZ2} and \ref{tab:invPSGkagZ2sf}. In Table~\ref{tab:U1}, the symmetric U(1) QSL (Nos.~1 - 6) and the Kalmeyer-Laughlin CSL (No.~7 - 12) states are shown. In Table~\ref{tab:U1sf}, the staggered flux phases are listed.

\begin{figure}
\includegraphics[width=.96\columnwidth]{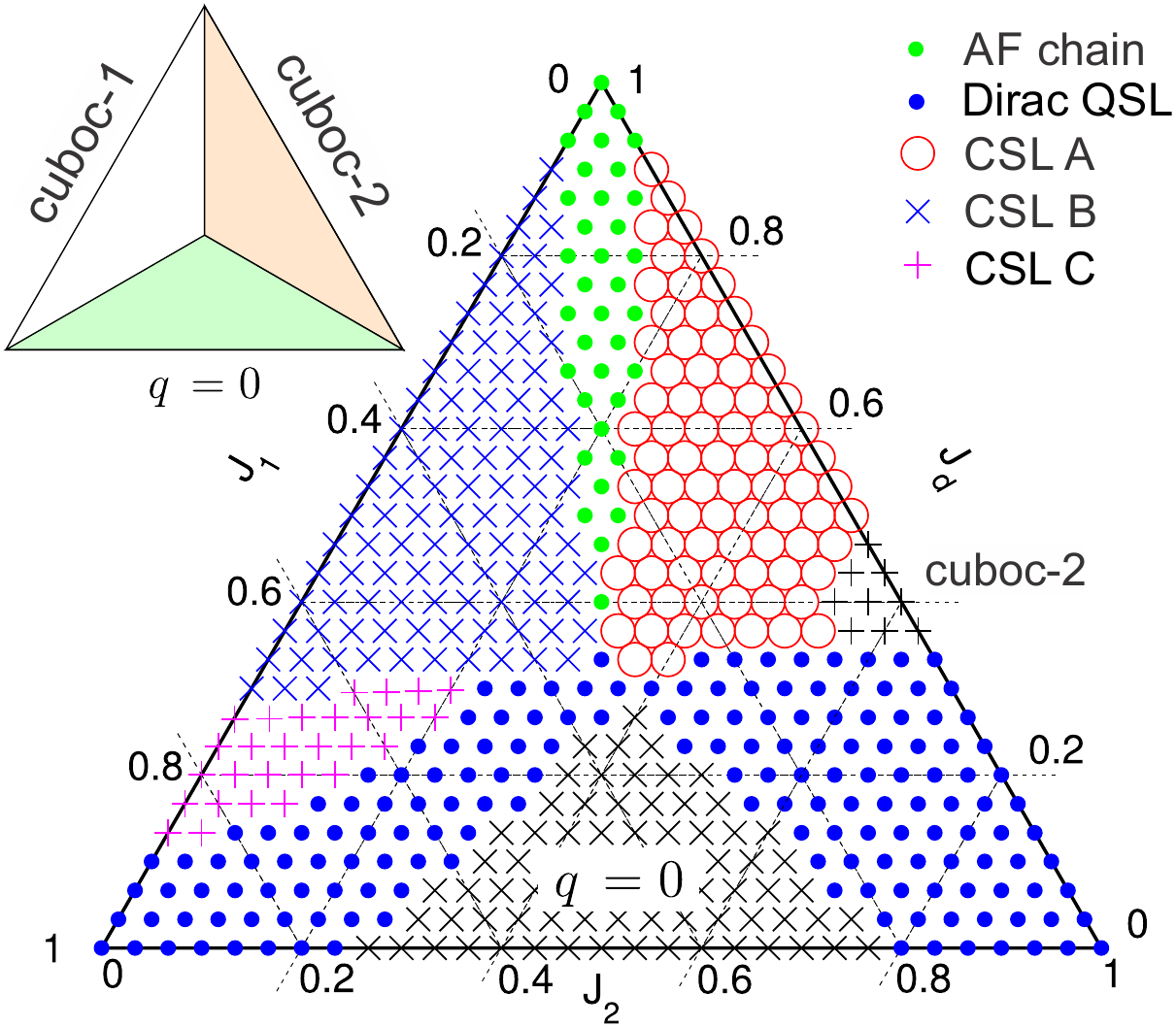}\caption{Variational phase diagram for the quantum $J_1$-$J_2$-$J_d$ Heisenberg model on the kagome lattice for the fully antiferromagnetic case. The parameters are normalized to $|J_1|+|J_2|+|J_d| = 1$. Black symbols are associated with robust N\'eel long-range orders. The left inset shows the classical analog.
\label{fig:quantumPD}}
\end{figure}

We calculate the variational energies of all U(1) QSL states in the model~\eqref{eq:model} after Gutzwiller projection on a periodic $3(8)^2$-site cluster, using the hopping amplitudes and phases that remain unrestricted by symmetry on first, second, and diagonal links as variational parameters. We disregard states that have a flat band at the Fermi energy, since it is not clear how to construct variational states in this case. In order to approach the correct limit for purely diagonal hopping (as $|\xi_1|, |\xi_2|\rightarrow 0$, $|\xi_d|\rightarrow 1$) towards the Shastry-Haldane resonating-valence-bond~(RVB) state of the Heisenberg spin chain,\cite{Shastry88_PRL.60.639, Haldane88_PRL.60.635} we introduce an additional complex phase $\pi/L$ for all diagonal hoppings ($L$ is the linear system size), but otherwise we use periodic spinon boundary conditions. This guarantees unbroken lattice rotation symmetry in the finite system.

In addition to the spin liquid wave functions, we compute the energies of correlated N\'eel states for regular magnetic orders that appear in the classical phase diagram Fig.~\ref{fig:classicalPD}. We incorporate quantum fluctuations in these product states via the Huse-Else construction, i.e., using spin Jastrow factors.\cite{Huse88_PRL_60_2531, Bieri12_PRB_86_224409, Bieri15_PRB.92.060407} The microscopic variational energies are then compared, and the resulting quantum phase diagram is presented in Fig.~\ref{fig:quantumPD}. In the following, we discuss the content of our phase diagram.

We should note that our variational investigation is restricted by a relatively small linear system size $L=8$, and energy accuracies of about $10^{-3}$. It is plausible that the energy differences between competing phases can sometimes be smaller than these error bars, especially close to phase boundaries. In a related study, Ref.~[\onlinecite{HuBeccaSheng_PRB.91.041124}] reported finite size effects that may require even larger systems. Therefore, although our quantum phase diagram is very rich and interesting, there is certainly room for improvement.

The N\'eel phases $q=0$ and {cuboc-2} survive in the quantum phase diagram Fig.~\ref{fig:quantumPD}, while we find that the {cuboc-1} phase disappears upon inclusion of quantum fluctuations. Several U(1) quantum spin liquids become lowest energy states.

The left corner of the ternary phase diagram in Fig.~\ref{fig:quantumPD} is the antiferromagnetic first-neighbor kagome Heisenberg model. Within the considered states and system size, we do not find an instability of the gapless Dirac state\cite{Hermele08_PRB.77.224413} towards one of our classified chiral U(1) phases. A consensus has been reached in the community that the ground state of the pure first-neighbor model is a spin liquid with unbroken spin rotation and translation symmetries. However, the nature of the QSL state is still under debate. In particular, the question whether it has gapless or gapfull spin excitations remains unsettled. While exact diagonalization is not fully conclusive due to small system sizes,\cite{SindzLhuillier09_EPL} density-matrix renormalization group (DMRG) computations suggest a symmetric $\mathbb{Z}_2$ liquid with a sizable spin gap.\cite{YanHuseWhite_03062011, Depenbrock12_PRL.109.067201, JiangWengSheng08_PRL.101.117203} On the other hand, large-scale variational improvements of the Dirac state using Lanczos steps found no evidence for such a gap, and a striking robustness of this state.\cite{Iqbal11_PRB.84.020407, Iqbal13_PRB.87.060405, Iqbal14_PRB.89.020407, Iqbal15_PRB.91.020402} Functional renormalization group calculations indicated exponential decay in spin correlations, giving thus support for a gapped liquid.\cite{Thomale14_PRB.89.020408} The promising coupled-cluster method has so far been inconclusive on this question.\cite{LiRichter12_PRB.86.214403} On the experimental side, herbertsmithite is believed to be described by a kagome Heisenberg model with a dominant first-neighbor exchange interaction. The majority of experiments seem to give strong evidence for a gapless QSL state in this material.\cite{MendelsBert10, *MendelsBert11} However, recent NMR measurements indicated, for the first time, the presence of a spin gap.\cite{FuImaiLee_Science05} In conclusion, more work is certainly necessary to reconciliate these results on the first-neighbor kagome Heisenberg model, both on theoretical and experimental fronts.

In our phase diagram, Fig.~\ref{fig:quantumPD}, as the diagonal interaction $J_d$ is increased, the U(1) Dirac spin liquid becomes unstable to spontaneous breaking of time reversal. This happens via threading of U(1) flux $\pi - 2\theta$ through the hexagon and $\theta$ through the small lattice triangles, thus breaking all reflection symmetries and giving a mass to the Dirac fermions. It corresponds to state No.~10 in Table~\ref{tab:U1} that we dub ``CSL~C''. A small complex second-neighbor hopping, and an even smaller (purely imaginary) diagonal hopping allowed by symmetry slightly lower the variational energy. Recently, CSL~C has independently been found and characterized by DMRG and also by parton methods.\cite{HuBeccaSheng_PRB.91.041124, Gong2014, Gong15_PRB.91.075112, SterLauchli15_PRB.92.125122} In agreement with Ref.~[\onlinecite{HuBeccaSheng_PRB.91.041124}], we find that the instability of the Dirac state towards CSL~C seems to require $J_d>J_2$.

The pure $J_2$ model at the right corner of the phase diagram in Fig.~\ref{fig:quantumPD} is three uncoupled kagome lattices, so the Dirac QSL state is again the lowest-energy state. The $q=0$ N\'eel order is stabilized in the intermediate region of $J_1$-$J_2$, in agreement with previous studies.\cite{Thomale14_PRB.89.020408, Gong15_PRB.91.075112, Iqbal15_PRB.91.020402}

As $J_d$ is increased beyond $\sim 1/3$ [i.e., $J_d\gtrsim (J_1+J_2)/2$], the optimized spinon hopping across the diagonal becomes stronger. The variational state then acquires a quasi-one-dimensional (1D) character with a gapless spinon spectrum. In the limit of pure $J_d$ at the top corner of Fig.~\ref{fig:quantumPD}, the model \eqref{eq:model} decouples into arrays of antiferromagnetic Heisenberg chains in three spatial directions. Coupling these chains by $J_1\simeq J_2$, the 1D character remains surprisingly robust, as shown by the green dots in the phase diagram. A similar effect was observed in the case of ferromagnetic first- and second-neighbor couplings.\cite{Bieri15_PRB.92.060407}

For $J_d\gtrsim 1/3$ and $J_1 > J_2$ or $J_1 < J_2$, we find that the chiral spin liquids Nos.~11 and 12 in Table~\ref{tab:U1} (CSL~B and CSL~A, respectively) are the lowest-energy states within the set of wave functions we considered. They have gapless spinon Fermi surfaces due to a strong diagonal hopping. The rotation representation $g_R$ is nontrivial in these states. The resulting staggering under rotation of the complex hopping phases on first- and second-neighbor links implies that the time-reversal breaking U(1) flux is introduced through the triangles of the lattice, while the hexagons maintain a trivial flux. (See Sec.~\ref{sec:invPSGkag})

\begin{figure*}
\includegraphics[width=\textwidth]{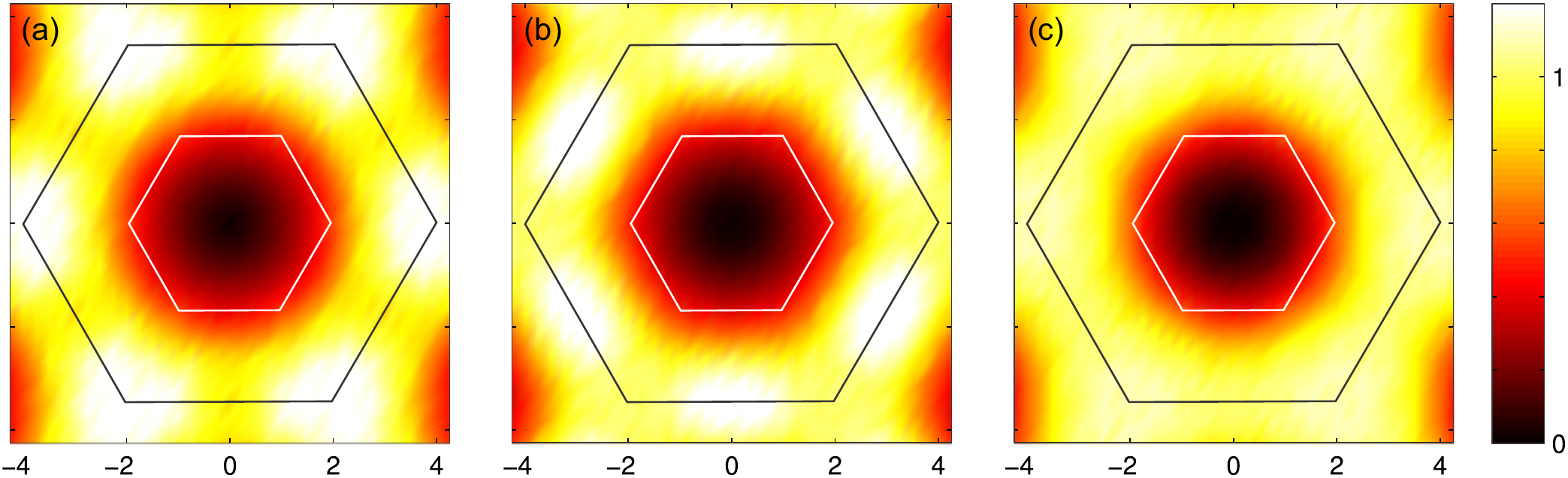}\caption{Static spin structure factor $N S({\bm q})$, Eq.~\eqref{eq:Sk}, in the phases (a)~FS QSL, (b)~Dirac QSL, and (c) CSL~C. Gutzwiller projection is done on a cluster of $N=3(12)^2$ sites, normalization is $\sum_{\bm q} S({\bm q}) = 1$.
\label{fig:projSk}}
\end{figure*}

Note that all QSL phases we find in Fig.~\ref{fig:quantumPD} have unbroken lattice rotation ($\tau_R = 0$), so they are either fully symmetric (Dirac) or Kalmeyer-Laughlin type (CSL A, B, and C). This is in contrast to the N\'eel state {cuboc-2} (and {cuboc-1}), where rotation symmetry is spontaneously broken. The only region where we find staggered-flux CSL states to be relatively low in energy is close to the 1D phase boundary in Fig.~\ref{fig:quantumPD}. Similarly, below the {cuboc-2 phase}, close to the line $J_1=0$, a Kalmeyer-Laughlin CSL appears to be low in energy. However, our limited precision and computational resources do not allow us to make stronger statements. More detailed investigations of U(1) QSL states, using higher accuracy, better minimization procedures, and larger system sizes would be interesting. Promising would also be a variational or mean-field study using the $\mathbb{Z}_2$ CSL states classified in Tables~\ref{tab:invPSGkagZ2} and \ref{tab:invPSGkagZ2sf}, a task which we leave for future work.

\subsection{Spin structure factors}

Neutron spectroscopy is a powerful tool to probe conventional and exotic phases in quantum magnets.\cite{deVries09_PRL.103.237201, Fak12_PRL_109_037208, Han2012, DallaPiazza2015, Boldrin_PRB.91.220408} In this experiment, static and dynamical spin structure factors can be measured, which provides valuable information on the nature of the phase. The structure factor is given by
\begin{equation}\label{eq:Sk}
  S({\bm q}) = N^{-1}\sum_{i, j} e^{-i {\bm q}\cdot ({\bm r}_i - {\bm r_j}) } \langle {\bm S}_i\cdot {\bm S}_j\rangle\,,
\end{equation}
where the sums go over all $N$ sites of the lattice. In N\'eel states with broken spin rotation symmetry, it exhibits strong intensities at the ordering wave vectors and soft goldstone modes.\cite{Messio11_PRB_83_184401} In quantum spin liquid phases, the structure factor is expected to be much broader, or even incoherent. 

In Ref.~[\onlinecite{Bieri15_PRB.92.060407}], we calculated the static (equal-time) structure factors for the CSL~A and CSL~B phases of Fig.~\ref{fig:quantumPD}. The quasi-one-dimensional character of these phases leads to lines of intensity, reminiscent of the uncoupled spin chains. However, important two-dimensional correlations still provide distinct features that have measurable consequences.

In Fig.~\ref{fig:projSk}, we present the static structure factors for the quantum spin liquid phases discussed in the last section. For comparison, we also include the symmetric state with a large Fermi surface~(FS) in Fig.~\ref{fig:projSk}(a); (First-neighbor hopping, No.~1 in Table~\ref{tab:U1}). Fig.~\ref{fig:projSk}(b) shows the structure factor of the Dirac QSL in the first-neighbor Heisenberg model, and Fig.~\ref{fig:projSk}(c) is the new CSL~C state at the point $J\simeq (0.63, 0.13, 0.24)$, where the hoppings are $|\xi| \simeq (0.75, 0.25, 0)$ and complex phases $\beta\simeq (0.08\pi, -0.61\pi, \pi/2)$; see also Ref.~[\onlinecite{Hu15_PRB.92.201105}] for an accurate determination of the optimal variational parameters. We display the structure factor in the first and extended Brillouin zones (white resp.\ black hexagons in Fig.~\ref{fig:projSk}). The numerical Gutzwiller projection is done on a $3 (12)^2$ site cluster. For smaller systems, the structure factor of the Dirac spin liquid has also been discussed in Refs.~[\onlinecite{Ran07_PRL.98.117205}, \onlinecite{Iqbal13_PRB.87.060405}, \onlinecite{Mei15_DSL}].

By inspection of Fig.~\ref{fig:projSk}, the Fermi surface QSL state has broad intensity maxima at the $K$ points of the extended Brillouin zone. In the Dirac state, the maxima are shifted to the $M$ points. Finally, the intensity is even broader in the CSL~C state, with very wide maxima at $K$ points. In this case, the maximum may be characterized as a ``ring'' on the boundary of the extended Brillouin zone. Note that the Dirac QSL structure factor is in good agreement with exact diagonalization results on the first-neighbor Kagome model for $N=36$ sites.\cite{LauchliLhuillier09_arxiv} On the other hand, the structure factor of this model obtained by DMRG seems to show more of a ``ring'' structure,\cite{Depenbrock12_PRL.109.067201} similar to CSL~C.

The static structure factor can be used to evidence these phases in neutron scattering experiments. However, more detailed information is provided by the {\it dynamical} spin structure factor. At low energy, it gives information about the excitation gap from the singlet ground state to spin $S=1$ (triplet) excitations. Here, the CSL~C phase is fully gapped while the other states displayed in Fig.~\ref{fig:projSk} are expected to be gapless. Furthermore, for gapless spinons, unique features due to the shape of the spinon Fermi surface are expected to show up in the structure factor at low energy. However, these effects are difficult to calculate beyond the quadratic theory for interacting spinons. Progress can be made by Gutzwiller projection of excitations,\cite{TanWang08_PRL.100.117004, DallaPiazza2015, Mei15_DSL} or by perturbative treatments of interaction.\cite{HalperinLeeRead93_PRB.47.7312, Lee09_PRB.80.165102, Metlizki15_PRB.91.115111, Punk_NatPhys14}

\subsection{Bulk spectra}\label{sect:spectra}

The symmetry group SG and its projective representations can impose constraints on the spinon spectrum of an invariant ansatz. Most importantly, it is sometimes possible to say, for a given PSG representation class, if the spectrum at certain points in the Brillouin zone~(BZ) is gapless, and which terms can potentially lead to a gap.

Here, we discuss the case of the kagome lattice, but the ideas apply in a similar way to the triangular and other lattices. One ingredient we use is that the PSG representations of the point group and the translation group factorize, Eq.~\eqref{eq:pRep}, as is the case for the considered lattices. For symmorphic space groups, we expect this generally to be the case.

First, we want to investigate the effect of projective symmetry transformations on the ansatz in Fourier space. For $\epsilon_2=1$, the unit cell is simply the primitive cell of the lattice, while for $\epsilon_2=-1$ it needs to be increased to accommodate the rotation representation, see Eq.~\eqref{eq:pRep}. We use translational symmetry in the uniform gauge and Fourier transform, $u_{nm}({\bm k}) = \sum_{i, j} u({\bm R}_i+{\bm r}_n, {\bm R}_j+{\bm r}_m) \exp \{i {\bm k}\cdot ({\bm R}_i - {\bm R}_j + {\bm r}_n - {\bm r}_m) \}$, where ${\bm k}$ is in the reduced BZ, and ${\bm r}_n$ labels sites in the unit cell. $u_{nm}({\bm k})$ are $2\times 2$ matrices that can be decomposed into $u_{nm} = u_{nm}^\mu \tau_\mu$ with $(\tau_\mu) = (i\mathbb{1}_2, \sigma_a)$. Similar to the discussion in Sec.~\ref{sec:algPSG} and Eq.~\eqref{eq:gt}, the projective representation acts in Fourier space as
\begin{equation}
  Q_x ( u ) = [(-)^{\tau_x} g_x({\bm k}) u( x^{-1} {\bm k} ) g_x({\bm k})^\dagger]\,.
\end{equation}
The unitary $g_x({\bm k})$ acts separately on spin and space indices, and we have
\begin{equation}\label{eq:gk}
  g_x({\bm k}) = G_x({\bm k}) \otimes g_x
\end{equation}
where $g_x$ is a constant SU(2) matrix, and $G_x({\bm k})$ is a $N\times N$ unitary. The transformation $G_x({\bm k})$ takes into account permutations of sublattice sites and translations that may accompany the symmetry $x$.

In Fourier space, the spinon Hamiltonian reduces to block diagonal form given by
\begin{equation}
  (H_{nm})_{\bm k} = u_{nm}({\bm k}) + [u_{mn}({\bm k})]^\dagger\, .
\end{equation}
For a unit cell of $N$ sites, the $H_{\bm k}$ is therefore a $2 N\times 2 N$ matrix. It may further be decomposed as $H_{\bm k} = H_{\bm k}^\mu\sigma_\mu$ with $(\sigma_\mu) = (\mathbb{1}_2, \sigma_a)$, and $H_{\bm k}^\mu$ are matrices of size $N\times N$.

From spin rotation symmetry and the discussion in Sec.~\ref{sec:h0}, it follows that the ansatz satisfies
\begin{equation}\label{eq:umk}
  u_{nm}(-{\bm k}) = [u_{nm}({\bm k})]^\dagger\,,
\end{equation}
or $u_{nm}^0(-{\bm k}) = -u_{nm}^0({\bm k})$, $u_{nm}^a(-{\bm k}) = u_{nm}^a({\bm k})$. In terms of the Hamiltonian components, we therefore have $H_{-{\bm k}}^0 = -H_{{\bm k}}^0$ and $H_{-{\bm k}}^a = H_{\bm k}^a$.

Special ``high symmetry'' points in the reduced BZ can give constraints on the spinon spectrum, or one can make statements about level degeneracies. If a symmetry $x$ leaves a point $\bm k$ in the Brillouin zone invariant (modulo backfolding by a reciprocal vector), i.e., $x({\bm k}) = {\bm k}$ mod ${\bm G}$, then the spinon Hamiltonian satisfies
\begin{equation}\label{eq:Hconstr}
  H_{\bm k} = (-)^{\tau_x} g_x({\bm k}) H_{\bm k} g_x({\bm k})^\dagger\,.
\end{equation}
Furthermore, when $x({\bm k}) = -{\bm k}$ mod ${\bm G}$, using the property \eqref{eq:umk}, gives a similar constraint on $H_{\bm k}$.

In the case of the kagome lattice, the matrices $g_x$ in Eq.~\eqref{eq:gk} are particularly simple and satisfy $(g_x)^2 = \pm \mathbb{1}_2$. The components $H_{\bm k}^\mu$ in Eq.~\eqref{eq:Hconstr} are therefore uncoupled, and the conjugation by $g_x = \sigma_\mu$ can only give additional signs for $H_{\bm k}^a$ with $a\neq\mu$. Equation \eqref{eq:Hconstr} then reduces to
\begin{equation}\label{eq:Hcommut}
  H_{\bm k}^\mu = (-)^{\tau_x^\mu} G_x({\bm k}) H_{\bm k}^\mu G_x({\bm k})^\dagger
\end{equation}
for every $N\times N$ block $H_{\bm k}^\mu$, $\mu = 0, 1, 2, 3$. If the sign $(-)^{\tau_x^\mu}$ is positive, then Eq.~\eqref{eq:Hcommut} can give information about level degeneracies. In the following, we consider the case when the sign is negative. In this case, and when the number of sites $N$ in the unit cell is odd, we can immediately conclude from Eq.~\eqref{eq:Hcommut} that $H_{\bm k}^\mu$ must have at least one zero eigenvalue, irrespective of the matrix $G_x({\bm k})$. 

As a first example, let us discuss the inversion symmetry $R^3$, which brings $\bm k \mapsto -\bm k$ for all wave vectors. Combined with the property \eqref{eq:umk}, this implies the flat bands at zero energy observed in the U(1) states Nos.~2, 3, 8, and 9 in Table~\ref{tab:U1}.

\begin{figure}
\includegraphics[width=\columnwidth]{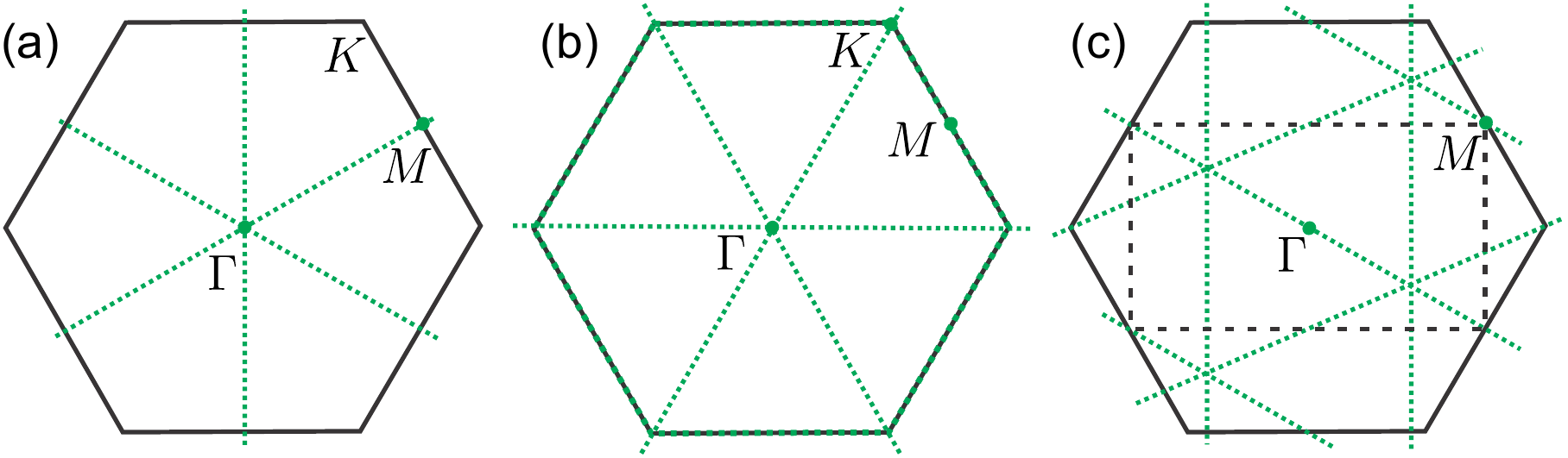}\caption{Lines and points of gapless Fermi surfaces (green dots online) in the U(1) states No.~11 [(a)], No.~17 [(b)] in Table~\ref{tab:U1sf}, and $\mathbb{Z}_2$ state No.~8 [(c)] in Table~\ref{tab:invPSGkagZ2}. Black hexagons are first BZs, the rectangle in (c) is the reduced BZ.
\label{fig:lineFS}}
\end{figure}

In general, lattice reflections leave straight lines in the Brillouin zone~(BZ) invariant. Let us consider the case of simple unit cell with $\epsilon_2=1$. On the one hand, reflection $\sigma$ and its rotations by $R$ and $R^2$ (see Fig.~\ref{fig:kagsym}) conserve the lines connecting the Brillouin zone center $\Gamma$ with the corners $K$. On the other hand, the reflection $R\sigma$ and its rotations leave the lines $\Gamma$--$M$ invariant. Combined with ${\bm k}\mapsto-{\bm k}$ inversion, they also leave the zone boundaries $K$--$K$ invariant. These symmetry considerations explain the lines of Fermi surfaces in the staggered-flux U(1) states 10, 11, 16, and 17 in Table~\ref{tab:U1sf}. In Fig.~\ref{fig:lineFS}, we give some examples of such symmetry-protected Fermi surface lines.

The symmetry analysis is more complicated in the case of $\epsilon_2=-1$, where the cell contains an even number of sites. The matrix representation $G({\bm k})$ of the symmetry is now even dimensional. For a negative sign in \eqref{eq:Hcommut}, the presence of zero eigenvalues of $H_{\bm k}$ can still be inferred from the form of $G({\bm k})$: if some of its eigenvalues do not come in pairs $\pm \lambda$, then $H_{\bm k}$ has zero eigenvalues. The lines of Fermi surfaces for the cases with $\epsilon_2=-1$ in Tables~\ref{tab:U1} and \ref{tab:U1sf} can be understood from such an analysis of reflection symmetries.\cite{hermanns}

\section{Discussion \& outlook}

In Sec.~\ref{sec:frac}, we present a general discussion of fermionic spinon fractionalization in quantum spin liquids. This fractionalization entails an emergent local SU(2) gauge symmetry in the enlarges spinon Hilbert space, and we discuss the construction of gauge invariant characterizations and spin order parameters. In Sec.~\ref{sec:theoryPSG}, we review how the emergent symmetry gives rise to classes of nontrivial representations of lattice symmetries in the gauge group, and how this leads to the projective symmetry group classification of quantum spin liquid phases. We systematically extend this classification to the case of chiral, i.e., time-reversal broken singlet quantum spin liquids. In particular, we distinguish between Kalmeyer-Laughlin and staggered-flux CSLs: the first conserving lattice rotations, the latter breaking them. In Secs.~\ref{sec:trgPSG} and \ref{sec:kagPSG}, we apply this general formalism to the case of triangular and kagome lattices, and we exhaustively list all projective symmetry representations, and the corresponding symmetric and chiral spin liquid ans\"atze for these lattices. Finally, in Sec.~\ref{sec:kagPSG}, we investigate the subset of U(1) QSL phases variationally for the $J_1$-$J_2$-$J_d$ Heisenberg model on the kagome lattice, and we discuss spin structure factors and symmetry constraints on the spinon spectrum in some of these phases.

In view of the recent renewal of interest in chiral spin liquids, the PSG classification can give valuable information on exotic states, fractional symmetry representation classes, and flux patterns that can potentially arise. Emergence of such phases is primarily expected in ground states of spin models on two- and three-dimensional lattices with strong geometric frustration. Interestingly and encouragingly, the number possible $\mathbb{Z}_2$ and U(1) QSL phases for triangular-based lattices (kagome, honeycomb, etc.) is limited and generally strongly reduced with respect to the square lattice, where this number is very large.\cite{Wen02_PRB.65.165113} It will be interesting to further investigate the phases classified in this paper by self-consistent or variational methods for microscopic spin models. Application of the general classification scheme presented in this paper to other lattices, or fractionalization of higher values of spin are further promising directions.

\begin{acknowledgments}
We thank Bernard~Bernu for discussions and related collaborations. SB gratefully acknowledges Gian~Michele Graf for helpful discussions, kind hospitality, and support. SB also thanks R.~Chitra, J\"urg~Fr\"ohlich, and Manfred~Sigrist for stimulating discussions. This work was supported in part by the Agence Nationale de la Recherche (France) under grant ANR-12-BS04-0021.
\end{acknowledgments}

\appendix

\section{Triangular PSG}\label{app:trgPSG}

\subsection{Algebraic PSG}\label{app:trgPSGalg}

Following the discussion in Sec.~\ref{sec:trgPTgroup}, it remains to derive the expressions \eqref{eq:pRep} for the gauge representations of the point group ($\mathbb{Z}_2$ PSG) of the triangular lattice. Let us first consider the reflection symmetry $\sigma$; (see definition in Fig.~\ref{fig:trgsym}). Using the translation representation~\eqref{eq:g12}, the algebraic relation \eqref{eq:pg1a} imposes the following constraint on the representation $g_\sigma(x,y)$:
\begin{equation}
  g_\sigma(x,y) = (\epsilon_{\sigma2}) (\epsilon_2)^{x} g_\sigma(x,y-1)\,,
\end{equation}
with $\epsilon_{\sigma2}=\pm 1$. This equation is solved by
\begin{equation}\label{eq:gsx}
  g_\sigma(x,y) = (\epsilon_{\sigma2})^y(\epsilon_2)^{x y} g_{\sigma}(x)\,.
\end{equation}
Furthermore, reflection has the property Eq.~\eqref{eq:pg2a}, i.e., $\sigma^2 = e$. Using Eq.~\eqref{eq:gsx}, this imposes the constraint
\begin{equation}
  (\epsilon_2)^{x y} (\epsilon_{\sigma2})^y g_{\sigma}(x) (\epsilon_2)^{x y} (\epsilon_{\sigma2})^x g_{\sigma}(y) = \pm \mathbb{1}_2\, .
\end{equation}
This implies that $g_\sigma(x) = (\epsilon_{\sigma2})^{x+y} (\epsilon_2)^{x y} g_{\sigma}$, where $g_\sigma$ is a constant matrix, and $(g_{\sigma})^2=\pm\mathbb{1}_2$. Keeping in mind that the staggered gauge transformations $g(x,y) = (-)^x$ and $g(x,y) = (-)^y$  leave the uniform gauges $g_\opx$ and $g_\opy$, Eq.~\eqref{eq:g12} invariant, we can use them to eliminate the sign $\epsilon_{\sigma2}$. The final result is
\begin{equation}\label{eq:gsapp}
  g_\sigma(x,y) = (\epsilon_2)^{x y} g_\sigma\,,
\end{equation}
as announced in Eq.~\eqref{eq:pReps}.

Next, we consider the rotation generator $R$. We start with the algebraic relations \eqref{eq:pg1b} and \eqref{eq:pg1c} involving the translations. Equation \eqref{eq:pg1b} imposes the constraint $(\epsilon_2)^y g_R(x-1,y) = (\epsilon_{R1}) g_R(x,y)$ on the rotation representation. It is solved by
\begin{equation}\label{eq:gR1}
  g_R(x, y) = (\epsilon_2)^{x y} (\epsilon_{R1})^x g_R(y)\,.
\end{equation}
Note that this constraint and its solution is also valid on the square lattice.\cite{Wen02_PRB.65.165113} In contrast, the relation \eqref{eq:pg1c} only applies to triangular-based lattices. It imposes the constraint $(\epsilon_2)^x g_R(x,y-1) = (\epsilon_{R2}) g_R(x,y) (\epsilon_2)^{y}$, implying
\begin{equation}\label{eq:gR2}
  g_R(x,y) = (\epsilon_{R2})^y (\epsilon_2)^{x y + y(y+1)/2} g_{R}(x)\,.
\end{equation}
Combining \eqref{eq:gR1} and \eqref{eq:gR2}, we find
\begin{equation}\label{eq:gR12}
g_R(x,y) = (\epsilon_{R1})^x (\epsilon_{R2})^y (\epsilon_2)^{x y + y(y+1)/2} g_{R}\,,
\end{equation}
where $g_R$ is a translation-invariant SU(2) matrix. The previous gauge choices still allow transformations of the form $g(x,y) = (-)^{x+y}$, which can be used to eliminate the sign $\epsilon_{R2}$ from Eq.~\eqref{eq:gR12}.

Using the group multiplication law \eqref{eq:Qmult}, we obtain the following representation of the reflection $R \sigma$ from Eqs.~\eqref{eq:gsapp} and \eqref{eq:gR12},
\begin{equation}
  g_{R\sigma}(x, y) = (\epsilon_{R1})^x (\epsilon_2)^{ y(y-1)/2 } g_R g_\sigma\, .
\end{equation}
The constraint Eq.~\eqref{eq:pg2b}, $(R \sigma)^2 = e$, now forces $\epsilon_{R1}=1$ and $(g_R g_\sigma)^2 = \pm \mathbb{1}_2$.

Finally, we have to enforce the constraint \eqref{eq:pg2c}, $R^6 = e$. One can check that the remaining sign $\epsilon_2$ in Eq.~\eqref{eq:gR12} goes through this equation, and the constant matrix $g_R$ must satisfy $(g_R)^6 = \pm\mathbb{1}_2$.

It remains to solve the equations for the constant (translation invariant) matrices,
\begin{subequations}\label{eq:constconstr}\begin{align}
(g_\sigma)^2 &= (\epsilon_{\sigma}) \mathbb{1}_2\,,\label{eq:consts2}\\
(g_R g_\sigma)^2 &= (\epsilon_{R \sigma}) \mathbb{1}_2\,,\label{eq:constRs2}\\
(g_R)^6 &= (\epsilon_{R}) \mathbb{1}_2\, .\label{eq:constR6}
\end{align}
\end{subequations}
Since the signs $\epsilon_{(\cdot)}$ in \eqref{eq:constconstr} provide gauge invariant characterisation of $\mathbb{Z}_2$ PSG representations, we may naively expect that they are sufficient for enumeration, and that we have $2^3 = 8$ representation classes. However, this expectation is incorrect for two reasons. First, one can easily see that not all combinations of signs lead to solutions in \eqref{eq:constconstr}. For example, $\epsilon_{\sigma}=\epsilon_{R\sigma}=1$ implies $g_\sigma = g_R = \mathbb{1}_2$, and we must necessarily have $\epsilon_{R} = 1$. It has been argued (see, e.g., Ref.~[\onlinecite{Mei15}]) that this fact is a shortcoming of the particular spin fractionalization scheme. While this point of view has some validity and may lead to interesting developments, we will not further pursue it here. Instead, we show that these signs are actually {\it not} sufficient to entirely characterize $\mathbb{Z}_2$ PSG classes on the triangular lattice.

The cases $\epsilon_{\sigma} = 1$ or $\epsilon_{R\sigma} = 1$ are relatively simple, as they immediately lead to the solutions $g_\sigma=\mathbb{1}_2$ or $g_R g_\sigma=\mathbb{1}_2$, respectively. In turn, Eq.~\eqref{eq:constR6} does not give any additional constraint, leading to the solutions No.~1, 3, and 4 in Table~\ref{tab:PSGtrg}.

The case $\epsilon_{\sigma} = \epsilon_{R\sigma} = -1$ is most interesting, as it gives rise to more complicated solutions. We can always choose a global gauge such that $g_\sigma = i\sigma_2$, solving Eq.~\eqref{eq:consts2}. The general solution to \eqref{eq:constRs2} is then $g_R g_\sigma = e^{i\beta\sigma_3}i\sigma_2$, where $\beta$ is a real parameter. Finally, plugging $g_R = e^{i\beta\sigma_3}$ into \eqref{eq:constR6}, we obtain $\beta = 0, \pi/6, \pi/3$, $\pi/2$, i.e., Nos.~2, 7, 6, and 5 in Table~\ref{tab:PSGtrg}.

Note that the solutions $\beta=0$, $\beta=\pi/3$ and $\beta=\pi/6$, $\beta=\pi/2$ imply the same sign in Eq.~\eqref{eq:constR6}, $\epsilon_R = +1$ and $\epsilon_R = -1$, respectively. That is, these phases cannot be distinguished by the space group fractionalization signs of the $\mathbb{Z}_2$ spin liquid. Since a pairing $\Delta$ in the corresponding ansatz transforms as $\Delta\mapsto\Delta e^{2i\beta}$ under $\pi/3$ lattice rotation, these representations potentially lead to $s$-, $p+i p$-, $d+i d$-, and $f$-wave spinon pairing symmetries on the triangular lattice.\cite{Grover10_PRB_81_245121, Mishmash13_PRL.111.157203} As it was first recognized in Ref.~[\onlinecite{Mishmash13_PRL.111.157203}], in the absence of a hopping, the pure $d+id$-wave paring state indeed represents a symmetric $\mathbb{Z}_2$ QSL phase that respects time reversal. For such pure pairing states, the SU(2) gauge flux through diamond plaquettes of the lattice is given by $\text{Tr} P = \cos 4\beta$, so we have $\text{Tr} P = \{ 1, -1/2, -1/2, 1\}$, respectively, for these representations $g_R$. We see that, even though the $\mathbb{Z}_2$ signs $\epsilon_{(\cdot)}$ are identical (e.g., for pure $s$-wave and in the $d+i d$ QBT state\cite{Mishmash13_PRL.111.157203}), the respective SU(2) fluxes differ, implying that they constitute distinct symmetric QSL phases.

Finally, we should mention that the ``complex'' solutions for the rotation representation, $\beta=\pi/6$ and $\beta=\pi/3$ above, were missed in recent PSG classification attempts.\cite{Mei15, Lu15} Similarly, this type of solutions seem to have been omitted in the original classification of symmetric QSLs on the square lattice.\cite{Wen02_PRB.65.165113} However, as we have shown here, these solutions are relevant and they can lead to distinct phases, even when we are only interested in the case of fully time-reversal symmetric quantum spin liquids.

\subsection{Invariant PSG}\label{app:trgPSGinv}

Here, we discuss the symmetry constraints on the ansatz for the triangular lattice. The field on links $u_1$, $u_2$, and $u_3$ are defined in Fig.~\ref{fig:trgsym}. The solutions to the symmetry constraints for the different PSG representations (Table~\ref{tab:PSGtrg}) and time-reversal signatures $\tau_\sigma, \tau_R$ are given in Tables~\ref{tab:invPSGtrgZ2} and \ref{tab:invPSGtrgZ2sf} of the main text.

The on-site field $\lambda$ must respect both generators of the point group, so we have
\begin{subequations}\label{eq:constr0tr}\begin{align}
  \lambda &= (-)^{\tau_\sigma} g_{\sigma} \lambda [g_{\sigma}]^\dagger\,,\\
  \lambda &= (-)^{\tau_R} g_{R} \lambda [g_{R}]^\dagger\,.\label{eq:constr0btr}
\end{align}
\end{subequations}
The first-neighbor link $u_1$ has the symmetry $\sigma$ keeping both sites fixed, and $T_\opx T_\opy R^3 \sigma$ exchanging sites. The corresponding constraint equations are
\begin{subequations}\label{eq:constr1tr}
\begin{align}
  u_1 &= (\epsilon_2) (-)^{\tau_\sigma} g_{\sigma} u_{1} [g_{\sigma}]^\dagger\,,\\
  [u_1]^\dagger &= (\epsilon_2) (-)^{\tau_R} (g_R)^3 u_1 [(g_R)^3]^\dagger\,.
\end{align}
\end{subequations}
For the second-neighbor link $u_2$, it is easiest to impose the symmetries $\sigma$ and $T_\opx T_\opy R^3 \sigma$, both exchanging sites:
\begin{subequations}
\begin{align}
  [u_2]^\dagger &= (-)^{\tau_\sigma} g_{\sigma} u_{2} [g_{\sigma}]^\dagger\,,\\
  [u_2]^\dagger &= (\epsilon_2) (-)^{\tau_R} (g_R)^3 u_{2} [(g_R)^3]^\dagger\,.
\end{align}
\end{subequations}
Finally, the third-neighbor link has the symmetry $\sigma$ keeping both sites fixed, and $R^3$ exchanging them, so
\begin{subequations}
\begin{align}
  u_3 &= (-)^{\tau_\sigma} g_{\sigma} u_{3} [g_{\sigma}]^\dagger\,,\\
  [u_3]^\dagger &= (-)^{\tau_R} (g_R)^3 u_3 [(g_R)^3]^\dagger\,.
\end{align}
\end{subequations}
As we see, first-, second-, and third-neighbor links respect two reflection symmetries of the triangular lattice, so they have the maximal number of two constraints in this case.

\section{Kagome PSG}\label{app:kagPSG}

\subsection{Algebraic PSG}

As we discussed in the main part of the text, the algebraic relations among symmetry generators of the space group are formally identical for kagome and triangular lattices. The relations \eqref{eq:Txy} and \eqref{eq:pg1} act independently on each of the three sublattice sites. We can therefore solve these constraints in exactly the same way as on the triangular lattice, for each sublattice site independently. The result is Eqs.~\eqref{eq:gsx} and \eqref{eq:gR12}, where the signs $\epsilon_2$ and $\epsilon_{R1}$, and especially the translation-invariant matrices $g_{s\sigma}$ and $g_{sR}$ are now sublattice dependent.

Next, we consider the point group relations \eqref{eq:pg2}. For the kagome lattice, these relations now couple the three sublattice sites. Equation~\eqref{eq:pg2c}, $R^6=e$, forces the translation signs $\epsilon_2$ to be the same on each sublattice. Again, Eq.~\eqref{eq:pg2a}, $(R\sigma)^2 = e$ implies $\epsilon_{R1} = +1$. Finally, the translation-invariant point group representations must satisfy the equations
\begin{subequations}\label{eq:kagomePSGeq}
\begin{align}
  g_{1\sigma} g_{3\sigma} &= (g_{2\sigma})^2\nonumber\\
   &= g_{3\sigma} g_{1\sigma} = (\epsilon_{\sigma}) \mathbb{1}_2\,,\label{eq:k1}\\
  (g_{1R} g_{3\sigma})^2 &= g_{2R} g_{1\sigma} g_{3R} g_{2\sigma}\nonumber\\
   &= g_{3R} g_{2\sigma} g_{2R} g_{1\sigma} = (\epsilon_{R\sigma}) \mathbb{1}_2 \,,\label{eq:k2}\\
  (g_{1R}g_{3R}g_{2R})^2 &= (g_{2R}g_{1R}g_{3R})^2\nonumber\\
   &= (g_{3R}g_{2R}g_{1R})^2 = (\epsilon_{R})\mathbb{1}_2\,,\label{eq:k3}
\end{align}\end{subequations}
where $s$ in $g_{s\sigma}$ and $g_{sR}$ is the sublattice index shown in Fig.~\ref{fig:kagsym}.

Before solving Eqs.~\eqref{eq:kagomePSGeq}, it is convenient to show that there is a canonical gauge where the rotation representations $g_{sR}$ do not depend on the sublattice site $s$. Under a global sublattice gauge transformation $g = [g_1, g_2, g_3]$, the point group representations of the kagome lattice transform as
\begin{subequations}\label{eq:kagomeGS1}
  \begin{align}
    g_{1\sigma} &\mapsto g_1 g_{1\sigma} g_3^\dagger\,,\\
    g_{2\sigma} &\mapsto g_2 g_{2\sigma} g_2^\dagger\,,\\
    g_{3\sigma} &\mapsto g_3 g_{3\sigma} g_1^\dagger\,,
  \end{align}
\end{subequations}
and
\begin{subequations}\label{eq:kagomeGS2}
  \begin{align}
    g_{1 R} &\mapsto g_1 g_{1 R} g_3^\dagger\,,\\
    g_{2 R} &\mapsto g_2 g_{2 R} g_1^\dagger\,,\\
    g_{3 R} &\mapsto g_3 g_{3 R} g_2^\dagger\,.
  \end{align}
\end{subequations}
Let us start in an arbitrary gauge where all $g_{sR}$ differ, and let us define $g_R$ as a root of the equation $(g_R)^3 = g_{3R} g_{2R} g_{1R}$. Then, by virtue of Eq.~\eqref{eq:kagomeGS2},  performing the change of gauge $g = [g_R^\dagger g_{3R} g_{2R}, g_{3R}, g_R]$ makes $(g_{1R}, g_{2R},g_{3R} )\mapsto(g_{R}, g_{R},g_{R})$, which completes the proof.

Next, we consider the solutions to Eq.~\eqref{eq:kagomePSGeq}, and we show that $g_{s\sigma}$ must also be sublattice independent in that case. Let us first discuss \eqref{eq:k3}, which is simply $(g_R)^6 = \epsilon_R$ in the canonical gauge. We again have four solutions, $g_R = e^{n\pi i\sigma_a/6}$, $n=0\ldots 3$, depending on the sign $\epsilon_R$. However, as discussed in Sec.~\ref{sec:kagPSG} of the main text, the sublattice gauge transformation \eqref{eq:sublGT} identifies them pairwise, and it is sufficient to consider the two solutions with $(g_R)^2 = \epsilon_R$.

The quadratic Eq.~\eqref{eq:k1} is formally solved by $g_{2\sigma} = \sqrt{\epsilon_\sigma}$, and Eq.~\eqref{eq:k2} by $g_{3\sigma} = g_R^\dagger\sqrt{\epsilon_{R\sigma}}$. Combining the nonquadratic parts of these equations, and using $(g_R)^2 = \epsilon_R$, one derives the relation $g_R\sqrt{\epsilon_{\sigma}} = \sqrt{\epsilon_{R\sigma}}$. Replacing these results back into $g_\sigma$, we obtain $g_\sigma = (\epsilon_\sigma\sqrt{\epsilon_{R\sigma}}^\dagger g_R, \sqrt{\epsilon_\sigma}, g_R^\dagger \sqrt{\epsilon_{R\sigma}}) = \sqrt{\epsilon_\sigma} (1, 1, 1)$. QED.

Once we have shown that both $g_{sR}$ and $g_{s\sigma}$ are sublattice independent in the canonical gauge, it is immediately clear that the solutions for the PSG classes are the same as for the triangular lattice, since Eqs.~\eqref{eq:kagomePSGeq} reduce to Eqs.~\eqref{eq:constconstr}. Omitting the gauge equivalent solutions as discussed in Sec.~\ref{sec:kagPSG}, the final result in Table~\ref{tab:PSGkag} is then readily obtained as before.

\subsection{Invariant PSG}\label{app:kagPSGinv}

Here, we consider an ansatz on the kagome lattice with first neighbor $u_1$, second neighbor $u_2$, and diagonal neighbor $u_3$ across the hexagons. For our choice of symmetry generators, it is convenient to impose the symmetry constraints on the links shown in Fig.~\ref{fig:kagsym}. The solution to the constraints are given in Tables~\ref{tab:invPSGkagZ2} and \ref{tab:invPSGkagZ2sf} of the main text.

For the on-site field, we choose to impose the constraints at sublattice site $s=2$$, \lambda = \lambda_2$. The symmetries leaving this site invariant are reflection $\sigma$ and rotation $R^3$ (up to irrelevant translations). The corresponding constraint equations are
\begin{subequations}\label{eq:constr0}\begin{align}
  \lambda &= (-)^{\tau_\sigma} g_{\sigma} \lambda [g_{\sigma}]^\dagger\,,\\
  \lambda &= (-)^{\tau_R} g_{R} \lambda [g_{R}]^\dagger\,.\label{eq:constr0b}
\end{align}
\end{subequations}
Here, we have used the fact that $(g_R)^2 = \pm 1$ for the kagome lattice.

Next, consider the mean field $u_1$ on the first-neighbor link. This link only has the reflection symmetry $\sigma R$, exchanging sublattice sites 1 and 2. Therefore, the constraint on the first-neighbor field $u_1$ is
\begin{equation}\label{eq:constr1}
  [u_1]^\dagger = (-)^{\tau_\sigma + \tau_R} (g_{\sigma} g_{R}) u_1 [g_{\sigma} g_{R}]^\dagger\, .
\end{equation}
The second-neighbor link in Fig.~\ref{fig:kagsym} has the symmetry $\sigma$ exchanging sublattice sites $1$ and $3$. Therefore, the constraint on the ansatz is
\begin{equation}\label{eq:constr2}
  [u_2]^\dagger = (-)^{\tau_\sigma} g_{\sigma} u_2 [g_{\sigma}]^\dagger\, .
\end{equation}
Finally, the diagonal link across the hexagon has reflection symmetry $\sigma$, and $R^3$ exchanging sites. The symmetry constraints on $u_3$ are therefore
\begin{subequations}\label{eq:constr3}\begin{align}
  u_3 &= (\epsilon_2) (-)^{\tau_\sigma} g_{\sigma} u_3 [g_{\sigma}]^\dagger\,,\label{eq:constr3a}\\
  [u_3]^\dagger &= (\epsilon_2) (-)^{\tau_R} g_{R} u_3 [g_{R}]^\dagger\,.\label{eq:constr3b}
\end{align}
\end{subequations}

Note that Eq.~\eqref{eq:constr0b} can be used to propagate the on-site field $\lambda$ to the other sublattice sites. Furthermore, the translation representations \eqref{eq:g12} do not affect the on-site field. Therefore, $\lambda$ is uniform in the chosen gauge. Similar to the on-site field in \eqref{eq:constr0b}, Eq.~\eqref{eq:constr3b} can be used to propagate $u_3$ by rotation.

We solve Eqs.~\eqref{eq:constr0} through \eqref{eq:constr3} for each of the 10 PSG representations and time-reversal signatures $\tau_\sigma$, $\tau_R$. The solutions for $u$ can be further simplified by choosing appropriate global gauges. Propagating the fields to the lattice as shown in Fig.~\ref{fig:dcellkag}, it turns out that some of the resulting ans\"atze $u$ are merely special (limiting) cases of others. This may be due to the limited range of mean fields we are taking into account. Here we are only concerned with ans\"atze $u$ that are distinct (i.e., gauge inequivalent) on the first three neighbors.

\section{U(1) flux operator}\label{app:U1flux}

To better understand the relation of the SU(2) gauge flux operator discussed in Sec.~\ref{sec:Pop} of the main text with the U(1) flux introduced in Ref.~[\onlinecite{WWZ89_PRB_39_11413}], let us start by discussing this approach. In the U(1) formalism, we only consider the U(1) subgroup of the local SU(2) symmetry,
\begin{equation}
  f_\alpha\mapsto e^{i\varphi} f_\alpha\,,
\end{equation}
and we disregard the particle-hole symmetry. A gauge and spin-rotation invariant loop operator is
\begin{equation}
  \hat P = \chi_{12}\chi_{23}\ldots\chi_{q1}\,,
\end{equation}
where $\chi_{ij} = {\bm f}_i^\dagger{\bm f}_j$ are singlet spinon hopping operators. Pairing terms $\eta_{ij} = {\bm f}^T_i\varepsilon{\bm f}_j$ are not considered in this formalism. In terms of the spinon operators, we have
\begin{equation}
  \hat P = -\text{Tr}[(1 - B_1) B_2\ldots B_q]\,,
\end{equation}
where $B_j$ are the matrices
\begin{equation}
  (B_{\alpha\beta}) = {\bm f} {\bm f}^\dagger = (f_\alpha f_\beta^\dagger)
\end{equation}
and the trace is over spin indices. Using the spin representation \eqref{eq:spinrep}, one finds that
\begin{equation}
  B = (1 - \frac{n}{2})\mathbb{1}_2 - S^a\sigma_a\, .
\end{equation}
Let us define $\overline{S} = S^a\sigma_a$. Enforcing the constraint {\it on every site} of the loop ($n_j = 1$), we find
\begin{equation}
  \hat P = \text{Tr}[(\frac{1}{2} + \overline{S}_1)(\frac{1}{2} - \overline{S}_2)\ldots (\frac{1}{2} - \overline{S}_q)]\,.
\end{equation}
Using $\varepsilon\overline{S}\varepsilon = S^a (\varepsilon \sigma_a\varepsilon) = S^a \sigma_a^* = \underline{S}$, we have
\begin{equation}\label{eq:u1flux}
  \hat P = \text{Tr}[(\frac{1}{2} - \underline{S}_1)(\frac{1}{2} + \underline{S}_2)\ldots (\frac{1}{2} + \underline{S}_q)]\, .
\end{equation}
In fact, without normal ordering, the U(1) flux, Eq.~\eqref{eq:u1flux}, is the same expression as the trace of the SU(2) flux Eq.~\eqref{eq:Pop}, and it depends on the base site of the loop. Note, however, that we have used the constraint on every site to get the U(1) flux \eqref{eq:u1flux}, while this is not necessary for calculating the trace of the SU(2) flux operator.

Similar to our discussion in Sec.~\ref{sec:Pop}, we must be concerned that the U(1) flux operator \eqref{eq:u1flux} depends on the base site of the loop. This is inconsistent with the mean-field flux, $\xi^\text{U(1)}_\mathcal{C} = \xi_{12}\ldots\xi_{q1}$, which is independent of base site. To fix this, we need to normal order the flux operator. Normal ordering has a different effect on the U(1) flux as it has on the SU(2) flux, because the particle number $n_j$ appears on every site in the first case, while it is absent in the latter case. Finally, the normal ordered U(1) flux operator turns out as
\begin{equation}\label{eq:u1fluxno}
  \normOrd{\hat P} = - \text{Tr}[(\underline{S}_1-\frac{1}{2})(\underline{S}_2-\frac{1}{2})\ldots (\underline{S}_q-\frac{1}{2})]\, .
\end{equation}
In terms of classical spins, this expression can be an arbitrary complex number, while the trace of the normal-ordered SU(2) flux, Eq.~\eqref{eq:trPhat}, is purely real or imaginary. These results are entirely consistent with the respective mean-field fluxes.

Note that, up to a sign, the normal-ordered U(1) flux for fermions is also given by the cyclic spin permutation operator $P_{123\ldots q} = P_{12} P_{23}\ldots P_{q 1}$, $P_{ij} = 2 {\bm S}_i\cdot{\bm S}_j + 1/2$. We have
\begin{equation}
  \normOrd{\hat P} = (-)^{q-1}P_{123\ldots q}\, .
\end{equation}
Dropping the sign, this is also the expression for the U(1) flux of bosonic spinons.

Comparing the expressions Eq.~\eqref{eq:trPhat} for the SU(2), and Eq.~\eqref{eq:u1fluxno} for the U(1) flux operator, we see that they are identical for classical spins when $S\gg 1$. For $q=3$, the imaginary part of the U(1) flux coincides with the SU(2) flux (= scalar chirality operator), but in general these expressions are different.


\begin{thebibliography}{208}
\bibitem [{par()}]{partons}%
  \BibitemOpen
\bibinfo {note} {{S}pinons are also called \enquote{partons}, and the formal
  spin fractionalization is referred to as ``parton construction''. This
  terminology makes reference to similar concepts used in the Standard Model of
  elementary particles.}\BibitemShut {Stop}%
\bibitem [{mps()}]{mps_spt}%
  \BibitemOpen
\bibinfo {note} {Conceptually similar, but distinct classification schemes have
  recently been discussed.\cite{ChenGuWen11_PRB.83.035107,
  YeWen13_PRB.87.195128, EssinHermele13_PRB.87.104406, Mesaros13_PRB.87.155115}
  These approaches generally try to go beyond the parton construction, but they
  tend to be less predictive and less adapted for concrete microscopic
  calculations. Still other classification schemes work directly with RVB wave
  functions.\cite{YangHong12_PRL.109.147209,
  Wildeboer12_PRL.109.147208}\EatDot.}\BibitemShut {Stop}%
\bibitem [{bfr()}]{bfrelation}%
  \BibitemOpen
\bibinfo {note} {Recently, it was argued that PSG classes for bosonic and
  fermionic fractionalization are related.\cite{LuChoVishwanath14,
  *ZaletelLuVishwanath15, QiFu15_PRB.91.100401, Yunoki06_PRB.74.014408}
  However, this relation still seems to be ill-understood.}\BibitemShut {Stop}%
\bibitem [{gau()}]{gaugeTransfBoson}%
  \BibitemOpen
\bibinfo {note} {The absence of particle-hole gauge symmetry in the bosonic
  fractionalization scheme implies that the action of time reversal cannot be
  factorised as in the fermionic case discussed here. This leads to some
  additional complications in the classification of bosonic chiral spin
  liquids.\cite{Messio13_PRB_87_125127}\EatDot.}\BibitemShut {Stop}%
\bibitem [{sum()}]{sumConvention}%
  \BibitemOpen
\bibinfo {note} {Repeated indices are always summed over in this
  paper.}\BibitemShut {Stop}%
\bibitem [{tri()}]{tripletU}%
  \BibitemOpen
\bibinfo {note} {Note that many of our results (in particular the
  classification of algebraic PSG) are also valid in the more general setting
  of complex $u^\mu_{ij}$ for triplet states.}\BibitemShut {Stop}%
\bibitem [{flu()}]{fluxLambda}%
  \BibitemOpen
\bibinfo {note} {The on-site field ${\bm \lambda}_j$ may be understood as a
  single-site SU(2) gauge flux. A nonzero on-site field inevitably reduces the
  IGG$_u$ on that site to U(1) rotations around the direction of the
  field.}\BibitemShut {Stop}%
\bibitem [{u1l()}]{u1liqsf}%
  \BibitemOpen
\bibinfo {note} {Here, we should note that the surprising possibility of a
  nontrivial flux through even-site loops in a time-reversal symmetric QSL is
  only possible when all SU(2) flux matrices on such loops commute. This point
  will be discussed in Sec.~\ref{sec:TRconstraint}.}\BibitemShut {Stop}%
\bibitem [{gTc()}]{gTchoice}%
  \BibitemOpen
\bibinfo {note} {For symmetric liquids it is well
  known\cite{Wen02_PRB.65.165113} that a nontrivial ansatz can only be obtained
  with a time-reversal representation that satisfies $(g_\tr)^2 = -1$. This is
  consistent with our initial gauge choice in Eq.~\eqref{eq:TRbare}, and the
  fact that time reversal must square to $-1$ on spin $S=1/2$
  particles.}\BibitemShut {Stop}%
\bibitem [{lan()}]{landauGauge}%
  \BibitemOpen
\bibinfo {note} {In analogy with the quantum Hall effect, the choice $g_\opx =
  \mathbb{1}_2$ may be called a ``Landau gauge''.}\BibitemShut {Stop}%
\bibitem [{tgu()}]{tgular_ineq}%
  \BibitemOpen
\bibinfo {note} {A simple way to verify that all triangular PSGs in
  Table~\ref{tab:PSGtrg} are inequivalent is to construct corresponding
  ans\"atze as described in Sec.~\ref{sec:invPSGgen} (e.g., in the symmetric
  case). The gauge-invariant SU(2) flux of the ansatz through the diamond
  plaquette can be used to differenciate PSG~2 and PSG~6, or PSG~5 and PSG~7.
  (See Sec.~\ref{sec:invPSGtrg} and Appendix~\ref{app:trgPSG})}\BibitemShut
  {NoStop}%
\bibitem [{mam()}]{mambriniNote}%
  \BibitemOpen
\bibinfo {note} {Spinon unit cell doubling, and the resulting sensitivity of
  lattice rotation depending on whether the linear system size is an even or
  odd multiple of two, may explain certain features observed in exact spectra
  of the triangular-lattice multiple spin exchange Heisenberg
  model.\cite{MisguichMambrini02_EPJB26} It could indicate a Kalmeyer-Laughlin
  CSL that emerges from the classical tetrahedra
  state.\cite{Messio11_PRB_83_184401} Further work in this direction would be
  interesting.}\BibitemShut {Stop}%
\bibitem [{kag()}]{kagSymPSG}%
  \BibitemOpen
\bibinfo {note} {If time reversal is added to the space group of the kagome
  lattice, the number of representation classes is doubled to 20 (symmetric
  PSG), consistent with Ref.~[\onlinecite{LuRanLee11_PRB.83.224413}]. Note that
  these authors use quite complicated (noncanoncal) gauges where calculations
  are less transparent. Among the ans\"atze Nos.~1 -- 8 in
  Table~\ref{tab:invPSGkagZ2}, only No.~6 is sensitive to the choice in
  time-reversal representation $g_\tr$.}\BibitemShut {Stop}%
\bibitem [{her()}]{hermanns}%
  \BibitemOpen
\bibinfo {note} {After completion of this work, we learned about a discussion
  of projective symmetry constraints on the spinon spectrum of
  three-dimensional Kitaev-Majorana spin liquids in
  Ref.~[\onlinecite{HermannsTrebst16_PRB.93.085101}], ressembling our
  discussion in Sec.~\ref{sect:spectra}.}\BibitemShut {Stop}%
\end{thebibliography}
\end{document}